\documentclass[11pt,a4paper]{article}
\usepackage{jheppub}
\pdfoutput=1
\usepackage[utf8]{inputenc}
\usepackage{dsfont}
\usepackage{IEEEtrantools}
\usepackage{amsfonts}
\usepackage{amsmath}
\allowdisplaybreaks[4]       
\usepackage{tensor} 
\usepackage{amssymb}
\usepackage{euscript}     
\usepackage{braket}
\usepackage{starfont}
     
\usepackage{tensor}        
\usepackage{amsthm}
\usepackage{graphicx}
\usepackage{slashed}
\usepackage{leftidx}
\usepackage{subfigure}
\usepackage{bbm}
\usepackage{color,soul}    
\definecolor{outerspace}{rgb}{0.25, 0.29, 0.3}
\definecolor{scarlet}{rgb}{1.0, 0.13, 0.0}
\usepackage[header,title,page,titletoc]{appendix}  
\definecolor{princetonorange}{rgb}{1.0, 0.56, 0.0}
\definecolor{WildStrawberry}{rgb}{1.0, 0.26, 0.64}
\definecolor{rossocorsa}{rgb}{0.83, 0.0, 0.0}
\definecolor{navyblue}{rgb}{0.0, 0.0, 0.5}
\usepackage[numbers,sort&compress]{natbib}  
\usepackage{float}

\usepackage{mathtools}

\def \hd {\hat{\partial}}
\def \D2R {\left( \frac{8 \partial^2 \mathcal{L}_E}{\partial {\rm Riem}^2} K^2 \right)}

\DeclareMathAlphabet{\pazocal}{OMS}{zplm}{m}{n}
\newcommand{\req}[1]{(\ref{#1})} %{Eq.\thinspace(\ref{#1})}  

\newcommand{\bea}{\begin{eqnarray}}
\newcommand{\diff}{\mathrm{d}}
\newcommand{\eea}{\end{eqnarray}}
\newcommand{\ba}{\begin{eqnarray}}
\newcommand{\ea}{\end{eqnarray}}

\newcommand{\be}{\begin{equation}}
\newcommand{\ee}{\end{equation} }
\newcommand{\beqa}{\begin{eqnarray}}
\newcommand{\eeqa}{\end{eqnarray}}
\newcommand{\beqar}{\begin{eqnarray*}}
\newcommand{\eeqar}{\end{eqnarray*}}

\renewcommand{\req}[1]{eq.~(\ref{#1})}
\def \bz {\bar{z}}
\def \Ls {L_{\star}}
\def \rd {\mathrm{d}}

\newcommand{\ssc}{\scriptscriptstyle}
\newcommand{\eg}{{\it e.g.,}\ }
\newcommand{\ie}{{\it i.e.,}\ }

\newcommand{\see}{S_{\ssc \rm EE}}

\newcommand{\ctt}{C_{\ssc T}}

\DeclareMathOperator{\tr}{tr}

\makeatletter
\newcommand{\dal}{\mathop{\mathpalette\dal@\relax}}
\newcommand{\dal@}[2]{%
  \begingroup
  \sbox\z@{$\m@th#1\square$}%
  \dimen0=\fontdimen8
    \ifx#1\displaystyle\textfont\else
    \ifx#1\textstyle\textfont\else
    \ifx#1\scriptstyle\scriptfont\else
    \scriptscriptfont\fi\fi\fi3
  \makebox[\wd\z@]{%
    \hbox to \ht\z@{%
      \vrule width \dimen0
      \kern-\dimen0
      \vbox to \ht\z@{
        \hrule height \dimen0 width \ht\z@
        \vss
        \hrule height 2\dimen0
      }%
      \kern-2.5\dimen0
      \vrule width 2.5\dimen0
    }%
  }%
  \endgroup
}
\makeatother

\usepackage{hyperref}
\hypersetup{
    colorlinks,
    citecolor=rossocorsa,
    filecolor=navyblue,
    linkcolor=navyblue,
    urlcolor=navyblue
}

\title{Holographic entanglement entropy\\ for perturbative higher-curvature gravities}
\author[{\text{\Zeus}}]{Pablo Bueno,}
\author[{\text{\Poseidon}}]{Joan Camps}
\author[{\text{\Apollon}},{\text{\Vulkanus}}]{and Alejandro Vilar L\'opez}
\affiliation[{\text{\Zeus}}]{Instituto Balseiro, Centro At\'omico Bariloche,\\
8400-S.C. de Bariloche, R\'io Negro, Argentina}
\affiliation[{\text{\Poseidon}}]{Department of Physics and Astronomy, University College London,\\
Gower Street, London WC1E 6BT, United Kingdom}
\affiliation[{\text{\Apollon}}]{Departamento de F\'isica de Part\'iculas, Universidade de Santiago de Compostela,\\
E-15782 Santiago de Compostela, Spain}
\affiliation[{\text{\Vulkanus}}]{Instituto Galego de F\'isica de Altas Enerx\'ias (IGFAE), Universidade de Santiago de Compostela,\\
E-15782 Santiago de Compostela, Spain}
\emailAdd{pablo.bueno@cab.cnea.gov.ar}
\emailAdd{alejandrovilar.lopez@usc.es}
\abstract{%
The holographic entanglement entropy functional for higher-curvature gravities involves a weighted sum whose evaluation, beyond quadratic order, requires a complicated theory-dependent splitting of the Riemann tensor components. Using the splittings of general relativity one can obtain unambiguous formulas perturbatively valid for general higher-curvature gravities. Within this setup, we perform a novel rewriting of the functional which gets rid of the weighted sum. The formula is particularly neat for general cubic and quartic theories, and we use it to explicitly evaluate the corresponding functionals. In the case of Lovelock theories, we find that the anomaly term can be written in terms of the exponential of a differential operator. We also show that order-$n$ densities involving $n_R$ Riemann tensors (combined with $n-n_R$ Ricci's) give rise to terms with up to $2n_R-2$ extrinsic curvatures. In particular, densities built from arbitrary Ricci curvatures combined with zero or one Riemann tensors have no anomaly term in their functionals. Finally, we apply our results for cubic gravities to the evaluation of universal terms coming from various symmetric regions in general dimensions. In particular, we show that the universal function characteristic of corner regions in $d=3$  gets modified in its functional dependence on the opening angle with respect to the Einstein gravity result.

}
\begin{document}
\maketitle

%\begin{titlepage}

%\begin{center}

%\phantom{ }
\vspace{3cm}

\section{Introduction}
\label{sec:Introduction}

%\comment{Somewhere explain that by ``universal'' we mean cutoff independent}

%\subsection{Conventions etc.}
%\begin{equation}
%\bot_{\mu \nu}= n_{\mu}^{a} n_{\nu}^{b} g_{ab}\, , \quad \epsilon_{\mu\nu}=  n_{\mu}^{a} n_{\nu}^{b} \epsilon_{ab}\, , \quad \epsilon_{\mu\nu} \epsilon_{\rho \sigma} =2  \bot_{\mu [\rho} \bot_{\nu | \sigma]} \, , \quad %\quad g^{\mu \nu} n_{\mu}^a n_{\nu}^b= g^{ab}
%g^{\mu \rho }\epsilon_{\mu\nu}\epsilon_{\rho\sigma}=\bot_{\nu \sigma}
%\, , \quad \epsilon_{\mu\nu}\epsilon^{\mu\nu}=2
%\end{equation}
%\begin{equation}
%T^a_{\, \, \, a} \equiv T^{\mu\nu} \bot_{\mu \nu}
%\end{equation}

In effective (super)gravity actions, higher-curvature terms appear as stringy and/or quantum corrections to the corresponding two-derivative actions ---see \eg \cite{Grisaru:1986px,Gross:1986iv,Gubser:1998nz}. In the AdS/CFT context \cite{Maldacena,Witten,Gubser}, the holographic duals of such modified actions are inequivalent to the ones defined by Einstein gravity (\eg the trace anomaly coefficients in four dimensions, $a$ and $c$, no longer coincide in general).
This extends beyond explicit top-down constructions and, in fact, particular higher-curvature models ---\eg with certain special properties which makes them more appealing--- can be used to probe interesting CFT physics \cite{Buchel:2009sk,Myers:2010jv,HoloECG,Camanho:2013pda,deBoer:2009gx}. In some cases, this approach has been used to unveil universal properties valid for completely general CFTs \cite{Myers:2010xs,Myers:2010tj,Kats:2007mq,Brigante:2007nu,Camanho:2010ru,Mezei:2014zla,Bueno1,Miao:2015dua,Bueno:2018yzo,Bueno:2020odt}.

An important entry in the holographic dictionary corresponds to entanglement entropy (EE), which for holographic theories dual to Einstein gravity (plus possible additional matter fields) can be computed using the Ryu-Takayanagi (RT) prescription \cite{Ryu:2006bv,Ryu:2006ef}. According to this, the EE for a region $A$ in the boundary CFT
is obtained as the area of the  bulk surface, $\Gamma_A$, which has the smallest area amongst all bulk surfaces which are homologous to $A$, divided by $4G$, \ie
\begin{equation}\label{RTaka}
S_{\rm \ssc HEE}^{\rm E} (A) = \frac{\mathcal{A} (\Gamma_A)}{4G}\, ,
\end{equation}
where  the ``E'' stands for Einstein gravity.
When the action includes higher-curvature terms, the area functional needs to be modified, similarly to the way the Bekenstein-Hawking black hole entropy formula \cite{Bekenstein:1973ur,Hawking:1974sw} is replaced by Wald's one \cite{Wald:1993nt,Iyer:1994ys}. The naive modification which would correspond to replacing \req{RTaka} by the same Wald functional fails for entanglement entropy \cite{Hung:2011xb}, and additional terms involving extrinsic curvatures of the generalized bulk surface are required. A hint of this is the fact that for Lovelock gravities, the result obtained from Wald's entropy differs from the alternative Jacobson-Myers functional \cite{Jacobson:1993xs} by terms of that type, which generically vanish for Killing horizons, but not for holographic entangling surfaces. The right expression for the holographic entanglement entropy (HEE) functional in the case of quadratic gravities was obtained in \cite{Fursaev:2013fta}. Building up on the generalized entropy methods of \cite{Lewkowycz:2013nqa}, a general formula (in principle) valid for theories involving arbitrary contractions of Riemann tensors and metrics was obtained in \cite{Dong:2013qoa,Camps:2013zua}. Schematically, it  has the form
\begin{equation}\label{dongf}
S^{\mathcal{L}_E({\rm Riemann})}_{\rm  \ssc HEE}(A)=S_{\rm   Wald} + S_{\rm   Anomaly}\, ,
\end{equation}
where, in addition to a Wald-like piece, there appears an extra ``anomaly'' term involving extrinsic curvatures of the generalized holographic surface. In adapted coordinates ---see subsection \ref{sec:Notation} below for our conventions--- these two terms read\footnote{Interesting additional developments and explorations include \cite{Bhattacharyya:2013gra,Bhattacharyya:2013jma,Bhattacharyya:2014yga,Chen:2013qma,Dong:2015zba,Harper:2018sdd,Huang:2015zua}.}\footnote{Generalizations of \req{dongf} to the case in which covariant derivatives of the Riemann appear in the action have also been presented \cite{Miao:2014nxa}.}
\begin{align}\label{waldano}
S_{\rm   Wald}&= 2 \pi \int_{\Gamma_A} \diff^{d-1} y \, \sqrt{h} \,  \frac{\partial \mathcal{L}_E}{\partial R_{z \bz z \bz}}\, , \\ \label{anomaly} S_{\rm   Anomaly}&= 2 \pi \int_{\Gamma_A} \diff^{d-1} y \, \sqrt{h} \, \sum_{\alpha} \left( \frac{\partial^2 \mathcal{L}_E}{\partial R_{z i z j} \partial R_{\bz k \bz l}} \right)_\alpha \frac{8 K_{zij} K_{\bz k l}}{(1+q_\alpha )}\, .
\end{align}
In principle, the generalized holographic surface $\Gamma_A$ should be obtained by extremizing the new functional \cite{Dong:2017xht}.
%Although not too complicated at first sight, the anomaly piece contains 
In the anomaly term, once the second derivative is performed, each of the Riemann tensor components appearing in the resulting expression has to be split into sums of pieces with different weights $q_{\alpha}$ according to some prescription.  That prescription depends on the way the conical defect appearing near the entangling region in the replica trick approach is regulated.   As observed and studied in \cite{Miao:2014nxa,Camps:2014voa,Miao:2015iba,Camps:2016gfs}, this procedure is non-unique, which leads to the so-called ``splitting problem''.\footnote{See subsection \ref{grsplit} for a more detailed summary of the discussion included in this paragraph and the following two.} While the choice of splittings does not affect $f(R)$, Lovelock or quadratic theories, it does play a crucial r\^ole for general theories involving $n\geq 3$ densities.\footnote{
The final form of the anomaly term once  the (some) splitting procedure and the sum over $\alpha$ are performed differs considerably from \req{anomaly}. In particular, for $n$-order densities, it may contain terms involving up to $2(n-1)$ extrinsic curvatures. This is evident from our new expression in \req{NewFunctional:FinalForm3}.}

The right splittings could in principle be identified for each particular theory by imposing that the relevant bulk geometry satisfies the corresponding equations of motion. In doing so, one would be left with a functional  ready to extremize, and the resulting on-shell evaluation would yield non-perturbative results for the HEE of the corresponding theory. Doing this in practice is a highly non-trivial task which has not been pursued for explicit higher-curvature theories so far. If one followed this approach, another relevant issue would arise. %when considering the functionals associated to higher-curvature gravities with non-perturbative couplings. 
For generic higher-curvature theories, the equations of motion implementing the extremization of the functional are not  second-order in derivatives, so it is not completely clear how to deal with the associated boundary value problem in those cases.

A different approach, which we follow here, entails considering holographic entangling surfaces which extremize the RT functional  (\ref{RTaka}) along with the splittings prescribed by Einstein gravity. By doing so, we avoid the boundary-value-problem issues associated to higher-order equations, and the results obtained are perturbatively valid at leading order in the higher-curvature couplings \cite{Camps:2016gfs}. Within this framework, we manage to get rid of the $\alpha$ sum in the anomaly piece (\ref{anomaly}) and obtain a general expression which can be compactly written as\footnote{See Section \ref{covid} for the covariant form.}
\begin{align} \label{NewFunctional:FinalForm3}
&S_{\rm   Anomaly}=32\pi \int_{\Gamma_A} \diff^{d-1} y \, \sqrt{h} \left[ \int_0^1 {\rm d}u \, u\,  {\rm e}^{-F(u)}  \left(\frac{\partial^2 \mathcal{L}_E}{\partial R_{z i z j} \partial R_{\bz k \bz l}}  K_{zij} K_{\bz k l} \right) \right] \, , \quad 
\end{align}
where the operator appearing in the exponential takes the form\footnote{As explained later, there is a normal  ordering prescription implicit in this expression which forces derivatives to act exclusively on the object in the parentheses ---see \req{NewFunctional:NormalOrdering} below.}
\begin{equation}
F(u)\equiv [(1-u^2)\mathcal{K}\indices{_{AI}} \hd^{AI}+ (1- u)\mathcal{K}\indices{_{BJ}} \hd^{BJ}]\, .
\end{equation}
and where $\mathcal{K}\indices{_{AI}} \hd^{AI}$ and $\mathcal{K}\indices{_{BI}} \hd^{BI}$ are differential operators involving derivatives with respect to particular Riemann tensor components contracted with extrinsic and Riemann curvature components. They appear defined in \req{NewFunctional:OperatorA} and \req{NewFunctional:OperatorB} respectively.  This new form of the functional becomes particularly simple for cubic and quartic theories ---see \req{AnomalyCubic:ExpandedExpression} and \req{AnomalyQuartic:ExpandedExpression} respectively--- and we use it to evaluate the explicit (covariant) HEE functionals for all cubic and quartic densities. The result for Lovelock theories is also rather suggestive ---see \req{OneComponent:LovelockExponential}.  Using our new results, we are also able to show that densities constructed exclusively from Ricci curvatures have a vanishing anomaly term, similarly to the well-known case of $f(R)$ gravities. This also extends to densities involving a single Riemann tensor contracted with Ricci curvatures. More generally, we prove that an order-$n$ density involving $n_R$ Riemann curvatures and $n-n_R$ Ricci curvatures can produce HEE functionals containing at most $2(n_R-1)$ extrinsic curvatures.  As an application of our results, we compute a variety of universal contributions to the EE coming from various symmetric regions in general dimensions for holographic theories dual to cubic gravities. Particularly interesting are the results for strips, for which no alternative interpretation of their coefficients exists beyond EE, and for corners, for which the functional form of the Einstein gravity function only starts to get modified at cubic order.
 %\comment{more last section}
 
The remainder  of the    paper goes as    follows. In subsection \ref{sec:Notation} we introduce our conventions and some notation. In Section \ref{grsplit}  we briefly review the construction that leads to the general form of the holographic entanglement entropy functional, the issue with the Riemann tensor splittings and the choice that allows us to obtain results perturbatively valid  for general higher-curvature theories. In Section \ref{sec:Rewriting} we derive a new formula for the anomaly piece of the HEE functional valid for perturbative higher-curvature corrections to Einstein gravity. We show how the formula gets considerably simplified in the cases of cubic, quartic and Lovelock densities. We also illustrate how our formula should be used in concrete cases by performing a detailed example for a term coming from quintic densities, verifying the match with the $\alpha$-expansion method. In Section \ref{sec:explicit} we present the explicit form of the HEE functionals for general: $f(R)$, Lovelock, quadratic, cubic, quartic, $\mathcal{L}({\rm Ricci})$ and $R_{\mu\nu\rho\sigma}T^{\mu\nu\rho\sigma}({\rm Ricci})$ densities  in covariant form. We also prove here that the functionals corresponding to densities involving $n-n_R$ Ricci tensors contain at most $2(n_R-1)$ extrinsic curvatures.
In Section \ref{unite} we evaluate, for general quadratic and cubic theories, the universal entanglement entropy coefficients characterizing spheres and strips in general dimensions, cylinders in $d=4$ and $d=6$ and corners in $d=3$. For the latter, we show that the functional dependence on the opening angle of the corner gets modified by the introduction of cubic densities with respect to the Einstein gravity result. We perform some comparisons of the result with free fields calculations, strengthening previously observed universal properties of this function. We conclude in Section \ref{finalc} with some final comments and directions.
Appendix \ref{formuls} contains the proof of a couple of identities which we use in our derivation of the new functional in Section \ref{sec:Rewriting}.

\subsection{Notation and conventions}
\label{sec:Notation}
%%%%%%%%%%%%%%%%%%%%%%

In the present paper we deal with various manifolds and metrics. Here we make some comments on our conventions and notation. We take indices in the $(d+1)$-dimensional bulk to be $\mu, \nu, \dots$, and the bulk metric is denoted by $g\indices{_{\mu \nu}}$. The entanglement entropy of a boundary region $A$ is computed as the integral of the entanglement functional on a spatial codimension-2 bulk surface homologous to $A$, which we call $\Gamma_A$. The induced metric on this surface is written as $h\indices{_{\mu \nu}}$, and we will often have to deal with its extrinsic curvature, $K\indices{^a_{\mu \nu}}$. This is defined considering two orthonormal vectors to the surface $n\indices{_a^{\mu}}$, where indices $a, b, \dots$ take values 1 and 2:
\begin{equation} \label{Notation:extrinsic_curvature}
K\indices{^a_{\mu \nu}} \equiv h\indices*{^{\rho}_{\mu}} h\indices*{^{\sigma}_{\nu}} \nabla\indices{_{\rho}} n\indices{^a_{\sigma}} \, , 
\end{equation}
and we assume an arbitrary extension of $n\indices{^a_{\mu}}$ to a neighborhood of the surface which keeps them normalized. Notice also that we work in Euclidean signature, which means $g\indices{_{\mu \nu}} n\indices{_a^{\mu}} n\indices{_b^{\nu}} = \delta\indices{_{a b}}$, and we define $n\indices{^a_{\mu}} = \delta\indices{^{a b}} g\indices{_{\mu \nu}} n\indices{_b^{\nu}}$. In particular, the induced metric can be written as 
\begin{equation} \label{Notation:inducedmetric}
h\indices{_{\mu \nu}} = g\indices{_{\mu \nu}} - n\indices{^a_{\mu}} n\indices{_{a \nu}}\, .
\end{equation}
We also introduce projectors 
\begin{equation} \label{Notation:projectors}
t_i^{\mu}\equiv \frac{ \partial x^{\mu}}{\partial y^i}\, ,
\end{equation}
where indices $i, j, \dots$ denote the tangent directions to the surface. Tensors with these kind of indices are always obtained by application of such projectors to their corresponding bulk tensors, \eg
\begin{equation}
h_{ij} \equiv t_i^{\mu} t_j^{\nu} h_{\mu \nu}\, , \quad \quad K_{ij}^a\equiv   t_i^{\mu} t_j^{\nu} K_{\mu\nu}^a \, .
\end{equation}
We also define the binormal to the surface and the normal projector, respectively, as
\begin{equation} \label{Covariant:BinormalNormalDefinitions}
\epsilon\indices{_{\mu \nu}} \equiv \epsilon\indices{_{a b}} n\indices{^a_{\mu}} n\indices{^b_{\nu}}  ~, \qquad \perp\indices{_{\mu \nu}} \equiv \delta\indices{_{a b}} n\indices{^a_{\mu}} n\indices{^b_{\nu}}  \, ,
\end{equation}
where $\epsilon\indices{_{a b}}$ is the two-dimensional Levi-Civita symbol. In particular, this means that when indices $a,b,\dots$ appear repeated in a tensorial structure the corresponding bulk tensor is contracted with the normal projector, namely 
\begin{equation}
V^a_{\, \, \, a} \equiv V^{\mu\nu} \bot_{\mu \nu}\, .
\end{equation}
The binormal and the normal projector satisfy the useful relations
\begin{equation}\label{relss}
\epsilon_{\mu\nu} \epsilon_{\rho \sigma} =2  \bot_{\mu [\rho} \bot_{\nu | \sigma]} \, ,\quad \quad %\quad g^{\mu \nu} n_{\mu}^a n_{\nu}^b= g^{ab}
g^{\mu \rho }\epsilon_{\mu\nu}\epsilon_{\rho\sigma}=\bot_{\nu \sigma}
\, , \quad \quad \epsilon_{\mu\nu}\epsilon^{\mu\nu}=2\, .
\end{equation}

When performing generic computations of the entanglement functional we  follow the conventions of \cite{Dong:2013qoa,Camps:2016gfs}. This means that we take a particular set of adapted coordinates for $\Gamma_A$ so that
\begin{equation} \label{Notation:adapted_metric}
\rd s^2 = \rd z \, \rd \bz + h\indices{_{i j}} \rd y^i \, \rd y^j  \, ,
\end{equation}
where $z \equiv \rho e^{i \tau}$, $\bz \equiv  \rho e^{-i \tau}$ are complex coordinates orthogonal to the surface. In these coordinates, the off-diagonal components $g\indices{_{z \bz}} = 1/2$ and $g\indices{^{z \bz}} = 2$ are the only non-vanishing part of the normal metric to the surface. %Indices $i, j, \dots$ are used to denote the tangent directions to the surface.

We take the cosmological constant to be negative throughout the paper, and write $-2\Lambda \equiv d(d-1)/L^2$, so that the action scale $L$ coincides with the AdS$_{d+1}$ radius, which we denote by $\Ls$, for Einstein gravity. For generic higher-curvature gravities, the equation which relates $L$ and $\Ls$ involves the corresponding higher-order couplings (it appears in \req{k00} below). Nevertheless, at leading order in the couplings ---which is the setup we consider here--- the two scales are equal to each other, $L=\Ls+\mathcal{O}(\alpha_i)$. We choose to present the results (mostly in Section \ref{unite}) in terms of the AdS radius $\Ls$.

\section{GR  splittings for perturbative higher-curvature theories }\label{grsplit}

%\comment{to be edited:}

Let us start considering the entanglement entropy for a region $A$ in some global state $\rho$ of some holographic CFT. This can be obtained as the $n \to 1$ limit of R\'enyi entropies $S_n(A)$, which in turn can be obtained via the replica trick as
\begin{equation} \label{SplittingProblem:ReplicaTrick}
S_n(A) = - \frac{1}{n - 1} \log {\rm Tr} \left( \rho_A^n \right) = - \frac{1}{n - 1} \left( \log \mathcal{Z}_n - n  \log \mathcal{Z}_1 \right) \, .
\end{equation}
In this expression, $n$ is a positive integer, $\rho_A$ is the reduced density matrix of region $A$, and $\mathcal{Z}_n$ is the partition function of the field theory in the $n$-fold cover. %This is an Euclidean manifold made out of $n$ copies of the original one which are sewn to each other along the region $A$ in a cyclical manner. 
In particular, $\mathcal{Z}_1$ is the partition function of the Euclidean manifold which, upon path integration, prepares the global state. In order to obtain the entanglement entropy $\see(A)$ as the limit $n \to 1$ of the previous expression, an analytic continuation in $n$ is also needed.

Following the argument of \cite{Lewkowycz:2013nqa}, when the field theory has a  gravity dual, in the saddle-point approximation it is possible to identify $\log \mathcal{Z}_n = - I_E[B_n]$, where $I_E[B_n]$ is the Euclidean action of the gravitational theory evaluated at the bulk solution $B_n$ which is dual to the $n$-fold cover. This boundary geometry has a $\mathbb{Z}_n$ symmetry which interchanges the $n$ copies and, if this is respected in the bulk, we can consider the quotient $\hat{B}_n = B_n / \mathbb{Z}_n$, which is regular everywhere except at the codimension-2 bulk surface $\mathcal{C}_n$ consisting of the fixed points of $\mathbb{Z}_n$. Furthermore, the replica symmetry also guarantees that
\begin{equation} \label{SplittingProblem:ReplicaAction}
I_E[B_n] = n I_E[\hat{B}_n] \, .
\end{equation} 
%
%where in $I_E[\hat{B}_n]$ we exclude contributions from the conical singularity $\mathcal{C}_n$ (since there are no such contributions in the left-hand side). 
We can now analytically continue to non-integer $n$ this construction, and obtain the entanglement entropy as:
\begin{equation} \label{SplittingProblem:EntanglementFromAction}
S_{\rm \ssc HEE}(A) = \lim_{n \to 1} \frac{n}{n - 1} \left( I_E[\hat{B}_n] - I_E[B_1] \right) = \left. \partial_n I_E[\hat{B}_n] \right|_{n = 1} \, .
\end{equation}
Since $I_E[B_1]$ is a bulk solution to the equations of motion, this variation away from $n = 1$ might seem to vanish. This is not the case because when we vary $n$ we change the opening angle of the conical defect at $\mathcal{C}_n$, and this region has to be excluded from the action integral, introducing a boundary where conditions change with $n$. Details of this procedure can be found \eg in \cite{Dong:2013qoa}. The relevant fact is that the computation of the entanglement entropy gets reduced to the evaluation of the on-shell Euclidean action of the gravitational theory in the conical defect $\mathcal{C}_n$. The opening angle of this defect is $2 \pi /n$, and after obtaining the contributions to the action we must take an $n$-derivative at $n = 1$.

In order to compute $S_{\rm \ssc HEE}(A)$ we need to evaluate the action of a given gravitational theory for  a bulk geometry which regulates the conical singularity. This is a rather technical task, but there is a key point which  was initially overlooked in \cite{Dong:2013qoa,Camps:2013zua}: there are many ways in which a conical defect can be regulated \cite{Miao:2014nxa,Camps:2014voa,Miao:2015iba,Camps:2016gfs}.  Different prescriptions produce different functionals. This ambiguity is usually called the ``splitting problem''. The particular gravitational theory of interest should determine the correct one  through its equations of motion \cite{Camps:2016gfs,Dong:2017xht}. %This is explained with great level of detail in \cite{Camps:2016gfs}, where it is also shown how one should proceed for a broad family of theories: those which include only perturbative higher curvature corrections to GR. 

When interested in perturbative higher-curvature corrections to Einstein gravity, the appropriate splittings were obtained in \cite{Camps:2016gfs}. At first order in the higher-order couplings one can simply regulate using Einstein's equations. This is so because the particular regularization does not affect the Einstein gravity term in \eqref{SplittingProblem:EntanglementFromAction} (it always produces the usual area law), and the higher curvature terms in the action are already first order in the couplings. As a consequence, corrections to the regulated geometry coming from modifications to the equations of motion are second order in the action.

All in all, the expression for the holographic entanglement entropy for a perturbative higher-curvature gravity with Euclidean action $\mathcal{L}_E(g_{\mu\nu},R_{\mu\nu\rho\sigma})$ is given by
%after having dealt with these issues, \cite{Camps:2016gfs} showed that the holographic entanglement entropy functional in a theory with Euclidean Lagrangian $\mathcal{L}_E \left( R_{\mu \nu \rho \sigma} \right)$ (depending only on contractions of the Riemann tensor) is:
%
\begin{equation} \label{SplittingProblem:GeneralFunctional}
S_{\rm \ssc HEE}(A) = 2 \pi \int_{\Gamma_A} d^{D-2} y \, \sqrt{h} \, \left[ \frac{\partial \mathcal{L}_E}{\partial R_{z \bz z \bz}} + \sum_{\alpha} \left( \frac{\partial^2 \mathcal{L}_E}{\partial R_{z i z j} \partial R_{\bz k \bz l}} \right)_\alpha \frac{8 K_{zij} K_{\bz k l}}{q_\alpha + 1} \right] \, ,
\end{equation}
where $\Gamma_A$ is just the RT surface and the prescription for the $\alpha$-sum is unambiguously determined ---see below. %The fact that we can use an holographic region which extremizes the Einstein gravity functional follows because the area term is stationary for it, which means that first-order variations will not change its value.
The area term in the previous equation ---coming from the Einstein gravity part of the action--- is stationary for the RT surface, and therefore first order variations of the surface will not change its value. On the other hand, contributions of higher-order terms to the previous functional will already be first-order in the couplings, and thus insensitive to first-order modifications of the surface.
%In principle, $\Gamma_A$ is a yet to determine codimension-2 surface but, in the case of a theory which contains first order perturbative corrections to GR, it can be taken to be the RT surface. This is so because the area term in the previous equation, which comes from the GR part of the action, is stationary for the RT surface, and therefore first order variations of the surface will not change its value. On the other hand, contributions of higher order terms to the previous functional will already be first order in the couplings, and thus insensitive to first order modifications of the surface.
%We still have to explain the meaning of the $\alpha$ sum in \eqref{SplittingProblem:GeneralFunctional}. This sum is usually called the ``anomaly term'' of the functional, in contrast with the first part of the expression which is the Wald term (analogous to the one obtained for black hole entropy in higher curvature theories, \cite{Wald:1993nt}). It appears only in theories containing higher curvature corrections to the Einstein-Hilbert action, and it is a direct consequence of the regularization of the conical singularity. 

As we mentioned before, there are in principle different ways to regulate the conical singularity, which give rise to different prescriptions for the $\alpha$ sum. %so let us present a description of the algorithm hidden in the $\alpha$ sum which is valid in general.\footnote{At least for the two prescriptions present in the literature, those of \cite{Dong:2013qoa} and \cite{Camps:2016gfs}.} 
On general grounds, the idea is the following. The second derivative of the Lagrangian will be a sum of terms which are monomials with different contractions of components of the Riemann tensor. These contractions are to be expanded in terms of their $z$ and $\bz$ indices, obtaining an expression of the second derivative of the Lagrangian involving only $R\indices{_{z \bz z \bz}}$, $R\indices{_{z \bz z i}}$, $R\indices{_{z \bz i j}}$, $R\indices{_{z i \bz j}}$, $R\indices{_{z i z j}}$, $R\indices{_{z i j k}}$, $R\indices{_{i j k l}}$, plus components related to these by complex conjugation of the indices.\footnote{Notice that components of the Ricci tensor and the Ricci scalar have to be expanded in terms of these basic objects as well. For instance, we would write
\begin{equation} \label{SplittingProblem:ExpansionRicci}
R\indices{_{z \bz}} = g\indices{^{\mu \nu}} R\indices{_{z \mu \bz \nu}} = - 2 R\indices{_{z \bz z \bz}} + g\indices{^{i j}} R\indices{_{z i \bz j}} \, .
\end{equation}}
After this is done, each regularization of the conical defect will provide a ``splitting'': a rule to divide each of the previous components of the Riemann tensor schematically as
\begin{equation} \label{NotationRewriting:GeneralSplitting}
R\indices{_{MI}} = \tilde{R}\indices{_{MI}} + \mathcal{K}\indices{_{MI}} \, .
\end{equation}
In this expression, $M$ labels the different components of the Riemann tensor enumerated before, while $I$ is a generalized index containing all the $i, j, k, \dots$ indices of the particular component under consideration (which might be none). This expansion has to be performed in all the components of the Riemann tensor, and once this is done, each of the resulting monomials is labelled by $\alpha$. The splitting provides also a value $q_{\alpha}$ for each $\mathcal{K}\indices{_{MI}}$. In each term we have a definite value of $q_{\alpha}$, given by the sum of the values of all the $\mathcal{K}\indices{_{MI}}$ in that monomial. Expression \eqref{SplittingProblem:GeneralFunctional} instructs us then to divide each term by $q_{\alpha} + 1$. Once this is done, we can eliminate the $\tilde{R}\indices{_{MI}}$ (which are auxiliary objects in this construction whose particular geometrical meaning is irrelevant as far as the functional construction is concerned) in favor of the Riemann tensor components by using \eqref{NotationRewriting:GeneralSplitting} again.

The particular example of \eqref{NotationRewriting:GeneralSplitting} relevant for our purposes comes from the  regularization of the conical defect imposed by Einstein's equations, which is valid for any theory containing perturbative corrections to Einstein gravity in the action. In such a case, the splittings take the form
\begin{align} \label{SplittingProblem:ExpansionsRiemann}
\nonumber R\indices{_{z \bz z \bz}} & = \tilde{R}\indices{_{z \bz z \bz}} - \frac{1}{8} K\indices{^{a i j}} K\indices{_{a i j}}  \, , \\
\nonumber R\indices{_{z \bz ij}} & = \tilde{R}\indices{_{z \bz i j}} - 2  K\indices{_{z [i|}^k} K\indices{_{\bz |j] k}} \, , \\
\nonumber R\indices{_{z i \bz j}} & = \tilde{R}\indices{_{z i \bz j}} - K\indices{_{z i}^k} K\indices{_{\bz j k}} \, , \\
R\indices{_{ijkl}} & = \tilde{R}\indices{_{ijkl}} - 2 K\indices{_{a i [k}} K\indices{^a_{l] j}} \, ,
\end{align} 
with the remaining components having a trivial splitting, \ie $\tilde{R}\indices{_{MI}} = 0$ for them. The values of $q_{\alpha}$ are: $q_{\alpha} = 1$ for any of the previous terms quadratic in extrinsic curvatures, $q_{\alpha} = 1$ for $R\indices{_{z i z j}}$ (and its complex conjugate), and $q_{\alpha} = 1/2$ for $R\indices{_{z i j k}}$ and $R\indices{_{z \bz z i}}$ (and their complex conjugates).

All in all, this complicated procedure is nothing but a way to generate contributions to the holographic entanglement entropy functional containing higher and higher powers of the extrinsic curvature. One of the main results in this paper will consist in reinterpreting  and rewriting this algorithm in a more transparent way, making manifest this generation of terms with an increasing number of powers of $K$.

\section{Rewriting the HEE functional}
\label{sec:Rewriting}
%%%%%%%%%%%%%%%%%%%%%%
In this section we perform a rewriting of the holographic entanglement entropy functional for higher-curvature gravities. We manage to write it completely in terms of explicit contractions of extrinsic curvatures and derivatives with respect to Riemann tensors,  getting rid of the weighted sum over $\alpha$ appearing in the anomaly piece. We do this for the Riemann tensor splittings corresponding to Einstein gravity, which allows us to produce a new general expression valid for arbitrary higher-curvature theories at leading order in the corresponding couplings. %\footnote{As we emphasize later, there is a priori no fundamental obstacle for similar ideas and formulas not to be obtainable for other splitting choices in a similar manner.} 
The structure of the expression is particularly simple for densities up to quartic order in curvature, and we provide new explicit formulas for cubic and quartic theories. Applied to the case of Lovelock theories, our formula for the corresponding anomaly piece can be suggestively written in terms of an exponential of the derivative of the only  component of the Riemann tensor which is relevant in that case, contracted with two extrinsic curvatures. We also perform a hopefully illustrative application of our formulas to a particular monomial coming from putative quintic densities showing how it agrees with the result obtained via the $\alpha$ sum.

%Let us try to develop a little bit more one of the ideas presented in the previous section: the possibility of rewriting the entanglement entropy functional completely in terms of derivatives, without any explicit mention to the $\alpha$-expansion. This is inspired by some results communicated privately by Joan. %{\color{purple}NOTE: Section not ready for the paper, it is only a collection of results which should eventually lead to the final version}.

%%%%%%%%%%%%%%%%%%%%%%
\subsection{Symmetry factors in derivatives and some notation}
\label{subsec:SymmetryFactors}
%%%%%%%%%%%%%%%%%%%%%%
Let us start by making a couple of comments regarding how to take derivatives with respect to Riemann tensor components and introducing some notation which we will be using throughout this section. 
%We will make ubiquitous use of derivatives with respect to curvature components in this section, and it will prove useful to clarify some aspects from the beginning. 

The issues discussed here arise due to the conventional definition of the derivative with respect to the Riemann tensor:
\begin{equation} \label{SymmetryFactors:RiemannDerivative}
\frac{\partial R\indices{_{\mu \nu \rho \sigma}}}{\partial R\indices{_{\alpha \beta \gamma \delta}}} \equiv \frac{1}{2} \left[ \delta\indices*{_{[\mu}^{\alpha}} \delta\indices*{_{\nu]}^{\beta}} \delta\indices*{_{[\rho}^{\gamma}} \delta\indices*{_{\sigma]}^{\delta}} + \delta\indices*{_{[\rho}^{\alpha}} \delta\indices*{_{\sigma]}^{\beta}} \delta\indices*{_{[\mu}^{\gamma}} \delta\indices*{_{\nu]}^{\delta}} \right] \, .
\end{equation}
This definition respects the symmetries of the Riemann tensor and, at the same time, it has the following nice (and expected) property, %
\begin{equation} \label{SymmetryFactors:TaylorFullRiemann}
R\indices{_{\alpha \beta \gamma \delta}} \frac{\partial R\indices{_{\mu \nu \rho \sigma}}}{\partial R\indices{_{\alpha \beta \gamma \delta}}} = R\indices{_{\mu \nu \rho \sigma}} \, ,
\end{equation}
which will be key when performing Taylor-like expansions of functions of the Riemann tensor.

Some care must be taken, however, when singling out specific components of the Riemann tensor. %, as it is the case with the $z$ and $\bz$ ones in the geometric construction that leads to the entanglement entropy functional. 
For instance, using the previous definition one finds
\begin{equation} 
\frac{\partial R\indices{_{z \bz i j}}}{\partial R\indices{_{z \bz k l}}} = \frac{1}{2} \left[ \delta\indices*{_{[z}^{z}} \delta\indices*{_{\bz]}^{\bz}} \delta\indices*{_{[i}^{k}} \delta\indices*{_{j]}^{l}} + \delta\indices*{_{[i}^{z}} \delta\indices*{_{j]}^{\bz}} \delta\indices*{_{[z}^{k}} \delta\indices*{_{\bz]}^{l}} \right] = \frac{1}{4} \delta\indices*{_{[i}^{k}} \delta\indices*{_{j]}^{l}} \, ,
\end{equation}
which leads to
\begin{equation} 
R\indices{_{z \bz k l}} \frac{\partial R\indices{_{z \bz i j}}}{\partial R\indices{_{z \bz k l}}} = \frac{1}{4} R\indices{_{z \bz i j}} \, .
\end{equation}
The factor  $1/4$ arises from the different positions in which we can put the $z$, $\bz$ indices using the symmetries of the Riemann tensor, $R\indices{_{z \bz k l}}$, $R\indices{_{\bz z k l}}$, $R\indices{_{k l z \bz}}$, and $R\indices{_{k l \bz z}}$. Something analogous happens for the rest of components of the Riemann tensor.  Hence, whenever performing Taylor-like expansions in terms of such components we will need to take these extra factors into account. In order to do so, it will prove useful to define a new derivative operator,  $\hat{\partial}$, which already includes them.
%This implies that we need to include extra factors when performing Taylor-like expansions in terms of the components of the Riemann tensor with $z$ and $\bz$ coefficients explicitly isolated. These factors have to do with the different positions in which we can write the $z$, $\bz$ indices by using the symmetries of the Riemann (e.g., 4 in the previous case, $R\indices{_{z \bz k l}}$, $R\indices{_{\bz z k l}}$, $R\indices{_{k l z \bz}}$, and $R\indices{_{k l \bz z}}$). On a practical level, they can be calculated both by counting those symmetries or via computations analogous to the previous one. We will define a new derivative operator, $\hat{\partial}$, which already includes these factors so that we do not have to carry them in our calculations. For completeness, we collect here 
The definitions for the different components read
%
%\begin{align} \label{SymmetryFactors:DerivativeFactorsDef}
%\nonumber \frac{\hd}{\hd R\indices{_{z \bz z \bz}}} \equiv & 4 \frac{\partial}{\partial R\indices{_{z \bz z \bz}}} \, , & \frac{\hd}{\hd R\indices{_{z \bz z i}}} \equiv & 8 \frac{\partial}{\partial R\indices{_{z \bz z i}}} \, , \\
%\nonumber \frac{\hd}{\hd R\indices{_{z \bz i j}}} \equiv & 4 \frac{\partial}{\partial R\indices{_{z \bz i j}}} \, , & \frac{\hd}{\hd R\indices{_{z i z j}}} \equiv & 4 \frac{\partial}{\partial R\indices{_{z i z j}}} \, , \\
%\nonumber \frac{\hd}{\hd R\indices{_{z i \bz j}}} \equiv & 8 \frac{\partial}{\partial R\indices{_{z i \bz j}}} \, , & \frac{\hd}{\hd R\indices{_{z i j k}}} \equiv & 4 \frac{\partial}{\partial R\indices{_{z i j k}}} \, , \\
%\frac{\hd}{\hd R\indices{_{i j k l}}} \equiv & \frac{\partial}{\partial R\indices{_{i j k l}}} \, ,
%\end{align}
%
\begin{align} \label{SymmetryFactors:DerivativeFactorsDef}
 \frac{\hd}{\hd R\indices{_{z \bz z \bz}}} \equiv & 4 \frac{\partial}{\partial R\indices{_{z \bz z \bz}}} \, , \quad \quad \frac{\hd}{\hd R\indices{_{z \bz z i}}} \equiv  8 \frac{\partial}{\partial R\indices{_{z \bz z i}}} \, , \quad \quad
\frac{\hd}{\hd R\indices{_{z \bz i j}}} \equiv  4 \frac{\partial}{\partial R\indices{_{z \bz i j}}} \, , \\  \frac{\hd}{\hd R\indices{_{z i z j}}} \equiv & 4 \frac{\partial}{\partial R\indices{_{z i z j}}} \, , \quad \quad
\frac{\hd}{\hd R\indices{_{z i \bz j}}} \equiv  8 \frac{\partial}{\partial R\indices{_{z i \bz j}}} \, , \quad \quad \frac{\hd}{\hd R\indices{_{z i j k}}} \equiv  4 \frac{\partial}{\partial R\indices{_{z i j k}}} \, , \\
\frac{\hd}{\hd R\indices{_{i j k l}}} \equiv&  \frac{\partial}{\partial R\indices{_{i j k l}}} \, .
\end{align}
The remaining ones can be obtained by complex conjugation.

%%%%%%%%%%%%%%%%%%%%%%
%\subsection{Compact notation for the components}
%\label{subsec:NotationRewriting}
%%%%%%%%%%%%%%%%%%%%%%

%There is some extra notation 
Below we will manipulate expressions involving multiple derivatives with respect to all these components of the Riemann tensor. In order to do that, it is convenient to introduce some notation which allows us to represent them in a compact form. Firstly, let us define 
%Notation can become quite cumbersome if we try to write explicitly every index and component of the Riemann tensor in what follows, so let us try to develop a more compact notation from the beginning. We define 
upper case latin indices $I, J, \dots$ to collect all $i, j, k, \dots$ indices that might appear in a given tensor. %For instance, if we are dealing with $R\indices{_{z \bz i j}}$, an index $I$ would refer to $ij$. 
Similarly, we introduce $M, N, \dots$ indices to represent the different Riemann tensor components involving $z$ and $\bz$ indices. In practice, we just want this notation to perform Taylor expansions, for which the relevant thing to keep in mind is the following compact definition
\begin{align} \label{NotationRewriting:FullTaylorOperator}
\nonumber R\indices{_{MI}} \hd^{MI} \equiv & +R\indices{_{z \bz z \bz}} \frac{\hd}{\hd R\indices{_{z \bz z \bz}}} + R\indices{_{z \bz i j}} \frac{\hd}{\hd R\indices{_{z \bz i j}}} + R\indices{_{z i \bz j}} \frac{\hd}{\hd R\indices{_{z i \bz j}}} + R\indices{_{i j k l}} \frac{\hd}{\hd R\indices{_{i j k l}}} \\
 & + \left[ R\indices{_{z \bz z i}} \frac{\hd}{\hd R\indices{_{z \bz z i}}} + R\indices{_{z i z j}} \frac{\hd}{\hd R\indices{_{z i z j}}} + R\indices{_{z i j k}} \frac{\hd}{\hd R\indices{_{z i j k}}} + {\rm c.c.} \right] \, ,
\end{align}
where c.c. stands for the complex conjugate components of the terms in the parentheses (which are the only ones that have a different number of $z$ and $\bz$ indices). This can be thought of as a sum over $M$ (the $z$ and $\bz$ indices) and then, for each $M$, an extra sum over tangent indices $I$. Note that for $R\indices{_{z \bz z \bz}}$ the second sum does not exist, and in that case $I$ represents an empty set of tangent indices.

As we explained in the previous section, different components of the Riemann tensor have different splitting structures. In general, any component splits as in \req{NotationRewriting:GeneralSplitting} 
%The final thing we need is a small modification of this compact notation, with the purpose of taking into account the different splitting structures that appear in \eqref{SplittingProblem:ExpansionsRiemann}. In general, any Riemann tensor component splits as:
%
%\begin{equation} \label{NotationRewriting:GeneralSplitting}
%R\indices{_{MI}} = \tilde{R}\indices{_{MI}} + \mathcal{K}\indices{_{MI}} ~ ,
%\end{equation}
%
where $\tilde{R}\indices{_{MI}}$ has $q_{\alpha} = 0$ and $\mathcal{K}\indices{_{MI}}$ has $q_{\alpha} \neq 0$. The  $q_{\alpha}$ for the $\mathcal{K}\indices{_{MI}}$ piece can take two values. Components $R_{z \bz z \bz}, R_{z \bz i j}, R_{z i \bz j}, R_{z i z j}, R_{\bz i \bz j}$, and $R_{i j k l}$ have $q_{\alpha} = 1$ for that part, and we will generically refer to them with labels $A, A', \dots$ On the other hand, components $R\indices{_{z i j k}}, R\indices{_{\bz i j k}}, R\indices{_{z \bz z i}}$, and $R\indices{_{\bz z \bz i}}$ have $q_{\alpha} = 1/2$ for the $\mathcal{K}\indices{_{MI}}$ part and we will refer to them with labels $B, B', \dots$. In terms of these, the operator \eqref{NotationRewriting:FullTaylorOperator} splits into two contributions:
\begin{align} \label{NotationRewriting:OperatorsAB}
 R\indices{_{AI}} \hd^{AI} \equiv & + R\indices{_{z \bz z \bz}} \frac{\hd}{\hd R\indices{_{z \bz z \bz}}} + R\indices{_{z \bz i j}} \frac{\hd}{\hd R\indices{_{z \bz i j}}} + R\indices{_{z i \bz j}} \frac{\hd}{\hd R\indices{_{z i \bz j}}} + R\indices{_{i j k l}} \frac{\hd}{\hd R\indices{_{i j k l}}}\\ \notag  & + \left[ R\indices{_{z i z j}} \frac{\hd}{\hd R\indices{_{z i z j}}} + {\rm c.c.} \right] ~ , \\
R\indices{_{BI}} \hd^{BI} \equiv & \left[ R\indices{_{z i j k}} \frac{\hd}{\hd R\indices{_{z i j k}}} + R\indices{_{z \bz z i}} \frac{\hd}{\hd R\indices{_{z \bz z i}}} + {\rm c.c.} \right] \,  .
\end{align}

%%%%%%%%%%%%%%%%%%%%%%
\subsection{New form of the HEE functional}
\label{subsec:NewFunctional}
%%%%%%%%%%%%%%%%%%%%%%
%\comment{HEREEE}
Equipped with this notation, we are ready to start rewriting the anomaly piece in the holographic entanglement entropy functional.
 %Equipped with the previous notation, we can start the actual rewriting of the entanglement entropy functional.
The $\alpha$ expansion appearing in that term is performed on the following object, for which we define the shorthand notation
\begin{equation} \label{NewFunctional:ShorthandSecondDerivative}
\D2R \equiv 8 \frac{\partial^2 \mathcal{L}_E}{\partial R\indices{_{z i z j}} \partial R\indices{_{\bz k \bz l}}} K\indices{_{z i j}} K\indices{_{\bz k l}} \, .
\end{equation}
This object is a complicated expression involving the different Riemann tensor components. Once we have it for a given theory, we have to apply an splitting of the form \eqref{NotationRewriting:GeneralSplitting} to each component, account for the $q_{\alpha}$ value of each monomial and divide it by $(1 + q_{\alpha})$. %We will try to follow these steps in an abstract way using derivatives. 

In order to understand the steps we will follow, %applying the splitting \eqref{NotationRewriting:GeneralSplitting} and counting the value of $q_{\alpha}$ for each monomial, 
it is illustrative to consider first a simplified version of the problem.  Suppose we have some function $f(x)$ and we want to substitute $x = \tilde{x} + k$ in a way such that we explicitly isolate monomials depending on the number of $k$ factors they have. A simple way to do this is to Taylor-expand $f(\tilde x+k)$ around $x=0$, namely,
\begin{equation} \label{NewFunctional:ExpansionFunction0}
f(\tilde{x} + k) = \sum_{n = 0}^{\infty} \frac{1}{n!} (\tilde{x} + k)^n f^{(n)}(0) \, ,
\end{equation}
and then apply the binomial theorem to $(\tilde{x} + k)^n$ to isolate terms with a definite number of $k$ factors. Notice also that, if we wish to avoid evaluating derivatives at $0$, we can also Taylor-expand the derivative around a general point $x$,
\begin{equation} \label{NewFunctional:ExpansionFunctionDerivative}
f^{(n)}(0) = \sum_{m = 0}^{\infty} \frac{1}{m!} (0 - x)^m f^{(n+m)}(x) = \sum_{m = 0}^{\infty} \frac{1}{m!} (- x)^m f^{(n+m)}(x) \, .
\end{equation}
Observe that, despite its appearance, this expression does not really depend on $x$.
%Contrary to what it might appear by looking at the right-hand side, does not depend on $x$ (it is just $f^{(n)}(0)$). 
Putting the pieces together, we see that counting the number of $k$'s in each monomial appearing in $f(\tilde{x} + k)$ amounts to expanding the binomial $(\tilde{x} + k)^n$ in the expression
\begin{equation} \label{NewFunctional:ExpansionFunction}
f(\tilde{x} + k) = \sum_{n,m = 0}^{\infty} \frac{1}{n!m!} (\tilde{x} + k)^n (-x)^m f^{(n+m)}(x) \, ,
\end{equation}
where we emphasize that the $x$'s in the right-hand side are not to be substituted by $\tilde{x}+k$. In the above expression we can  %as we argued %(in fact, the previous expression already has no $x$ dependence, as the left-hand side shows). 
%One final comment might be useful to understand better what we are about to do in the case of a function of the Riemann tensor components instead of a single variable $x$. We can 
pair each of the $n+m$ derivatives with the factors $(\tilde{x} + k)$ and $x$ provided we introduce some ordering convention. The idea is to impose that derivatives only act on $f(x)$, and not on explicit $x$ factors,
\begin{equation} \label{NewFunctional:ExpansionFunction}
f(\tilde{x} + k) = \left[ : \sum_{n,m = 0}^{\infty} \frac{1}{n!m!} \left[(\tilde{x} + k) \partial_x \right]^n (-x \partial_x)^m : \right] f (x) ~ .
\end{equation}
This notation will turn out to be convenient when dealing with the analogous expressions involving Riemann tensor components. 

Now, with some care, the idea presented above can be extended to functions of several variables. In the case of interest here, these variables will be Riemann tensor components. Roughly speaking, $f(x)$ will be replaced by the object defined in (\ref{NewFunctional:ShorthandSecondDerivative}) and  $x=\tilde x + k$ will be the splitting of each component, $R_{MI}=\tilde R_{MI}+\mathcal{K}_{MI}$.  The first step is the expansion around $0$, which gives an expression in which such splitting %\eqref{NotationRewriting:GeneralSplitting} 
is already applied
\begin{align} \label{NewFunctional:ExpansionAt0}
\D2R = \sum_{n=0}^{\infty} \frac{1}{n!} & \left( \tilde{R}\indices{_{M_1 I_1}} + \mathcal{K}\indices{_{M_1 I_1}} \right) \dots \left( \tilde{R}\indices{_{M_n I_n}} + \mathcal{K}\indices{_{M_n I_n}} \right)  \\ \nonumber 
& \times \left[ \frac{\hd}{\hd R\indices{_{M_1 I_1}} \dots \hd R\indices{_{M_n I_n}}} \D2R \right]_{{\rm Riem} = 0} \, .
\end{align}
Just like for $f^{(n)}(0)$ above, the derivatives piece at zero can be traded by derivatives at a general value of the Riemann tensor as
\begin{align} \label{NewFunctional:ExpansionDerivativeAt0}
& \left[ \frac{\hd}{\hd R\indices{_{M_1 I_1}} \dots \hd R\indices{_{M_n I_n}}} \D2R \right]_{{\rm Riem} = 0} = \\
\nonumber  & \sum_{m = 0}^{\infty} \frac{1}{m!} (-R\indices{_{N_1 J_1}}) \dots (-R\indices{_{N_m J_m}}) \frac{\hd}{\hd R\indices{_{N_1 J_1}} \dots \hd R\indices{_{N_m J_m}} \hd R\indices{_{M_1 I_1}} \dots \hd R\indices{_{M_n I_n}}} \D2R \, .
\end{align}
The two previous expressions can be combined into a single and simpler one if we introduce again a sort of normal ordering prescription for derivatives. By this we mean:
\begin{equation} \label{NewFunctional:NormalOrdering}
:\left( R\indices{_{MI}} \hd^{MI} \right)^n: \, \equiv R\indices{_{M_1 I_1}} \dots R\indices{_{M_n I_n}} \frac{\hd}{\hd R\indices{_{M_1 I_1}} \dots \hd R\indices{_{M_n I_n}}} ~,
\end{equation}
so that derivatives only act on the object completely to the right of the expression. Then we have
\begin{equation} \label{NewFunctional:ExpansionAtR}
 \D2R = \left[ : \sum_{n,m=0}^{\infty} \frac{1}{n!m!} \left(\left(\tilde{R}\indices{_{MI}} + \mathcal{K}\indices{_{MI}} \right) \hd^{MI} \right)^n \left( - R\indices{_{NJ}} \hd^{NJ} \right)^m : \right] \D2R \, .
\end{equation}
From now on, we will work with the operator between brackets alone, since it contains all we need, namely, the explicit dependence on the $\mathcal{K}\indices{_{MI}} $. We will also implicitly assume the normal ordering convention for derivatives. 

Now, let us use the following useful identity
\begin{equation} \label{NewFunctional:SwitchSums}
\sum_{n,m=0}^{\infty} f(n,m) = \sum_{S=0}^{\infty} \sum_{n=0}^{S} f(n, S-n) \, ,
\end{equation}
to collect terms in the sums depending on the total number of derivatives they have, $S = n+m$. This gives:
\begin{align} \label{NewFunctional:ExpressionBeforeQSplit}
\nonumber & \sum_{S=0}^{\infty} \sum_{n=0}^{S} \frac{1}{n!(S-n)!} \left(\left(\tilde{R}\indices{_{MI}} + \mathcal{K}\indices{_{MI}} \right) \hd^{MI} \right)^n \left( - R\indices{_{NJ}} \hd^{NJ} \right)^{S-n}  \\
& =\sum_{S=0}^{\infty} \frac{1}{S!} \left(\left(\tilde{R}\indices{_{MI}} + \mathcal{K}\indices{_{MI}}  - R\indices{_{MI}} \right) \hd^{MI} \right)^S ~, 
\end{align}
where we have applied the binomial theorem.\footnote{For this to be valid, we need the elements inside the parentheses to commute with each other. This is guaranteed by the normal ordering prescription for derivatives.} Let us pause for a moment and look at \req{NewFunctional:ExpressionBeforeQSplit}. Here we could be tempted to use $\tilde{R}\indices{_{MI}} + \mathcal{K}\indices{_{MI}}  - R\indices{_{MI}} = 0$, which would mean that the previous operator is simply the identity (because of the $S=0$ term). This is not a contradiction. As a matter of fact, the only thing we have done so far is applying the identity in an elaborated way. But we have achieved our goal, since we have isolated the appearances of $\mathcal{K}\indices{_{MI}} $ in the $\alpha$-expansion: all these factors are the ones explicitly appearing in the previous expression.

From now on, we will have to deal separately with the two types of Riemann tensor components: those we called type $A$ (with $q_{\alpha} = 1$ for the corresponding $\mathcal{K}\indices{_{MI}}$) and those we called type $B$ (with $q_{\alpha} = 1/2$ for the corresponding $\mathcal{K}\indices{_{MI}}$). This can be easily done from the previous expression,
\begin{align}
& \sum_{S=0}^{\infty} \frac{1}{S!} \left(\left(\tilde{R}\indices{_{AI}} + \mathcal{K}\indices{_{AI}}  - R\indices{_{AI}} \right) \hd^{AI} + \left(\tilde{R}\indices{_{BI}} + \mathcal{K}\indices{_{BI}}  - R\indices{_{BI}} \right) \hd^{BI} \right)^S = \\
\nonumber & \sum_{S = 0}^{\infty} \sum_{T = 0}^S \frac{1}{T! (S-T)!} \left[\left(\tilde{R}\indices{_{AI}} + \mathcal{K}\indices{_{AI}}  - R\indices{_{AI}} \right) \hd^{AI}\right]^T \left[\left(\tilde{R}\indices{_{BJ}} + \mathcal{K}\indices{_{BJ}}  - R\indices{_{BJ}} \right) \hd^{BJ}\right]^{S-T}
\, . 
\end{align}
The next step is to isolate the number of $\mathcal{K}$'s of each type, to prepare for the $(1 + q_{\alpha})$ division,
\begin{align}
\sum_{S = 0}^{\infty} \sum_{T = 0}^S \sum_{\lambda_1 = 0}^T \sum_{\lambda_2 = 0}^{S-T} & \frac{1}{T! (S-T)!} \frac{T!}{\lambda_1! (T - \lambda_1)!} \left[\mathcal{K}\indices{_{AI}} \hd^{AI}\right]^{\lambda_1} \left[\left(\tilde{R}\indices{_{A'I'}} - R\indices{_{A'I'}} \right) \hd^{A'I'}\right]^{T-\lambda_1} \\ \nonumber 
 & \frac{(S-T)!}{\lambda_2! (S - T - \lambda_2)!} \left[\mathcal{K}\indices{_{BJ}} \hd^{BJ}\right]^{\lambda_2} \left[\left(\tilde{R}\indices{_{B'J'}} - R\indices{_{B'J'}} \right) \hd^{B'J'}\right]^{S-T-\lambda_2} ~ .
\end{align}
In this expression, it is manifest that we have $\lambda_1$ components $\mathcal{K}_{AI}$, which contribute 1 to $q_{\alpha}$, and $\lambda_2$ components $\mathcal{K}_{BJ}$, which contribute $1/2$ to $q_{\alpha}$. Hence, we are ready to divide by $(1 + q_{\alpha} )=( 1 + \lambda_1 + \lambda_2/2)$, obtaining
\begin{align}
\nonumber \sum_{S = 0}^{\infty} \sum_{T = 0}^S \sum_{\lambda_1 = 0}^T \sum_{\lambda_2 = 0}^{S-T} & \frac{2}{(2 + 2 \lambda_1 + \lambda_2)} \frac{1}{\lambda_1! (T - \lambda_1)!} \left[\mathcal{K}\indices{_{AI}} \hd^{AI}\right]^{\lambda_1} \left[\left(\tilde{R}\indices{_{A'I'}} - R\indices{_{A'I'}} \right) \hd^{A'I'}\right]^{T-\lambda_1} \\
 & \frac{1}{\lambda_2! (S - T - \lambda_2)!} \left[\mathcal{K}\indices{_{BJ}} \hd^{BJ}\right]^{\lambda_2} \left[\left(\tilde{R}\indices{_{B'J'}} - R\indices{_{B'J'}} \right) \hd^{B'J'}\right]^{S-T-\lambda_2} ~ .
\end{align}
At this point, the $\alpha$-sum has been performed, and we do not need to explicitly keep the $\mathcal{K}$ dependence isolated. We can also rewrite the $\tilde{R}$  back in terms of conventional Riemann tensor components. Using $\tilde{R}\indices{_{MI}} = R\indices{_{MI}} - \mathcal{K}\indices{_{MI}}$ we have
\begin{equation}
\sum_{S = 0}^{\infty} \sum_{T = 0}^S \sum_{\lambda_1 = 0}^T \sum_{\lambda_2 = 0}^{S-T} \frac{2}{(2 + 2 \lambda_1 + \lambda_2)} \frac{(-1)^{S-\lambda_1-\lambda_2}}{\lambda_1! (T - \lambda_1)!} \frac{1}{\lambda_2! (S - T - \lambda_2)!} \left(\mathcal{K}\indices{_{AI}} \hd^{AI}\right)^{T} \left(\mathcal{K}\indices{_{BJ}} \hd^{BJ}\right)^{S-T} ~ .
\end{equation}
At this point we proceed to perform the $\lambda_1$ and $\lambda_2$ sums, which do not affect the derivative operators. Let us start with the $\lambda_2$ one. It is possible to show that
\begin{align} \label{NewFunctional:SumLambda1Expression}
 \sum_{\lambda=0}^{S-T} \frac{(-1)^{\lambda}}{(2+2\lambda_1+\lambda )} \frac{1}{\lambda! (S-T - \lambda)!} & =  \frac{(2\lambda_1+1)!}{(2\lambda_1+2+S-T)!} \, .
\end{align}
Detailed derivations of this identity as well as of \req{NewFunctional:SumLambda2Expression} are included in appendix \ref{formuls}.
%
%where the first term inside the brackets does not survive after $n-1$ derivatives evaluated at $x = 1$ because of the factor $(1-x)^{\tilde{T}+n}$, and we have used:
%
%\begin{equation}
%\partial_x^{n-1} \left( \frac{1}{x} \right) = \frac{(-1)^{n-1} (n-1)!}{x^n} ~ .
%\end{equation}
%
After performing this sum, the operator becomes:
\begin{equation} \label{NewFunctional:OperatorInProgress}
2 \sum_{S = 0}^{\infty} \sum_{T = 0}^S \sum_{\lambda_1 = 0}^T \frac{(-1)^{S-\lambda_1}}{\lambda_1! (T - \lambda_1)!} \frac{(2 \lambda_1 + 1)!}{(2\lambda_1 + 2 + S - T)!} \left(\mathcal{K}\indices{_{AI}} \hd^{AI}\right)^{T} \left(\mathcal{K}\indices{_{BJ}} \hd^{BJ}\right)^{S-T} \, .
\end{equation}
We can now try to do the $\lambda_1$ one. We find the following integral representation of the sum,\footnote{This can be explicitly written in terms of Gauss' hypergeometric function as \begin{equation}  \sum_{\lambda=0}^{T} \frac{(-1)^{\lambda}}{\lambda! (T - \lambda)!} \frac{(2 \lambda + 1 )!}{(2 \lambda + 2+S-T)!} = \frac{2+S- (S-T)\, {}_{2}F_1 \left[1,-T ;3+S;-1\right]}{2(1+S)(2+S)(S-T)! T!}\, ,\end{equation} but the integral form turns out to be more useful for our purposes.}
\begin{align} \label{NewFunctional:SumLambda2Expression}
\sum_{\lambda=0}^{T} \frac{(-1)^{\lambda}}{\lambda! (T - \lambda)!} \frac{(2 \lambda + 1 )!}{(2 \lambda + 2+S-T)!} & = \frac{1}{T! (S-T)!} \int_0^1 {\rm d} u \, u (1 - u^2)^T (1-u)^{S-T} \, .
%\\ &=\frac{{}_{3}F_2 \left[ \left\{1,\frac{3}{2},-T\right\}, \left\{\frac{3+S-T}{2},\frac{4+S-T}{2}\right\},1 \right]}{T! \Gamma[3+S-T]} \\ &= \frac{2+S- (S-T)\, {}_{2}F_1 \left[1,-T ;3+S;-1\right]}{2(1+S)(2+S)(S-T)! T!}
\end{align}

%where ${}_{3}F_2$ is a generalized hypergeometric function. 
%\\
%\comment{HEEEEREEE}\\
%
%There are only a couple of manipulations left to arrive to the final form of the operator. 
Continuing from \eqref{NewFunctional:OperatorInProgress}, we have %(recall that $m = 2 + S - T$):
\begin{align}
\nonumber & \int_0^1 {\rm d}u \, 2u \sum_{S = 0}^{\infty} \sum_{T = 0}^S \frac{(-1)^S}{T! (S - T)!} \left((1-u^2) \mathcal{K}\indices{_{AI}} \hd^{AI}\right)^{T} \left((1- u)\mathcal{K}\indices{_{BJ}} \hd^{BJ}\right)^{S-T}  \\
& = \sum_{S = 0}^{\infty} \frac{(-1)^S}{S!} \int_0^1 {\rm d}u \, 2u  \left[(1-u^2) \mathcal{K}\indices{_{AI}} \hd^{AI} + (1- u)\mathcal{K}\indices{_{BJ}} \hd^{BJ}\right]^{S} \, .
\end{align}
This is our final result. Let us collect everything here, including the definitions needed to interpret it. We have found that the anomaly term in the holographic entanglement entropy functional can be written as
\begin{align} \label{NewFunctional:FinalForm}
&\sum_{\alpha} \frac{1}{1 + q_{\alpha}}  \left. \D2R \right._{\alpha}  \\
\nonumber & = \sum_{S = 0}^{\infty} \frac{1}{S!} \int_0^1 {\rm d}u \, 2u  :\left[- (1-u^2) \mathcal{K}\indices{_{AI}} \hd^{AI} - (1- u)\mathcal{K}\indices{_{BJ}} \hd^{BJ}\right]^{S}:  \D2R \, ,
\end{align}
where we emphasized again that derivatives have to be taken after normal ordering and
\begin{align} \label{NewFunctional:OperatorA}
\mathcal{K}\indices{_{AI}} \hd^{AI} \equiv & - \frac{1}{2} K\indices{^{a i j}} K\indices{_{a i j}} \frac{\partial}{\partial R\indices{_{z \bz z \bz}}} - 8 K\indices{_{z i}^{k}} K\indices{_{\bz jk}} \frac{\partial}{\partial R\indices{_{z \bz i j}}} - 8 K\indices{_{z i}^k} K\indices{_{\bz j k}} \frac{\partial}{\partial R\indices{_{z i \bz j}}} \\
\nonumber & - 2 K\indices{_{a i k}} K\indices{^a_{l j}} \frac{\partial}{\partial R\indices{_{i j k l}}} + \left( 4 R\indices{_{z i z j}} \frac{\partial}{\partial R\indices{_{z i z j}}} + {\rm c.c.} \right) ~ , \\\label{NewFunctional:OperatorB}
\mathcal{K}\indices{_{BJ}} \hd^{BJ} \equiv & \left( 4 R\indices{_{z i j k}} \frac{\partial}{\partial R\indices{_{z i j k}}} + 8 R\indices{_{z \bz z i}} \frac{\partial}{\partial R\indices{_{z \bz z i}}} + {\rm c.c.} \right) ~ .
\end{align}
Observe that the sum in \req{NewFunctional:FinalForm} can be formally performed, allowing us to write the result in an exponential form
\begin{align} \label{NewFunctional:FinalForm2}
&\sum_{\alpha} \frac{1}{1 + q_{\alpha}}  \left. \D2R \right._{\alpha}=\int_0^1 {\rm d}u \, 2u\,  {\rm e}^{-F(u)}  \D2R \, , 
\end{align}
where
\begin{equation}
F(u)\equiv [(1-u^2)\mathcal{K}\indices{_{AI}} \hd^{AI}+ (1- u)\mathcal{K}\indices{_{BJ}} \hd^{BJ}]\, .
\end{equation}
%In fact, the above integral can be also performed formally,%\footnote{Explicitly, one finds \comment{BLAAAh}
%%\begin{equation}
%%\int_0^1 {\rm d}z \, 2z\,  {\rm e}^{[(1-z^2)a+ (1-z)b]} = \frac{{\rm e}^{a+b} -1} {a}+\frac{\sqrt{\pi} b}{2 a^{3/2}}  {\rm e}^{[a+b+\frac{b^2}{4a}]} \left({\rm Erf} \left[ \frac{b}{2\sqrt{a}}\right]- {\rm Erf} \left[ \frac{2a + b}{2\sqrt{a}}\right]\right)\, .
%%\end{equation}
%}
% giving a not very illuminating result \comment{in terms of exponentials and error functions of  combinations of $\mathcal{K}\indices{_{AI}} \hd^{AI}$ and $\mathcal{K}\indices{_{BI}} \hd^{BI}$}. %\begin{equation}
%\int_0^1 {\rm d}z \, 2z\,  {\rm exp}^{F(z)} = {\rm exp}^{  \mathcal{K}\indices{_{AI}} \hd^{AI}+ \mathcal{K}\indices{_{BI}} \hd^{BI}  }
%\end{equation}
%
In subsection \ref{covid} below we present a covariant version of these new formulas. Observe that even though the anomaly term naively involves the contraction of intrinsic curvatures with two extrinsic curvatures, it is manifest from our formula  that the sum over $\alpha$ hides possible contractions with an arbitrary (even) number of extrinsic curvatures ---in particular, order $n$ densities will produce terms involving up to $2(n-1)$ extrinsic curvatures. 

There are some obvious particular cases in which the above expression simplifies considerably. Firstly, if no $B$ type terms appear in the second derivative of the Lagrangian, we can write
%Let us mention as a concluding remark that, if we know that only one of the types of Riemann tensor components appears in the second derivative of the Lagrangian, the previous integral can be explicitly performed. If we only have type A terms, we use:
%
%\begin{equation} \label{NewFunctional:IntegralTypeA}
%\int_0^1 {\rm d} z \, 2z (1-z^2)^S = \frac{1}{S + 1} ~ ,
%\end{equation}
%
%to obtain:
%
\begin{align} \label{NewFunctional:OnlyTypeA}
\sum_{\alpha} \frac{1}{1 + q_{\alpha}} \left. \D2R \right._{\alpha} &= \sum_{S = 0}^{\infty} \frac{1}{(S+1)!} \left( - \mathcal{K}\indices{_{AI}} \hd^{AI} \right)^{S} \D2R \\ \notag &= \left[\mathcal{K}\indices{_{AI}} \hd^{AI} \right]^{-1} \left[1- {\rm e}^{-\mathcal{K}\indices{_{AI}} \hd^{AI}} \right]  \D2R \, .
\end{align}
Similarly, if only type $B$ terms were present, %we can use:
%
%\begin{equation} \label{NewFunctional:IntegralTypeB}
%\int_0^1 {\rm d} z \, 2z (1-z)^S = \frac{2}{(S+2)(S+1)} ~ ,
%\end{equation}
%
%to obtain:
%
the result would simplify to
\begin{align} \label{NewFunctional:OnlyTypeB}
\sum_{\alpha} \frac{1}{1 + q_{\alpha}} \left. \D2R \right._{\alpha} & = \sum_{S = 0}^{\infty} \frac{2}{(S+2)!} \left( - \mathcal{K}\indices{_{BI}} \hd^{BI} \right)^{S} \D2R \\ \notag &=-2 \left[\mathcal{K}\indices{_{BI}} \hd^{BI} \right]^{-2}  \left[1-\mathcal{K}\indices{_{BI}} \hd^{BI}-{\rm e}^{-\mathcal{K}\indices{_{BI}} \hd^{BI}}  \right] \D2R \, .
\end{align}
As we will see in a moment, there is at least one important case for which only type $A$ terms appear, namely, Lovelock theories. It is harder to imagine how only type $B$ terms could appear. Nevertheless, the result obtained here will prove to be useful for presenting the explicit form of the anomaly term for cubic and quartic theories.

Before closing this subsection, let us mention that, while our new formulas have been obtained assuming a particular splitting for the Riemann tensor components ---namely, the one valid for perturbative higher-curvature gravities summarized in \req{NotationRewriting:GeneralSplitting}--- an analogous procedure to the one presented here should allow to produce similar expressions for other possible splittings. 

%%%%%%%%%%%%%%%%%%%%%%
\subsection{Anomaly term in Lovelock theories}
\label{subsec:AnomalyLovelock}
%%%%%%%%%%%%%%%%%%%%%%
Lovelock gravities \cite{Lovelock1,Lovelock2} are special in many respects ---see also subsection \ref{love1} below. In particular, as argued in \cite{Dong:2013qoa}, the object \eqref{NewFunctional:ShorthandSecondDerivative} only contains a single kind of Riemann tensor component for them, namely, $R\indices{_{i j k l}}$.  
%It is a well known fact that Lovelock theories, due to their particular structure, only contain one kind of Riemann tensor component in the object \eqref{NewFunctional:ShorthandSecondDerivative}: $R\indices{_{i j k l}}$. We can quickly review the argument presented in \cite{Dong:entanglement_entropy} which shows this to be true. 
The Lovelock density of order $n$ is defined by 
\begin{equation} \label{AnomalyLovelock:LovelockLagrangian}
\mathcal{X}_{2n}(R) \equiv \frac{1}{2^n} \delta_{\nu_1 \nu_2 \cdots \nu_{2n-1}\nu_{2n}}^{\mu_1 \mu_2\cdots \mu_{2n-1}\mu_{2n}} R^{\nu_1 \nu_2}_{\mu_1 \mu_2} \cdots R^{\nu_{2n-1} \nu_{2n}}_{\mu_{2n-1}\mu_{2n}}\, ,
\end{equation}
where $\delta_{\nu_1 \nu_2 \cdots \nu_{2n-1}\nu_{2n}}^{\mu_1 \mu_2\cdots \mu_{2n-1}\mu_{2n}}$  is the totally antisymmetric   product of $2n$ Kronecker deltas.
Now, since we have that 
\begin{equation}
\frac{\partial R^{\mu \nu}_{\rho \sigma}}{\partial R\indices{_{z i z j}}} = \frac{1}{2} \left( g^{z[\mu} g^{\nu] i} \delta_{[\rho}^{z} \delta_{\sigma]}^{j} + g^{z[\mu} g^{\nu] j} \delta_{[\rho}^{z} \delta_{\sigma]}^{i} \right) = 2 \delta_{\bz}^{[\mu} \delta_{m}^{\nu]} \delta_{[\rho}^{z} \delta_{\sigma]}^{(i} g^{j) m}\, ,
\end{equation}
and a similar result for the derivative with respect to $R\indices{_{\bz k \bz l}}$, the second derivative contracted with $K^2$ appearing in the anomaly term is of the form:
\begin{equation} \label{AnomalyLovelock:SecondDerivative}
 \frac{8\partial^2 \mathcal{X}_{2n}}{\partial R\indices{_{z i z j}} \partial R\indices{_{\bz k \bz l}}} K\indices{_{z i j}} K\indices{_{\bz k l}} =  \frac{- 8n (n-1)}{2^{n-2}} \delta^{z \bz i j \mu_1 \mu_2 \dots \mu_{2n - 5} \mu_{2n - 4}}_{z \bz k l \nu_1 \nu_2 \dots \nu_{2n - 5} \nu_{2n - 4}} R^{\nu_1 \nu_2}_{\mu_1 \mu_2} \cdots R^{\nu_{2n-5} \nu_{2n-4}}_{\mu_{2n-5}\mu_{2n-4}} K\indices{_{z i}^k} K\indices{_{\bz j}^l} \, .
\end{equation}
Due to the completely antisymmetric character of the generalized delta, none of the indices $\mu_n$ or $\nu_n$ can be $z$ or $\bz$. This forces all components of the Riemann tensor to be of the type $R^{j_1 j_2}_{i_1 i_2}$, as anticipated.\footnote{Something similar happens with the Wald term. As a result, the entanglement entropy functional for Lovelock theories can be written in terms of intrinsic curvatures to the surface ---see \req{jm} below.}
%This proves our previous claim that only $R\indices{_{i j k l}}$ components appear for Lovelock theories after the two derivatives of the Lagrangian are taken as indicated in the anomaly term of the entanglement entropy functional.
 Therefore, we only have to take into account the part proportional to $\partial / \partial R\indices{_{i j k l}}$ in \eqref{NewFunctional:OperatorA}. Using the result \eqref{NewFunctional:OnlyTypeA} valid when only type $A$ terms are present we find
\begin{align} \label{OneComponent:LovelockExponential}
& \sum_{\alpha} \frac{1}{1 + q_{\alpha}} \left. \left( \frac{8 \partial^2 \mathcal{L}^{\rm Lovelock}_E}{\partial {\rm Riem}^2} K^2 \right) \right._{\alpha} \\
= & \sum_{S=0}^{\infty} \frac{1}{(S+1)!} \left( 2 K\indices{_{a i k}} K\indices{^a_{l j}} \frac{\partial}{\partial R\indices{_{i j k l}}} \right)^{S}\left( \frac{8 \partial^2 \mathcal{L}^{\rm Lovelock}_E}{\partial {\rm Riem}^2} K^2 \right)  \\
= & \left[ 2 K\indices{_{a i k}} K\indices{^a_{l j}} \frac{\partial}{\partial R\indices{_{i j k l}}} \right]^{-1} \left[ \exp \left( 2 K\indices{_{a i k}} K\indices{^a_{j l}} \frac{\partial}{\partial R\indices{_{i j k l}}} \right) - 1 \right] \left( \frac{8 \partial^2 \mathcal{L}^{\rm Lovelock}_E}{\partial {\rm Riem}^2} K^2 \right)  \, .
\end{align} 

This is a rather suggestive expression. On the other hand, we know that for Lovelock theories the combination of the anomaly and Wald terms must reduce to the so-called Jacobson-Myers (JM) functional  ---see \req{jm} below. Let us see how this works when the anomaly term is written as in \req{OneComponent:LovelockExponential}. First of all, notice that the extrinsic curvatures in the second derivative  can be written covariantly using the antisymmetry of the generalized delta as
\begin{equation} \label{AnomalyLovelock:SecondDerivativeCovariant}
\left( \frac{8 \partial^2 \mathcal{X}_{2n}}{\partial {\rm Riem}^2} K^2 \right) =  \frac{n (1-n)}{2^{n-3}} \delta^{i_1 j_1 \dots i_{n-1} j_{n-1}}_{k_1 l_1 \dots k_{n-1} l_{n-1}} R^{k_1 l_1}_{i_1 j_1} \cdots R^{k_{n-2} l_{n-2}}_{i_{n-2} j_{n-2}} K\indices{_{a i_{n-1}}^{k_{n-1}}} K\indices{^a_{j_{n-1}}^{l_{n-1}}} \, ,
\end{equation}
where we have also reduced the generalized delta eliminating the $z$ and $\bz$ indices. Applying $S$ times the differential operator is now straightforward:
\begin{align} \label{AnomalyLovelock:SOperators}
 & \sum_{S=0}^{\infty} \frac{1}{(S+1)!} \left( 2 K\indices{_{a i k}} K\indices{^a_{l j}} \frac{\partial}{\partial R\indices{_{i j k l}}} \right)^{S} \left( \frac{8 \partial^2 \mathcal{X}_{2n}}{\partial {\rm Riem}^2} K^2 \right)  \\
\nonumber = & - \sum_{S=0}^{n-2} \frac{1}{(S+1)!} \frac{n (n-1) \cdots (n-1-S)}{2^{n-3 - S}} \delta^{i_1 j_1 \cdots i_{n-1} j_{n-1}}_{k_1 l_1 \cdots k_{n-1} l_{n-1}} R^{k_1 l_1}_{i_1 j_1} \cdots R^{k_{n-2-S} l_{n-2-S}}_{i_{n-2-S} j_{n-2-S}}  \\
\nonumber & \quad \times K\indices{_{a_{n-1-S} i_{n-1-S}}^{k_{n-1-S}}} K\indices{^{a_{n-1-S}}_{j_{n-1-S}}^{l_{n-1-S}}} \dots K\indices{_{a_{n-1} i_{n-1}}^{k_{n-1}}} K\indices{^{a_{n-1}}_{j_{n-1}}^{l_{n-1}}} \\
\nonumber = & - n \sum_{S=1}^{n-1} \frac{1}{2^{n-2 - S}} \binom{n-1}{S} \delta^{i_1 j_1 \cdots i_{n-1} j_{n-1}}_{k_1 l_1 \dots k_{n-1} l_{n-1}} R^{k_1 l_1}_{i_1 j_1} \cdots R^{k_{n-1-S} l_{n-1-S}}_{i_{n-1-S} j_{n-1-S}}  \\
\nonumber & \quad \times K\indices{_{a_{n-S} i_{n-S}}^{k_{n-S}}} K\indices{^{a_{n-S}}_{j_{n-S}}^{l_{n-S}}} \cdots K\indices{_{a_{n-1} i_{n-1}}^{k_{n-1}}} K\indices{^{a_{n-1}}_{j_{n-1}}^{l_{n-1}}} \, .
\end{align}
Furthermore, the Wald term reads
\begin{equation} \label{AnomalyLovelock:WaldTerm}
\frac{\partial \mathcal{X}_{2n}}{\partial R\indices{_{z \bz z \bz}}} = - \frac{n}{2^{n-2}} \delta^{i_1 j_1 \dots i_{n-1} j_{n-1}}_{k_1 l_1 \dots k_{n-1} l_{n-1}} R^{k_1 l_1}_{i_1 j_1} \dots R^{k_{n-1} l_{n-1}}_{i_{n-1} j_{n-1}} \, .
\end{equation}
This can be combined with \eqref{AnomalyLovelock:SOperators}, acting as the $S = 0$ term of the sum. When this is included, the binomial coefficient and the $2^{-S}$ factor in each term  can be employed to write the full functional as
\begin{align} \label{AnomalyLovelock:CombinationWaldAnomaly}
 & \frac{\partial \mathcal{X}_{2n}}{\partial R\indices{_{z \bz z \bz}}} + \sum_{S=0}^{\infty} \frac{1}{(S+1)!} \left( 2 K\indices{_{a i k}} K\indices{^a_{l j}} \frac{\partial}{\partial R\indices{_{i j k l}}} \right)^{S} \left( \frac{8 \partial^2 \mathcal{X}_{2n}}{\partial {\rm Riem}^2} K^2 \right) \\
\nonumber = &  -\frac{n}{2^{n-2}} \delta^{i_1 j_1 \dots i_{n-1} j_{n-1}}_{k_1 l_1 \dots k_{n-1} l_{n-1}} \left(R^{k_1 l_1}_{i_1 j_1}  + 2 K\indices{_{a_1 i_1}^{k_1}} K\indices{^{a_1}_{j_1}^{l_1}}\right) \\  &\notag \dots \left( R^{k_{n-1} l_{n-1}}_{i_{n-1} j_{n-1}} + 2 K\indices{_{a_{n-1} i_{n-1}}^{k_{n-1}}} K\indices{^{a_{n-1}}_{j_{n-1}}^{l_{n-1}}} \right) \, ,
\end{align}
where we used the fact that the binomial factor is the number of ways we can pick 
%Since the binomial factor is the number of ways we can pick 
$S$ squared extrinsic curvature factors and $(n-1-S)$ Riemann tensors from the previous product (and the antisymmetric delta can be used to rewrite all possible combinations as essentially the same). The final observation is that $\tilde{R}\indices{_{i j k l}}$ is actually the intrinsic curvature tensor of the surface  \cite{Dong:2013qoa}, which we denote $\mathcal{R}\indices{_{i j k l}}$. Then, comparing with \req{SplittingProblem:ExpansionsRiemann}, it follows that we can write the HEE functional for a given order-$n$ Lovelock density as
%A final bit of information is then needed. We mentioned when discussing the splitting that the $\tilde{R}$ objects in \eqref{SplittingProblem:ExpansionsRiemann} were only auxiliary, and that their geometric significance was not relevant. This is not completely true at this point: it can be shown that $\tilde{R}\indices{_{i j k l}}$ is actually the intrinsic curvature tensor of the surface, which we will call $\mathcal{R}\indices{_{i j k l}}$ \cite{Dong:entanglement_entropy}. 
%It is clear then that we can rewrite the entanglement entropy functional for Lovelock theories as Jacobson and Myers did:
%
\begin{equation} \label{AnomalyLovelock:JMFunctional}
S^{{\mathcal{X}_{2n}}}_{\rm HEE} = - 4 \pi n \int_{\Gamma_A} {\rm d}^{d-1}y \, \sqrt{h} \, \mathcal{X}_{2(n-1)}(\mathcal{R}) \, ,
\end{equation}
which is the JM form \cite{Jacobson:1993xs,Hung:2011xb}.  This has the interesting property of being fully determined in terms of intrinsic curvatures associated to the holographic entangling surface.

%Naturally, in the case of Lovelock gravities, the Jacobson-Myers functional (\ref{jm}) 
%
%where we are using the formal inverse of the operator and the exponential expansion to compactly write the series in a suggestive manner.

%%%%%%%%%%%%%%%%%%%%%%
\subsection{Anomaly term for cubic gravities}
\label{subsec:AnomalyCubic}
%%%%%%%%%%%%%%%%%%%%%%

Our new formula for the anomaly term in \eqref{NewFunctional:FinalForm} gets notably simplified for cubic theories. This is a consequence of the second derivative of the Lagrangian being linear in curvatures for these theories, which implies that only $S = 0, 1$ terms need to be included in the sum. In addition, the object \eqref{NewFunctional:ShorthandSecondDerivative} is ``neutral'' in $z$ and $\bz$ indices ---\ie it has and equal number of $z$'s and $\bz$'s\footnote{This is a consequence of the scalar character of the Lagrangian, which guarantees that, when written with lower indices, Riemann tensor components are contracted with metrics $g^{\mu \nu}$. The only non-vanishing component in the $z$, $\bz$ indices is $g^{z \bz} = 2$, so for each $z$ there must be one and only one $\bz$. After the two derivatives are taken following \eqref{NewFunctional:ShorthandSecondDerivative}, the number of $z$'s and $\bz$'s in Riemann tensor components decreases by two, but it is still equal for both types of indices.}--- so no components with a different number of $z$ and $\bz$ indices can appear inside it. In particular, there are no type $B$ terms, and the last term appearing in \eqref{NewFunctional:OperatorA} is also missing. Therefore, we can write the anomaly term for cubic theories as
\begin{align} \label{AnomalyCubic:ExpandedExpression}
&\sum_{\alpha} \frac{1}{1 + q_{\alpha}} \left. \left( \frac{8 \partial^2 \mathcal{L}^{\rm Riem^3}_E}{\partial {\rm Riem}^2} K^2 \right) \right._{\alpha}  =   \left[ 1 + \frac{1}{4} K\indices{^{a i j}} K\indices{_{a i j}} \frac{\partial}{\partial R\indices{_{z \bz z \bz}}} + 4 K\indices{_{z i}^{k}} K\indices{_{\bz jk}} \frac{\partial}{\partial R\indices{_{z \bz i j}}} \right. \\
\nonumber & \quad \left. + 4 K\indices{_{z i}^k} K\indices{_{\bz j k}} \frac{\partial}{\partial R\indices{_{z i \bz j}}} + K\indices{_{a i k}} K\indices{^a_{j l}} \frac{\partial}{\partial R\indices{_{i j k l}}} \right]  \left( \frac{8 \partial^2 \mathcal{L}^{\rm Riem^3}_E}{\partial {\rm Riem}^2} K^2 \right) \, .
\end{align}
In the explicit expressions for the functionals presented in the following section we have obtained the corresponding functionals using both the  $\alpha$-expansion procedure and this new derivative expression, finding perfect agreement.
%%%%%%%%%%%%%%%%%%%%%%
\subsection{Anomaly term for quartic gravities}
\label{subsec:AnomalyQuartic}
%%%%%%%%%%%%%%%%%%%%%%

Although slightly more complicated than the cubic ones, quartic theories are still simple enough to deserve an independent discussion. In this case, the second derivative of the Lagrangian is quadratic in curvature tensors, so we have to include $S = 0, 1, 2$ in \eqref{NewFunctional:FinalForm}. However, the neutral character in $z$'s and $\bz$'s of \eqref{NewFunctional:ShorthandSecondDerivative} allows us to simplify the general expression. In the expansion of the second derivative in terms of the basic components of the Riemann tensor,  %characterized by their $z$ and $\bz$ indices, we observe that 
each of the resulting monomials must be neutral in $z$ and $\bz$. %(when considering all indices to be lowered, as we are always doing). 
The first consequence of this fact is that components $R\indices{_{z \bz z \bz}}$, $R\indices{_{z \bz i j}}$, $R\indices{_{z i \bz j}}$, and $R\indices{_{i j k l}}$ cannot appear paired with the remaining ones, so we can drop all terms that involve mixed second derivatives between these two sets. Furthermore, by the same argument, $R\indices{_{z i z j}}$ can only appear paired with $R\indices{_{\bz i \bz j}}$ and thus, at second order in derivatives, type $B$ components do not mix with the type $A$ ones. %This allows us to use the results \eqref{NewFunctional:IntegralTypeA} and \eqref{NewFunctional:IntegralTypeB} for all terms relevant in quartic theories, in particular for $S=2$. 
Also, the last term (in parentheses) in \eqref{NewFunctional:OperatorA} does not mix with the remaining part of that operator when taking the square. All this means that the $S = 2$ term of \eqref{NewFunctional:FinalForm} for quartic theories will be:
\begin{align}\notag
& \frac{1}{3!} \left( \frac{1}{2} K\indices{^{a i j}} K\indices{_{a i j}} \frac{\partial}{\partial R\indices{_{z \bz z \bz}}} + 8 K\indices{_{z i}^{k}} K\indices{_{\bz j k}} \frac{\partial}{\partial R\indices{_{z \bz i j}}} + 8 K\indices{_{z i}^k} K\indices{_{\bz j k}} \frac{\partial}{\partial R\indices{_{z i \bz j}}} + 2 K\indices{_{a i k}} K\indices{^a_{j l}} \frac{\partial}{\partial R\indices{_{i j k l}}} \right)^2 \\
& + \frac{1}{3!} \left( - 4 R\indices{_{z i z j}} \frac{\partial}{\partial R\indices{_{z i z j}}} + {\rm c.c.} \right)^2 + \frac{2}{4!} \left( - 4 R\indices{_{z i j k}} \frac{\partial}{\partial R\indices{_{z i j k}}} - 8 R\indices{_{z \bz z i}} \frac{\partial}{\partial R\indices{_{z \bz z i}}} + {\rm c.c.} \right)^2 ~ ,
\end{align}
where, although not explicitly written, recall that all derivatives are to be understood under the normal ordering prescription, so they do not act on any of the Riemann tensor components appearing explicitly in the previous expression. This can be simplified a little bit more by using once again the fact that all terms in the second derivative of the Lagrangian have to be neutral in $z$ and $\bz$. Thus, in the second term, only the mixed derivative $\partial^2 /( \partial R\indices{_{z i z j}} \partial R\indices{_{\bz k \bz l}})$ contributes. Something similar happens in the last term, where only globally neutral combinations contribute. All in all, including also the $S = 0$ and $S = 1$ parts of the anomaly term, we can write, for quartic theories:
\begin{align}\label{AnomalyQuartic:ExpandedExpression}
& \sum_{\alpha} \frac{1}{1 + q_{\alpha}} \left. \left( \frac{8 \partial^2 \mathcal{L}^{\rm Riem^4}_E}{\partial {\rm Riem}^2} K^2 \right) \right._{\alpha}  \\
\nonumber & =  \left[ 1 + \frac{1}{2} \left[ \frac{1}{2} K\indices{^{a i j}} K\indices{_{a i j}} \frac{\partial}{\partial R\indices{_{z \bz z \bz}}} + 8 K\indices{_{z i}^{k}} K\indices{_{\bz j k}} \frac{\partial}{\partial R\indices{_{z \bz i j}}} + 8 K\indices{_{z i}^k} K\indices{_{\bz j k}} \frac{\partial}{\partial R\indices{_{z i \bz j}}} + 2 K\indices{_{a i k}} K\indices{^a_{j l}} \frac{\partial}{\partial R\indices{_{i j k l}}} \right. \right. \\
\nonumber & \quad \left. \left( - 4 R\indices{_{z i z j}} \frac{\partial}{\partial R\indices{_{z i z j}}} + {\rm c.c.} \right) \right] + \frac{2}{3!} \left( - 4 R\indices{_{z i j k}} \frac{\partial}{\partial R\indices{_{z i j k}}} - 8 R\indices{_{z \bz z i}} \frac{\partial}{\partial R\indices{_{z \bz z i}}} + {\rm c.c.} \right) \\
\nonumber & \quad + \frac{1}{3!} \left( \frac{1}{2} K\indices{^{a i j}} K\indices{_{a i j}} \frac{\partial}{\partial R\indices{_{z \bz z \bz}}} + 8 K\indices{_{z i}^{k}} K\indices{_{\bz j k}} \frac{\partial}{\partial R\indices{_{z \bz i j}}} + 8 K\indices{_{z i}^k} K\indices{_{\bz j k}} \frac{\partial}{\partial R\indices{_{z i \bz j}}} + 2 K\indices{_{a i k}} K\indices{^a_{j l}} \frac{\partial}{\partial R\indices{_{i j k l}}} \right)^2 \\
\nonumber & \quad + \frac{32}{3!} R\indices{_{z i z j}} R\indices{_{\bz k \bz l}} \frac{\partial^2}{\partial R\indices{_{z i z j}} \partial R\indices{_{\bz k \bz l}}} + \frac{2}{4!} \left( 128 R\indices{_{z \bz z i}} R\indices{_{\bz z \bz j}} \frac{\partial^2}{\partial R\indices{_{z \bz z i}} \partial R\indices{_{\bz z \bz j}}}  \right. \\
\nonumber & \quad \left. \left.+ 64 R\indices{_{z \bz z i}} R\indices{_{\bz j k l}} \frac{\partial^2}{\partial R\indices{_{z \bz z i}} \partial R\indices{_{\bz j k l}}} + 64 R\indices{_{\bz z \bz i}} R\indices{_{z j k l}} \frac{\partial^2}{\partial R\indices{_{\bz z \bz i}} \partial R\indices{_{z j k l}}} + 32 R\indices{_{z i j k}} R\indices{_{\bz l m n}} \frac{\partial^2}{\partial R\indices{_{z i j k}} \partial R\indices{_{\bz l m n}}} \right) \right]    \\
\nonumber &  \left( \frac{8 \partial^2 \mathcal{L}^{\rm Riem^4}_E}{\partial {\rm Riem}^2} K^2 \right) \, .
\end{align}
When computing the 26 functionals corresponding to independent quartic densities in subsection \ref{quarticc} we have made use of this expression, which turns out to be much faster than performing the corresponding $\alpha$ expansions. We have nonetheless verified in a few cases that both procedures yield the same results.

%%%%%%%%%%%%%%%%%%%%%%
\subsection{An example mixing type $A$ and type $B$ terms}
\label{subsec:MixedTypes}
%%%%%%%%%%%%%%%%%%%%%%

In the previous subsections involving Lovelock, cubic and quartic densities, we found that it was possible to treat separately type $A$ and $B$ terms. In this subsection we provide a simple example of a situation in which this separation is not possible. The previous arguments show that this happens for densities which are at least fifth-order in the Riemann tensor. In order to avoid unnecessary complications, let us assume that one of these densities produces a term mixing type $A$ and $B$ with the following form
\begin{equation} \label{MixedTypes:ExampleTerm}
\D2R \supset C(K^2) R\indices{_{z i j k}} R\indices{_{\bz z \bz}^k} R\indices{_z^i_{\bz}^j} \, ,
\end{equation}
where $C(K^2) \equiv c K\indices{_{z}^{l m}} K\indices{_{\bz l m}}$ with $c$ a constant, and the $\supset$ symbol means that this is only one of many terms that would appear when expanding the second derivative in terms of the different $z$ and $\bz$ components of the curvature tensor for an actual quintic (or higher order) density. We have not checked whether or not a term like this arises from a concrete fifth order Lagrangian, but it certainly could.\footnote{Notice that the term is globally neutral in $z$ and $\bz$, as it should.} In any case, it will serve as an example of how one should proceed if a different combination of type $A$ and $B$ terms arises.

%Since for this term it is particularly simple, 

Let us first obtain the result by means of the $\alpha$ sum, which in this case turns out to be particularly simple. Applying the splitting rules, this term becomes
\begin{equation} \label{MixedTypes:AlphaSum}
\sum_{\alpha} \left. \D2R \right._{\alpha} \supset C(K^2) \left( R\indices{_{z i j k}} R\indices{_{\bz z \bz}^k} \tilde{R}\indices{_z^i_{\bz}^j} - R\indices{_{z i j k}} R\indices{_{\bz z \bz}^k} K\indices{_z^{i l}} K\indices{_{\bz}^j_l} \right) \, .
\end{equation}
The first term has $q_{\alpha} = 1$, while the second has $q_{\alpha} = 2$. Then, dividing by $1 + q_{\alpha}$, we get
\begin{align} \notag
\sum_{\alpha} \frac{1}{1 + q_{\alpha}} \left. \D2R \right._{\alpha} & \supset C(K^2) \left( \frac{1}{2} R\indices{_{z i j k}} R\indices{_{\bz z \bz}^k} \tilde{R}\indices{_z^i_{\bz}^j} - \frac{1}{3} R\indices{_{z i j k}} R\indices{_{\bz z \bz}^k} K\indices{_z^{i l}} K\indices{_{\bz}^j_l} \right) \\
\label{MixedTypes:AlphaDivided}  & = C(K^2) \left( \frac{1}{2} R\indices{_{z i j k}} R\indices{_{\bz z \bz}^k} R\indices{_z^i_{\bz}^j} + \frac{1}{6} R\indices{_{z i j k}} R\indices{_{\bz z \bz}^k} K\indices{_z^{i l}} K\indices{_{\bz}^j_l} \right) \, ,
\end{align}
where we have rewritten $\tilde{R}\indices{_z^i_{\bz}^j}$ in terms of the Riemann tensor component again in the last line.

Let us now obtain the same result by means of the derivative expression, \eqref{NewFunctional:FinalForm}. We need to take into account terms up to $S = 3$ in the series, but fortunately not every type $A$ or $B$ component appears in the piece of the Lagrangian we are considering. This means we can define new operators including only the relevant parts:
\begin{equation} \label{MixedTypes:ReducedOperators}
\partial_A \equiv - 8 K\indices{_{z i}^l} K\indices{_{\bz j l}} \frac{\partial}{\partial R\indices{_{z i \bz j}}} ~ , \qquad \partial_B \equiv 4 R\indices{_{z i j k}} \frac{\partial}{\partial R\indices{_{z i j k}}} + 8 R\indices{_{\bz z \bz k}} \frac{\partial}{\partial R\indices{_{\bz z \bz k}}} \, .
\end{equation}
Now, the $S = 0$ term is just the original \eqref{MixedTypes:ExampleTerm}. For the $S = 1$ term we apply the operator:
\begin{equation} \label{MixedTypes:OperatorS1}
- \frac{1}{2!} \partial_A - \frac{2}{3!} \partial_B \, ,
\end{equation}
which produces:
\begin{equation} \label{MixedTypes:ResultS1}
s_1 \equiv C(K^2) \left( - \frac{2}{3} R\indices{_{z i j k}} R\indices{_{\bz z \bz}^k} R\indices{_z^i_{\bz}^j} + \frac{1}{2} R\indices{_{z i j k}} R\indices{_{\bz z \bz}^k} K\indices{_z^{i l}} K\indices{_{\bz}^j_l} \right) \, .
\end{equation}
For the $S = 2$ term operator we already find mixing between $\partial_A$ and $\partial_B$. Solving the integral expression given in \eqref{NewFunctional:FinalForm}
\begin{equation} \label{MixedTypes:OperatorS2}
\frac{1}{2!} \int_0^1 \rd u\,2u : \left(- (1-u^2) \partial_A - (1- u)\partial_B \right)^{2} : \; = \; : \left( \frac{1}{6} \partial_A^2 + \frac{7}{30} \partial_A \partial_B + \frac{1}{12} \partial_B^2 \right) :  \, .
\end{equation}
We stress once again that normal ordering means that derivatives do not act on curvature components appearing in the operators \eqref{MixedTypes:ReducedOperators}, only on those components in the second derivative object \eqref{MixedTypes:ExampleTerm}. This makes $\partial_A$ and $\partial_B$ commuting objects (inside a normal ordered expression). Furthermore, having only a single type $A$ component, the $\partial_A^2$ term in the previous expression will not contribute. The last two terms produce
\begin{equation} \label{MixedTypes:ResultS2}
s_2 \equiv C(K^2) \left( \frac{1}{6} R\indices{_{z i j k}} R\indices{_{\bz z \bz}^k} R\indices{_z^i_{\bz}^j} - \frac{7}{15} R\indices{_{z i j k}} R\indices{_{\bz z \bz}^k} K\indices{_z^{i l}} K\indices{_{\bz}^j_l} \right) \, .
\end{equation}
Finally, let us consider the $S = 3$ term. The operator is
\begin{equation*} \label{MixedTypes:OperatorS3}
\frac{1}{3!} \int_0^1 \rd u\,2u : \left(- (1-u^2) \partial_A - (1- u)\partial_B \right)^{3} : \; = - : \left( \frac{\partial_A^3}{24} + \frac{19 \partial_A^2 \partial_B}{210} + \frac{\partial_A \partial_B^2}{15} + \frac{\partial_B^3}{60} \right) :  \, .
\end{equation*}
In this case, since \eqref{MixedTypes:ExampleTerm} has one type $A$ and two type $B$ terms, the third piece of this operator is the only one giving a non-vanishing contribution. Its value is
\begin{equation} \label{MixedTypes:ResultS3}
s_3 \equiv \frac{2}{15} C(K^2) R\indices{_{z i j k}} R\indices{_{\bz z \bz}^k} K\indices{_z^{i l}} K\indices{_{\bz}^j_l} \, .
\end{equation}
We can finally combine all contributions, $s_0$ (which is just the original term \eqref{MixedTypes:ExampleTerm}), $s_1$, $s_2$, and $s_3$ to obtain
\begin{equation}\label{MixedTypes:ResultDerivatives}
\hspace{-.2cm}\sum_{\alpha} \frac{1}{1 + q_{\alpha}} \left. \D2R \right._{\alpha} \supset C(K^2) \left( \frac{1}{2} R\indices{_{z i j k}} R\indices{_{\bz z \bz}^k} R\indices{_z^i_{\bz}^j} + \frac{1}{6} R\indices{_{z i j k}} R\indices{_{\bz z \bz}^k} K\indices{_z^{i l}} K\indices{_{\bz}^j_l} \right) \, ,
\end{equation}
which coincides with \eqref{MixedTypes:AlphaDivided}, as it should.

\subsection{Covariant form of the new HEE formula}
\label{covid}
%%%%%%%%%%%%%%%%%%%%%%
So far we have presented all our expressions in the particular set of adapted coordinates $(z, \bz, y^i)$. Here we will rewrite our general formulas in covariant form, which is more useful for explicit applications (like the ones in Section \ref{unite}). In order to do that, we first write the metric as in \req{Notation:inducedmetric},
%Up until now we have been presenting all the expressions in the particular set of adapted (complex) coordinates to the surface, $(z, \bz, y^i)$. It is nevertheless possible -- and useful for particular applications, since normally we do not have the adapted coordinates at our disposal -- to rewrite everything in a covariant manner. For that we write the metric as:
%
\begin{equation} \label{Covariant:MetricNormalVectors}
g\indices{_{\mu \nu}} =h_{\mu\nu}+ \delta\indices{_{a b}} n\indices{^a_{\mu}} n\indices{^b_{\nu}}  \, ,
\end{equation}
so that in the adapted coordinates $n\indices{^a_i} = 0$, and $h\indices{_{\mu \nu}}$ is non-vanishing only for tangent components ($h\indices{_{z z}} = h\indices{_{z \bz}} = h\indices{_{\bz \bz}} = 0$). 
%We also define the binormal to the surface and the normal projector, respectively, as
%
%\begin{equation} \label{Covariant:BinormalNormalDefinitions}
%\epsilon\indices{_{\mu \nu}} \equiv \epsilon\indices{_{a b}} n\indices{^a_{\mu}} n\indices{^b_{\nu}}  ~, \qquad \perp\indices{_{\mu \nu}} \equiv \delta\indices{_{a b}} n\indices{^a_{\mu}} n\indices{^b_{\nu}}  \, ,
%\end{equation}
%
%where $\epsilon\indices{_{a b}}$ is the two-dimensional Levi-Civita symbol. 
It is easy to check that, in the adapted coordinates, the binormal to the surface and the normal projector, defined in \req{Covariant:BinormalNormalDefinitions}, satisfy $\epsilon\indices{_{z \bz}} = - \epsilon\indices{_{z \bz}} = i/2$, $\perp\indices{_{z z}} = \perp\indices{_{\bz \bz}} = 0$,\footnote{There is an ordering assumption in the value of $\epsilon\indices{_{z \bz}}$, the normal vectors $n\indices{^1}$ and $n\indices{^2}$ are defined so that we get $\epsilon\indices{_{z \bz}} = i/2$} and $\perp\indices{_{z \bz}} = 1/2$. The following identities can then be shown to hold for the adapted coordinates
\begin{align} \label{Covariant:DeltasInZ}
\delta\indices*{_{\mu}^z} \delta\indices*{_{\nu}^{\bz}} & = \perp\indices{_{\mu \nu}} - i \epsilon\indices{_{\mu \nu}} \, , 
& \delta\indices*{^{\mu}_z} \delta\indices*{^{\nu}_{\bz}} & = \frac{1}{4} \left( \perp\indices{^{\mu \nu}} + i \epsilon\indices{^{\mu \nu}}\right) \, , \\
\delta\indices*{_{\mu}^z} \delta\indices*{^{\nu}_{z}} & = \frac{1}{2} \left( \perp\indices{_{\mu}^{\nu}} - i \epsilon\indices{_{\mu}^{\nu}} \right) \, , 
& \delta\indices*{_{\mu}^{\bz}} \delta\indices*{^{\nu}_{\bz}} & = \frac{1}{2} \left( \perp\indices{_{\mu}^{\nu}} + i \epsilon\indices{_{\mu}^{\nu}}\right) \, . 
\end{align}
These are all different forms of the same identity, related by raising or lowering the $z$ and $\bz$ indices, but the different forms are useful in different contexts. In particular, they can be used to write in a covariant form the different terms appearing in the entanglement entropy functional. 

Let us start with the Wald term,
\begin{equation} \label{Covariant:WaldTerm}
\frac{\partial \mathcal{L}_E}{\partial R\indices{_{z \bz z \bz}}} = \delta\indices*{_{[\mu}^z} \delta\indices*{_{\nu]}^{\bz}} \delta\indices*{_{[\rho}^z} \delta\indices*{_{\sigma]}^{\bz}} \frac{\partial \mathcal{L}_E}{\partial R\indices{_{\mu \nu \rho \sigma}}} = - \epsilon\indices{_{\mu \nu}} \epsilon\indices{_{\rho \sigma}} \frac{\partial \mathcal{L}_E}{\partial R\indices{_{\mu \nu \rho \sigma}}} ~ .
\end{equation}
The last form, which is the familiar one for this piece \cite{Wald:1993nt,Iyer:1994ys}, is fully covariant, as desired. Similar manipulations can be applied to the anomaly term. For the second derivative of the Lagrangian contracted with two extrinsic curvatures we get\footnote{Notice that we take \eqref{Notation:extrinsic_curvature} as defining a spacetime tensor, $K\indices{^{\lambda}_{\mu \nu}} \equiv K\indices{^{a}_{\mu \nu}} n\indices{_a^{\lambda}}$. This tensor satisfies, in adapted coordinates, $K\indices{_{\lambda \mu \nu}} V^{\nu} = K\indices{_{\lambda \mu i}} V^{i}$ for any vector $V^{\mu}$.}
\begin{align} \label{Covariant:SecondDerivative}
\nonumber \D2R = & 2 \left[ \perp\indices{^{\lambda_1 \lambda_2}} \left( \perp\indices{_{\mu_1 \mu_2}} \perp\indices{_{\nu_1 \nu_2}} - \epsilon\indices{_{\mu_1 \mu_2}} \epsilon\indices{_{\nu_1 \nu_2}} \right) + \epsilon\indices{^{\lambda_1 \lambda_2}} \left( \perp\indices{_{\mu_1 \mu_2}} \epsilon\indices{_{\nu_1 \nu_2}} + \epsilon\indices{_{\mu_1 \mu_2}} \perp\indices{_{\nu_1 \nu_2}} \right) \right] \\ 
& \times \frac{\partial^2 \mathcal{L}_E}{\partial R\indices{_{\mu_1 \rho_1 \nu_1  \sigma_1}} \partial R\indices{_{\mu_2\rho_2 \nu_2  \sigma_2}}} K\indices{_{\lambda_1 \rho_1 \sigma_1}} K\indices{_{\lambda_2 \rho_2 \sigma_2}} \, .
\end{align}
The operator for the type $A$ terms \eqref{NewFunctional:OperatorA} becomes
\begin{align} \label{Covariant:OperatorA}
\nonumber \mathcal{K}\indices{_{A I}} \hd\indices{^{A I}} = & \bigg[ \frac{1}{2}\perp\indices{^{\lambda_1 \lambda_2}} h\indices{^{\tau_1 \tau_2}} h\indices{^{\omega_1 \omega_2}} \epsilon\indices{_{\mu \nu}} \epsilon\indices{_{\rho \sigma}} - 2 \epsilon\indices{^{\lambda_1 \lambda_2}} h\indices*{^{\tau_1}_{\rho}} h\indices*{^{\tau_2}_{\sigma}} h \indices{^{\omega_1 \omega_2}} \epsilon\indices{_{\mu \nu}} - 2 \perp\indices{^{\lambda_1 \lambda_2}} h\indices*{^{\tau_1}_{\mu}} h\indices*{^{\omega_1}_{\rho}} h\indices*{^{\tau_2}_{\nu}} h\indices*{^{\omega_2}_{\sigma}}  \\
\nonumber & \; - 2 \big( \perp\indices{^{\lambda_1 \lambda_2}} \perp\indices{_{\mu \rho}} + \epsilon\indices{^{\lambda_1 \lambda_2}} \epsilon\indices{_{\mu \rho}} \big) h\indices*{^{\tau_1}_{\nu}} h\indices*{^{\tau_2}_{\sigma}} h\indices{^{\omega_1 \omega_2}} \bigg] K\indices{_{\lambda_1 \tau_1 \omega_1}} K\indices{_{\lambda_2 \tau_2 \omega_2}} \frac{\partial}{\partial R\indices{_{\mu \nu \rho \sigma}}} \\
 & \; + 2 \left( \perp\indices{_{\mu_2}^{\mu_1}} \perp\indices{_{\rho_2}^{\rho_1}} - \epsilon\indices{_{\mu_2}^{\mu_1}} \epsilon\indices{_{\rho_2}^{\rho_1}} \right) h\indices*{^{\nu_1}_{\nu_2}} h\indices*{^{\sigma_1}_{\sigma_2}} R\indices{_{\mu_1 \nu_1 \rho_1 \sigma_1}} \frac{\partial}{\partial R\indices{_{\mu_2 \nu_2 \rho_2 \sigma_2}}} \, ,
\end{align}
while that of type $B$ terms reads
\begin{equation} \label{Covariant:OperatorB}
\mathcal{K}\indices{_{B I}} \hd\indices{^{B I}} =4 \left[  \perp\indices*{_{\mu_2}^{\mu_1}} h\indices*{^{\nu_1}_{\nu_2}} h\indices*{^{\rho_1}_{\rho_2}} h\indices*{^{\sigma_1}_{\sigma_2}} +   \perp\indices*{^{\mu_1}_{\mu_2}} \perp\indices*{^{\nu_1}_{\nu_2}} \perp\indices*{^{\rho_1}_{\rho_2}} h\indices*{^{\sigma_1}_{\sigma_2}} \right] R\indices{_{\mu_1 \nu_1 \rho_1 \sigma_1}} \frac{\partial}{\partial R\indices{_{\mu_2 \nu_2 \rho_2 \sigma_2}}} \, .
\end{equation}
Note that, since they always appear in pairs, all the binormals in these expressions could be replaced by normal projectors via the first identity appearing in \req{relss}, %
%\begin{equation} \label{Covariant:IdentityBinormal}
%\epsilon\indices{_{\mu \nu}} \epsilon\indices{_{\rho \sigma}} = 2 \perp\indices{_{\mu [\rho}} \perp\indices{_{\sigma] \nu}} \, ,
%\end{equation}
so the whole thing would be written exclusively in terms of contractions of $h_{\mu\nu}$ and $\perp_{\mu\nu}$ with  curvature tensors. 

The covariant form of the full holographic entanglement entropy functional can be finally written as
%Let us finally conclude by rewriting the full functional in this covariant form. The entanglement entropy would be given by:
%
\begin{align} \label{Covariant:FullFunctional}
&S_{\rm \ssc HEE}(A) = -2 \pi  \int_{\Gamma_A} d^{D-2} y \, \sqrt{h} \, \Bigg[  \epsilon\indices{_{\mu \nu}} \epsilon\indices{_{\rho \sigma}}\frac{\partial \mathcal{L}_E}{\partial R\indices{_{\mu \nu \rho \sigma}}}  \\
\nonumber & - \sum_{S = 0}^{\infty} \frac{1}{S!} \int_0^1 {\rm d}u \, 2u : \left(- (1-u^2) \mathcal{K}\indices{_{AI}} \hd^{AI} - (1- u)\mathcal{K}\indices{_{BJ}} \hd^{BJ}\right)^{S} : \D2R \Bigg] ~ ,
\end{align}
where derivatives are to be taken respecting the normal ordering prescription introduced in \eqref{NewFunctional:NormalOrdering}, and the covariant form of the objects appearing in the last line are given in \eqref{Covariant:SecondDerivative}--\eqref{Covariant:OperatorB}.

\section{Explicit covariant form of the functionals}\label{sec:explicit}
In this section we present the explicit holographic entanglement entropy functionals for various classes of higher-curvature theories. Like in the rest of the paper, our approach  here is to consider such terms as perturbative corrections to Einstein gravity, so that entanglement entropies are computed by the on-shell evaluation of the corresponding Ryu-Takayanagi surfaces on the corrected functionals obtained using the Einstein gravity splitting. We start with a review of the previously known cases of $f(R)$, Lovelock and quadratic theories, for which the splitting problem plays no r\^ole (and hence the functionals can be also used non-perturbatively). Then, we present new functionals valid for general cubic and quartic theories at leading order in the couplings. We also show that for theories constructed from general contractions of the Ricci tensor and the metric, the anomaly piece vanishes at leading order in the couplings. We observe that the same happens for densities involving a single Riemann tensor, and make general comments on the structure of the perturbative functionals as a function of the number of Riemann tensors.

\subsection{$f(R)$ gravities}
Let us start with $f(R)$ theories. These are the simplest modifications of the Einstein-Hilbert action within the pure-metric class. For an action of the form
\begin{equation}
I_E^{ f(R)}=-\frac{1}{16\pi G} \int \rd^{d+1}x \sqrt{|g|} \left[\frac{d(d-1)}{L^2}+ R+ f(R) \right]\, ,
\end{equation}
the HEE functional only contains a Wald-like piece and is simply given by \cite{Dong:2013qoa}
\begin{equation}
S_{\rm \ssc HEE}^{ f(R)}=\frac{\mathcal{A}(\Gamma_A)}{4G}+\frac{1}{4 G} \int_{\Gamma_A} \rd^{d-1}y \sqrt{h} f'(R)\, .
\end{equation}
Since there is no anomaly piece, this expression can be used non-perturbatively in the putative $f(R)$ couplings by extremizing the full functional.

\subsection{Lovelock gravities}\label{love1}
Let us move to Lovelock theories \cite{Lovelock1,Lovelock2,Padmanabhan:2013xyr}. These are the most general diffemorphism-invariant pure-metric theories of gravity which possess covariantly-conserved second-order equations of motion. The general Euclidean action in $d+1$ dimensions reads
\begin{equation}\label{lovel}
I_E^{\rm Lovelock}=-\frac{1}{16\pi G} \int \rd^{d+1}x \sqrt{|g|} \left[\frac{d(d-1)}{L^2}+ R+\sum_{n=2}^{\lfloor \frac{d+1}{2} \rfloor } \lambda_n L^{2(n-1)} \mathcal{X}_{2n}(R) \right]\, ,
\end{equation}
where $\lfloor x \rfloor$ is the integer part of $x$, the $\lambda_n$ are dimensionless couplings and the order-$n$ invariants $\mathcal{X}_{2n}$ were defined in \req{AnomalyLovelock:LovelockLagrangian} above.
%\begin{equation}
% \mathcal{X}_{2n}(R) \equiv \frac{1}{2^n} \delta_{\nu_1 \nu_2 \cdots \nu_{2n-1}\nu_{2n}}^{\mu_1 \mu_2\cdots \mu_{2n-1}\mu_{2n}} R^{\nu_1 \nu_2}_{\mu_1 \mu_2} \cdots R^{\nu_{2n-1}}_{\mu_{2n-1}\mu_{2n}}\, ,
%\end{equation}
%where $\delta_{\nu_1 \nu_2 \cdots \nu_{2n-1}\nu_{2n}}^{\mu_1 \mu_2\cdots \mu_{2n-1}\mu_{2n}}$  is the totally antisymmetric   product of $2n$ Kronecker deltas. 
$\mathcal{X}_{2n}$ becomes the Euler density of compact manifolds when evaluated in $2n$ dimensions. The simplest Lovelock theories (besides Einstein gravity) correspond to the Gauss-Bonnet and cubic densities, which read respectively
\begin{align}
\mathcal{X}_4=&+ R^2-4R_{\mu\nu}R^{\mu\nu}+R_{\mu\nu\rho\sigma}R^{\mu\nu\rho\sigma}\, ,\\
\mathcal{X}_6=&+R^3-12R_{\mu}^{\nu} R_{\nu}^{\mu} R +16 R_{\mu}^{\nu}R_{\nu}^{\rho}R_{\rho}^{\mu}+24 R_{\mu\nu\rho\sigma}R^{\mu\rho}R^{\nu\sigma}+3 R R_{\mu\nu\rho\sigma}R^{\mu\nu\rho\sigma} \\ &-24 R_{\mu\nu\rho\sigma}R^{\mu\nu\rho}\,_{\gamma}R^{\sigma\gamma}-8 R\indices{_{\mu}^{\rho}_{\nu}^{\sigma}} R\indices{_{\rho}^{\gamma}_{\sigma}^{\delta}} R\indices{_{\gamma}^{\mu}_{\delta}^{\nu}}+ 4 R\indices{_{\mu\nu}^{\rho\sigma}} R\indices{_{\rho\sigma}^{\gamma\delta}} R\indices{_{\gamma\delta}^{\mu\nu}}\, .
\end{align}

As we mentioned before, for theories beyond quadratic order, the splitting problem challenges the construction of general entanglement entropy functionals. However, the special structure of Lovelock theories makes them unaffected by the splittings choice \cite{Camps:2014voa,Miao:2015iba,Camps:2016gfs}. The entanglement entropy is then unambiguously given by the JM functional \cite{Jacobson:1993xs,Hung:2011xb} previously mentioned. This reads, for a general Lovelock theory,
\begin{equation}\label{jm}
S_{\rm \ssc HEE}^{\rm Lovelock}=\frac{\mathcal{A}(\Gamma_A)}{4G}+ \sum_{n=2}^{\lfloor \frac{d+1}{2} \rfloor} \frac{L^{2(n-1)}}{4G} \int_{\Gamma_A} \rd^{d-1}y \sqrt{h}   \lambda_n  \Delta_n^{\rm Lovelock} \, , \end{equation}
where
\begin{equation}
\Delta_n^{\rm Lovelock}= n \mathcal{X}_{2(n-1)}(\mathcal{R})\, ,
\end{equation}
where the lower-order densities are computed with respect to the induced metric $h_{ij}$. %Observe that the functional is fully constructed in terms of intrinsic curvatures associated to the holographic entangling surface. %As we saw in the previous section, an alternative form of this functional 

 %Eq.(\ref{jm}) has been used to evaluate entanglement entropy non-perturbatively in the $\lambda_n$ in previous works \cite{}.

%\comment{new exponential Joan formula?}

\subsection{Quadratic gravities}
Next we consider theories involving up to four derivatives of the metric. The most general action can be written as
\begin{equation}\label{quaact}
I_E^{\rm Riem^2}=-\frac{1}{16\pi G} \int \rd^{d+1}x \sqrt{|g|} \left[\frac{d(d-1)}{L^2}+R+L^2 \sum_{i=1}^3 \alpha_i \mathcal{L}_i^{(2)} \right]\, ,
\end{equation}
where
\begin{equation}
\mathcal{L}_1^{(2)}\equiv R^2\, , \quad \mathcal{L}_2^{(2)}\equiv R_{\mu\nu}R^{\mu\nu}\, , \quad \mathcal{L}_3^{(2)}\equiv R_{\mu\nu\rho\sigma}R^{\mu\nu\rho\sigma}\, .
\end{equation}
The HEE functional for this class of theories was first obtained in \cite{Fursaev:2013fta}. It reads
\begin{equation}\label{seerie2}
\see^{\rm Riem^2}=\frac{\mathcal{A}(\Gamma_A)}{4G}+ \frac{L^2}{4G} \int_{\Gamma_A} \rd^{d-1}y \sqrt{h} \sum_{i=1}^3\alpha_i \Delta_i^{(2)}\, ,
\end{equation}
where
\begin{equation}
\Delta_1^{(2)}=2 R\, , \quad \Delta_2^{(2)}= R^a_a-\frac{1}{2}K^a K_a \, , \quad \Delta_3^{(2)}= 2 \left(R^{ab}_{ab}-K_{aij}K^{aij} \right)\, .
\end{equation}
Just like for $f(R)$, and Lovelock theories, there is no splitting problem in this case as the expressions only involve terms quadratic in extrinsic curvatures. Consequently, \req{seerie2} can be trusted at all orders in $\alpha_i$. %Once again, obtaining reliable results beyond the perturbative regime requires finding the corresponding surface extremizing the full $\see^{\rm Riem^2}$ functional, and then evaluating it on the solution. Since the corresponding equations will be now higher order on the parametrization, finding a solution will require imposing additional boundary conditions in a way which is not very transparent. 
%\comment{ How did you deal with this in your paper, Alejandro? }
 When the terms are considered as perturbative corrections to Einstein gravity, the above expressions get slightly simplified, namely
\begin{equation}\label{seerie22}
S_{\rm \ssc HEE}^{\rm Riem^2}=\frac{\mathcal{A}(\Gamma_A)}{4G}+ \frac{L^2}{4G} \int_{\Gamma_A} \rd^{d-1}y \sqrt{h} \sum_{i=1}^3\alpha_i \Delta_i^{(2)} + \mathcal{O}(\alpha_i^2)\, ,
\end{equation}
where now
\begin{equation}
\Delta_1^{(2)}=2 R\, , \quad \Delta_2^{(2)}= R^a_a \, , \quad \Delta_3^{(2)}= 2 \left(R^{ab}_{ab}-K_{aij}K^{aij} \right)\, .
\end{equation}
The difference with respect to the nonperturbative case is the fact that, in this case, the functional that needs to be extremized is the RT one, whose equation of motion reads $K^a=0$. We can then remove all the traces of extrinsic curvatures appearing in the higher-curvature functionals when looking for expressions valid at leading order in the couplings.

\subsection{Cubic gravities}
Let us now move to the cubic case. At this order, there are eight independent cubic invariants 
\begin{equation}\label{cubic}
I_E^{\rm Riem^3}=-\frac{1}{16\pi G} \int \rd^{d+1}x \sqrt{|g|} \left[\frac{d(d-1)}{L^2}+R+L^4 \sum_{i=1}^8 \beta_i \mathcal{L}_i^{(3)} \right] \, .
\end{equation}
We label our basis of densities as follows
\begin{align}
&\mathcal{L}_1^{(3)} \equiv R\indices{_{\mu}^{\rho}_{\nu}^{\sigma}} R\indices{_{\rho}^\delta_\sigma^\gamma} R\indices{_\delta^\mu_\gamma^\nu}    \, , &&\mathcal{L}_2^{(3)} \equiv R\indices{_{\mu\nu}^{\rho\sigma}}R\indices{_{\rho\sigma}^{\delta\gamma}} R\indices{_{\delta\gamma}^{\mu\nu}}\, ,   \\ \notag
&\mathcal{L}_3^{(3)}\equiv R_{\mu\nu\rho\sigma}R\indices{^{\mu\nu\rho}_\delta} R^{\sigma\delta}\, ,  &&\mathcal{L}_4^{(3)} \equiv R_{\mu\nu\rho\sigma} R^{\mu\nu\rho\sigma}R \, ,  \\ \notag
& \mathcal{L}_5^{(3)}\equiv R_{\mu\nu\rho\sigma} R^{\mu \rho} R^{\nu\sigma}\, , && \mathcal{L}_6^{(3)}\equiv R_{\mu}^{\nu} R_{\nu}^{\rho}R_{\rho}^{\mu}\, ,  \\ \notag
& \mathcal{L}_7^{(3)} \equiv R_{\mu\nu}R^{\mu\nu} R\, ,   && \mathcal{L}_8^{(3)}\equiv R^3\, .  
 \end{align}
 Using our new formula in \req{AnomalyCubic:ExpandedExpression} for the anomaly piece, we
% Following the procedure outlined in the previous section,
 find the following expression for the functional corresponding to a general cubic Lagrangian of the form (\ref{cubic}),
% For the sake of clarity, we will present the HEE functionals one by one, turning off all the others in each case. The full $\see^{\rm Riem^3}$ will be just the addition of all the different terms. We have
\begin{equation}\label{cubicfun}
S_{\rm \ssc HEE}^{\rm Riem^3}=\frac{\mathcal{A}(\Gamma_A)}{4G}+\frac{L^4}{4G} \int_{\Gamma_A} \rd^{d-1}y \sqrt{h}   \sum_{i=1}^8\beta_i \Delta_i^{(3)} + \mathcal{O}(\beta_i^2)\, ,
\end{equation} 
 where the new terms read
 \begin{align}\label{cubic1}
\Delta_1^{(3)}=& +\frac{3}{2} \left( R^{a \nu}{}_{a \mu} R^{b \mu}{}_{b \nu} - R^{a \nu b \mu} R_{a \mu b \nu} \right)\\ \notag &-\frac{3}{2} R^{ijkl} K_{a i k} K^{a}{}_{jl} - 3 R^{a b i j} K_{a i}{}^{k} K_{b j k} + \frac{3}{4} R^{ab}{}_{ab} K^{c i j} K_{c i j} - 
 \frac{3}{8} K^{a i j} K_{a i j} K^{b k l} K_{b k l}\\ \notag & + \frac{9}{4} K_{a i}{}^j K_{bj}{}^k K^a{}_{k}{}^l K^b{}_l{}^i - \frac{3}{2} K_{a i}{}^j K^a{}_{j}{}^k K_{b k}{}^l K^b{}_l{}^i 
- \frac{3}{4} K_{aij} K_{b k l} K^{bij} K^{akl}\, ,\\
\Delta_2^{(3)}=&+3 R^{a b \rho \sigma} R_{a b \rho \sigma} \\ \notag &-6 K_{a i}{}^k K_{b j k} \left( R^{a i b j} - R^{b i a j} \right) -6 K_{a i k} K^{a j k} R^{b i}{}_{bj}  +3 K_{a i}{}^j K_{bj}{}^k K^a{}_{k}{}^l K^b{}_l{}^i \\ \notag &-6 K_{a i}{}^j K^a{}_{j}{}^k K_{b k}{}^l K^b{}_l{}^i  \, , \\
\Delta_3^{(3)}=& +\frac{1}{2} R^{a \mu \nu \rho} R_{a \mu \nu \rho} +2 R^{a \lambda} R^b{}_{a b \lambda} \\ \notag &
- K^{a}{}_{i}{}^k K_{a j k} R^{ij} -\frac{1}{2} K^{aij} K_{aij} R^b{}_b \, ,\\
 \Delta_4^{(3)}=&+ R_{\mu \nu \rho \sigma} R^{\mu \nu \rho \sigma} +2 R R^{ab}{}_{ab}   \\ \notag &  -2 K^{a i j} K_{a i j}  R    \, , \\
\Delta_5^{(3)}=&+ R_\mu{}^\nu R^{a \mu}{}_{a \nu} -\frac{1}{2} R^{ab} R_{ab} +\frac{1}{2} R^a{}_a R^b{}_b  \, ,\\
  \Delta_6^{(3)}=& +\frac{3}{2} R^{a \mu} R_{a \mu}  \, , \\
\Delta_7^{(3)}=& +R_{\mu \nu} R^{\mu \nu} +R^a{}_a R  
 \, ,\\
  \Delta_8^{(3)}=& +3R^2 \, .
\end{align}
In each case, the first line corresponds to the Wald-like piece, whereas the rest come from the anomaly one. In the above expressions we have already made use of the RT on-shell condition $K^a=0$. If they were to be used nonperturbatively (including extremization of the whole functionals, etc.), additional terms would appear \cite{Caceres:2020jrf}. However, in that case one would need to find first the right splittings in each case and the whole functionals would (most likely) change completely ---although the results at $\mathcal{O}(\beta_i)$ will have to reduce to the ones found using the perturbatively valid ones presented here. 

We observe that the first two functionals, which are the only ones involving chains of three Riemann tensors, have the most complicated expressions. On the other hand, $\Delta_3^{(3)}$ and $\Delta_4^{(3)}$, which involve pairs of Riemann tensors are simpler but still have pieces coming from the anomaly part. Finally, densities with a single Riemann or none have a vanishing anomaly piece and their HEE functionals at leading order are just given by the corresponding Wald-entropy expressions. We will see later that this hierarchy in the level of complication of the functionals as a function of the number of Riemann tensors involved actually extends to general-order densities.

Besides the cubic Lovelock densities, there are other interesting theories one can consider and whose HEE functionals can be straightforwardly obtained by replacing the corresponding combinations of $\beta_i$ in \req{cubicfun}.
Below, when computing EE universal terms, we will also make explicit the results for a couple of such theories in $d=4$ and $d=3$, respectively. The first is five-dimensional Quasi-topological gravity \cite{Quasi2,Quasi,Oliva:2011xu,Oliva:2012zs}, whose action reads
\begin{equation}
I_E^{\rm QTG}=-\frac{1}{16\pi G} \int \diff^{5}x \sqrt{g} \left[\frac{12}{L^2}+R+\frac{7\mu_{\rm QTG} L^4}{4} \mathcal{Z}_5\right]\, ,
\end{equation}
where
\begin{equation}
\mathcal{Z}_5\equiv \mathcal{L}_1^{(3)}-\frac{9}{7} \mathcal{L}_3^{(3)}+\frac{3}{8} \mathcal{L}_4^{(3)}  +\frac{15}{7} \mathcal{L}_5^{(3)} + \frac{18}{7} \mathcal{L}_6^{(3)} -\frac{33}{14}  \mathcal{L}_7^{(3)}+\frac{15}{56} \mathcal{L}_8^{(3)}\, ,
\end{equation}
and where we have omitted the usual Gauss-Bonnet density which is usually included in the action. The second is four-dimensional Einsteinian cubic gravity \cite{PabloPablo,Hennigar:2016gkm,PabloPablo2}, whose action is given by
\begin{equation}\label{cubic2}
I^{\rm  ECG}_E=- \frac{1}{16\pi G} \int \diff ^{4}x \sqrt{g} \left[\frac{6}{L^2}+R- \frac{\mu_{\rm ECG} L^4}{8} \mathcal{P} \right]\, , 
\end{equation}
where
\begin{equation}
 \mathcal{P}\equiv  12 \mathcal{L}_{1}^{(3)}+\mathcal{L}_{2}^{(3)}-12\mathcal{L}_{5}^{(3)}+8\mathcal{L}_{6}^{(3)} \, .
\end{equation}
These theories define holographic toy models of non-supersymmetric CFTs in $d=4$ and $d=3$, respectively. Various holographic aspects of such models have been explored before \eg in \cite{Myers:2010jv,Myers:2010xs,Myers:2010tj,HoloECG,Bueno:2020odt,Bueno:2018yzo,Mir:2019ecg}. The special properties of Quasi-topological and Einsteinian cubic gravities include the fact that they possess second-order equations on maximally symmetric backgrounds, that they allow for generalizations of the Schwarzschild solution with a single function, \ie satisfying $g_{tt}g_{rr}=-1$,  as well as the fact that the associated thermodynamic properties can be computed fully analytically \cite{Hennigar:2017ego,Bueno:2017sui,Bueno:2019ycr}. 
%The second is four-dimensional Einsteinian cubic gravity \comment{something about properties of ECG and QT:  The special properties of ECG make it a natural toy model of a non-supersymmetric three-dimensional CFT suitable for holographic studies, similarly to Quasi-topological gravity in one additional dimension \cite{Myers:2010jv}. Various holographic aspects of ECG were explored in \cite{HoloECG}.

%\comment{particular theories}

\subsection{Quartic gravities}\label{quarticc}
At the following order, quartic in curvature, there are 26 independent densities one can write ---see \eg \cite{0264-9381-9-5-003,Aspects},
\begin{equation}\label{quartic}
I_E^{\rm Riem^4}=-\frac{1}{16\pi G} \int \rd^{d+1}x \sqrt{|g|} \left[\frac{d(d-1)}{L^2}+R+L^6 \sum_{i=1}^{26} \gamma_i \mathcal{L}_i^{(4)} \right]\, ,
\end{equation}
%where
%\begin{align}
%&\mathcal{L}_1^{(4)} \equiv   \, , &&\mathcal{L}_2^{(4)} \equiv \, ,   \\ \notag
%&\mathcal{L}_3^{(4)}\equiv  \, ,  &&\mathcal{L}_4^{(4)} \equiv \, ,  \\ \notag
%& \mathcal{L}_5^{(4)}\equiv \, , && \mathcal{L}_6^{(4)}\equiv \, ,  \\ \notag
%& \mathcal{L}_7^{(4)} \equiv \, ,   && \mathcal{L}_8^{(4)}\equiv \, ,  \\ \notag
%& \mathcal{L}_9^{(4)} \equiv \, ,   && \mathcal{L}_{10}^{(4)}\equiv \, ,  \\ \notag
%& \mathcal{L}_{11}^{(4)} \equiv \, ,   && \mathcal{L}_{12}^{(4)}\equiv \, ,  \\ \notag
%& \mathcal{L}_{13}^{(4)} \equiv \, ,   && \mathcal{L}_{14}^{(4)}\equiv \, ,  \\ \notag
%& \mathcal{L}_{15}^{(4)} \equiv \, ,   && \mathcal{L}_{16}^{(4)}\equiv \, ,  \\ \notag
%& \mathcal{L}_{17}^{(4)} \equiv \, ,   && \mathcal{L}_{18}^{(4)}\equiv \, ,  \\ \notag
%& \mathcal{L}_{19}^{(4)} \equiv \, ,   && \mathcal{L}_{20}^{(4)}\equiv \, ,  \\ \notag
%& \mathcal{L}_{21}^{(4)} \equiv \, ,   && \mathcal{L}_{21}^{(4)}\equiv \, ,  \\ \notag
%& \mathcal{L}_{23}^{(4)} \equiv \, ,   && \mathcal{L}_{24}^{(4)}\equiv \, ,  \\ \notag
%& \mathcal{L}_{25}^{(4)} \equiv \, ,   && \mathcal{L}_{26}^{(4)}\equiv \, ,  \\ \notag
%& \mathcal{L}_{27}^{(4)} \equiv \, ,   
 %\end{align}
where we choose our basis to be 
\begin{align}
&\mathcal{L}_{26}^{(4)} \equiv R^4 \, , &&\mathcal{L}_{25}^{(4)} \equiv R^2 R\indices{_{\mu \nu}} R\indices{^{\mu \nu}}  \, ,   \\ \notag
&\mathcal{L}_{24}^{(4)} \equiv R R\indices{_{\mu}^{\nu}} R\indices{_{\nu}^{\rho}} R\indices{_{\rho}^{\mu}} \, , &&\mathcal{L}_{23}^{(4)} \equiv R\indices{_{\mu \nu}} R\indices{^{\mu \nu}} R\indices{_{\rho \sigma}} R\indices{^{\rho \sigma}} \, ,   \\ \notag
&\mathcal{L}_{22}^{(4)} \equiv R\indices{_{\mu}^{\nu}} R\indices{_{\nu}^{\rho}} R\indices{_{\rho}^{\sigma}} R\indices{_{\sigma}^{\mu}} \, , &&\mathcal{L}_{21}^{(4)} \equiv R R\indices{_{\mu \nu \rho \sigma}} R\indices{^{\mu \rho}} R\indices{^{\nu \sigma}} \, ,   \\ \notag
&\mathcal{L}_{20}^{(4)} \equiv R\indices{^{\mu \nu}} R\indices{_{\mu \rho \nu \sigma}} R\indices{^{\delta \rho}} R\indices{_{\delta}^{\sigma}} \, , &&\mathcal{L}_{19}^{(4)} \equiv R^2 R\indices{_{\mu \nu \rho \sigma}} R\indices{^{\mu \nu \rho \sigma}} \, ,   \\ \notag
&\mathcal{L}_{18}^{(4)} \equiv R R\indices{_{\mu \nu \rho \sigma}} R\indices{^{\mu \nu \rho}_{\delta}} R\indices{^{\sigma \delta}} \, , &&\mathcal{L}_{17}^{(4)} \equiv R\indices{_{\delta \gamma}} R\indices{^{\delta \gamma}} R\indices{_{\mu \nu \rho \sigma}} R\indices{^{\mu \nu \rho \sigma}}  \, ,   \\ \notag
&\mathcal{L}_{16}^{(4)} \equiv R\indices{^{\mu \nu}} R\indices{_{\nu}^{\rho}} R\indices{^{\sigma \delta \gamma}_{\mu}} R\indices{_{\sigma \delta \gamma \rho}} \, , &&\mathcal{L}_{15}^{(4)} \equiv R\indices{^{\mu \nu}} R\indices{^{\rho \sigma}} R\indices{^{\delta \gamma}_{\mu \rho}} R\indices{_{\delta \gamma \nu \sigma}} \, ,   \\ \notag
&\mathcal{L}_{14}^{(4)} \equiv R\indices{^{\mu \nu}} R\indices{^{\rho \sigma}} R\indices{^{\delta}_{\mu}^{\gamma}_{\nu}} R\indices{_{\delta \rho \gamma \sigma}} \, , &&\mathcal{L}_{13}^{(4)} \equiv R\indices{^{\mu \nu}} R\indices{^{\rho \sigma}} R\indices{^{\delta}_{\mu}^{\gamma}_{\rho}} R\indices{_{\delta \nu \gamma \sigma}} \, ,   \\ \notag
&\mathcal{L}_{12}^{(4)} \equiv R R\indices{_{\mu \nu}^{\rho \sigma}} R\indices{_{\rho \sigma}^{\delta \gamma}} R\indices{_{\delta \gamma}^{\mu \nu}} \, , &&\mathcal{L}_{11}^{(4)} \equiv R R\indices{_{\mu}^{\rho}_{\nu}^{\sigma}} R\indices{_{\rho}^{\delta}_{\sigma}^{\gamma}} R\indices{_{\delta}^{\mu}_{\gamma}^{\nu}} \, ,   \\ \notag
&\mathcal{L}_{10}^{(4)} \equiv R\indices{^{\mu \nu}} R\indices{_{\mu}^{\rho}_{\nu}^{\sigma}} R{_{\delta \gamma \xi \rho}} R\indices{^{\delta \gamma \xi}_{\sigma}} \, , &&\mathcal{L}_{9}^{(4)} \equiv R\indices{^{\mu \nu}} R\indices{^{\rho \sigma \delta \gamma}} R\indices{_{\rho \sigma}^{\xi}_{\mu}} R\indices{_{\delta \gamma \xi \nu}} \, ,   \\ \notag
&\mathcal{L}_{8}^{(4)} \equiv R\indices{^{\mu \nu}} R\indices{^{\rho \sigma \delta \gamma}} R\indices{_{\rho}^{\xi}_{\delta \mu}} R\indices{_{\sigma \xi \gamma \nu}} \, , &&\mathcal{L}_{7}^{(4)} \equiv R\indices{_{\mu \nu \rho \sigma}} R\indices{^{\mu \nu \rho \sigma}} R\indices{_{\delta \gamma \xi \chi}} R\indices{^{\delta \gamma \xi \chi}} \, ,   \\ \notag
&\mathcal{L}_{6}^{(4)} \equiv R\indices{^{\mu \nu \rho \sigma}} R\indices{_{\mu \nu \rho}^{\delta}} R\indices{_{\gamma \xi \chi \sigma}} R\indices{^{\gamma \xi \chi}_{\delta}} \, , &&\mathcal{L}_{5}^{(4)} \equiv R\indices{^{\mu \nu \rho \sigma}} R\indices{_{\mu \nu}^{\delta \gamma}} R\indices{_{\delta \gamma}^{\chi \xi}} R\indices{_{\rho \sigma \chi \xi}} \, ,   \\ \notag
&\mathcal{L}_{4}^{(4)} \equiv R\indices{^{\mu \nu \rho \sigma}} R\indices{_{\mu \nu}^{\delta \gamma}} R\indices{_{\rho \delta}^{\chi \xi}} R\indices{_{\sigma \gamma \chi \xi}} \, , &&\mathcal{L}_{3}^{(4)} \equiv R\indices{^{\mu \nu \rho \sigma}} R\indices{_{\mu \nu}^{\delta \gamma}} R\indices{_{\rho}^{\chi}_{\delta}^{\xi}} R\indices{_{\sigma \chi \gamma \xi}} \, , \\ \notag
&\mathcal{L}_{2}^{(4)} \equiv R\indices{^{\mu \nu \rho \sigma}} R\indices{_{\mu}^{\delta}_{\rho}^{\gamma}} R\indices{_{\delta}^{\chi}_{\gamma}^{\xi}} R\indices{_{\nu \chi \sigma \xi}} \, , &&\mathcal{L}_{1}^{(4)} \equiv R\indices{^{\mu \nu \rho \sigma}} R\indices{_{\mu}^{\delta}_{\rho}^{\gamma}} R\indices{_{\delta}^{\chi}_{\nu}^{\xi}} R\indices{_{\gamma \chi \sigma \xi}} \, .  
\end{align}
Using our formula in \req{AnomalyQuartic:ExpandedExpression}, we find the explicit functional for the above densities to be given by
\begin{equation}
S_{\rm \ssc HEE}^{\rm Riem^4}=\frac{\mathcal{A}(\Gamma_A)}{4G}+\frac{L^6}{4G} \int_{\Gamma_A} \rd^{d-1}y \sqrt{\gamma}   \sum_{i=1}^{26}\gamma_i \Delta_i^{(4)} + \mathcal{O}(\gamma_i^2)\, ,
\end{equation}
where now we find
\begin{align}\label{quartic1}
\Delta_{26}^{(4)}=& + 4 R^3  
 \, ,\\
\Delta_{25}^{(4)}=& + 2 R R\indices{_{\mu \nu}} R\indices{^{\mu \nu}} + R^2 R\indices{^a_a}  
 \, ,\\
\Delta_{24}^{(4)}=& + R\indices{_{\mu}^{\nu}} R\indices{_{\nu}^{\rho}} R\indices{_{\rho}^{\mu}} + \frac{3}{2} R R^{a \mu} R_{a \mu}  
 \, ,\\
\Delta_{23}^{(4)}=& + 2 R\indices{^a_a} R\indices{_{\mu \nu}} R\indices{^{\mu \nu}}  
 \, ,\\
\Delta_{22}^{(4)}=& + 2 R\indices{^{a \mu}} R\indices{_a^{\nu}} R\indices{_{\mu \nu}}  
 \, ,\\
\Delta_{21}^{(4)}=& + R\indices{_{\mu \nu \rho \sigma}} R\indices{^{\mu \rho}} R\indices{^{\nu \sigma}} + R R\indices{^{a \mu}_{a \nu}} R\indices{^{\nu}_{\mu}} - \frac{1}{2} R R\indices{^{a b}} R\indices{_{a b}} + \frac{1}{2} R R\indices{^a_a} R\indices{^b_b}  
 \, ,\\
\Delta_{20}^{(4)}=& - R\indices{_{a \mu \nu \rho}} R\indices{^{a \rho}} R\indices{^{\mu \nu}} + \frac{1}{2} R\indices{^{a \mu}_{a \nu}} R\indices{_{\mu}^{\rho}} R\indices{_{\rho}^{\nu}} - \frac{1}{2} R\indices{_{\mu}^a} R\indices{_a^b} R\indices{_b^{\mu}} + \frac{1}{2} R\indices{^a_a} R\indices{_{\mu b}} R\indices{^{\mu b}}  
 \, ,\\
\Delta_{19}^{(4)}=& + 2 R R\indices{_{\mu \nu \rho \sigma}} R\indices{^{\mu \nu \rho \sigma}} + 2 R^2 R\indices{^{a b}_{a b}} \\ \notag
 & - 2 R^2 K\indices{^{a i j}} K\indices{_{a i j}} 
 \, ,\\
\Delta_{18}^{(4)}=& + R\indices{_{\mu \nu \rho \sigma}} R\indices{^{\mu \nu \rho}_{\delta}} R\indices{^{\sigma \delta}} + \frac{1}{2} R R\indices{_{a \mu \nu \rho}} R\indices{^{a \mu \nu \rho}} + 2 R R\indices{^{a b}_{a \mu}} R\indices{^{\mu}_b} \\ \notag
 & - \frac{1}{2} R R\indices{^a_a} K\indices{^{b i j}} K\indices{_{b i j}} - R R\indices{^{ij}} K\indices{_{a i k}} K\indices{^a_j^k}
 \, ,\\
\Delta_{17}^{(4)}=& + R\indices{^a_a} R\indices{_{\mu \nu \rho \sigma}} R\indices{^{\mu \nu \rho \sigma}} + 2 R\indices{^{a b}_{a b}} R\indices{_{\mu \nu}} R\indices{^{\mu \nu}} \\ \notag
 & + 2 K\indices{^{a i j}} K\indices{_{a i j}} \left( R\indices{_{b k}} R\indices{^{b k}} + \frac{2}{3} R\indices{_{b c}} R\indices{^{b c}} - R\indices{_{\mu \nu}} R\indices{^{\mu \nu}} - \frac{1}{3} R\indices{^b_b} R\indices{^c_c} \right)
 \, ,\\
\Delta_{16}^{(4)}=& + R\indices{_{\mu}^a} R\indices{_{a \nu \rho \sigma}} R\indices{^{\mu \nu \rho \sigma}} + 2 R\indices{_{\mu}^a} R\indices{^b_{a b \nu}} R\indices{^{\mu \nu}} \\ \notag
 & - \frac{1}{2} K\indices{^{a i j}} K\indices{_{a i j}} \left( \frac{1}{2} R\indices{_{b k}} R\indices{^{b k}} + \frac{1}{3} R\indices{_{b c}} R\indices{^{b c}} + \frac{1}{3} R\indices{^b_b} R\indices{^c_c} \right) - K\indices{^a_i^k} K\indices{_{a j k}} \left( R\indices{^{i l}} R\indices{^j_l} + \frac{1}{2} R\indices{^{i b}} R\indices{^j_b} \right)
 \, ,\\
\Delta_{15}^{(4)}=& + R\indices{_{a \mu \rho \sigma}} R\indices{^a_{\nu}^{\rho \sigma}} R\indices{^{\mu \nu}} + 2 R\indices{^{a \mu}} R\indices{^{b \nu}} R\indices{_{a b \mu \nu}} \\ \notag
 & - K\indices{^a_i^k} K\indices{_{a j k}} \left( R\indices{^b_b} R\indices{^{i j}} - \frac{1}{2} R\indices{^{i b}} R\indices{^j_b} \right) - K\indices{^a_i^k} K\indices{^b_{j k}} R\indices{^{[i}_a} R\indices{^{j]}_b}
 \, ,\\
\Delta_{14}^{(4)}=& + R\indices{^a_{\mu a \rho}} R\indices{_{\nu \sigma}} R\indices{^{\mu \nu \rho \sigma}} + R\indices{^a_a} R\indices{^b_{\mu b \nu}} R\indices{^{\mu \nu}} - R\indices{_{a \mu b \nu}} R\indices{^{a b}} R\indices{^{\mu \nu}} \\ \notag
 & + \frac{1}{24} K\indices{^{a i j}} K\indices{_{a i j}} \left( R\indices{^b_b} R\indices{^c_c} - 2 R\indices{_{b c}} R\indices{^{b c}} \right) - \frac{1}{4} K\indices{^a_i^k} K\indices{_{a j k}} R\indices{^{i b}} R\indices{^j_b} - \frac{1}{2} K\indices{^a_i^k} K\indices{^b_{j k}} R\indices{^{[i}_a} R\indices{^{j]}_b} \\ \notag
 & - \frac{1}{2} K\indices{^a_{i j}} K\indices{_{a k l}} R\indices{^{i j}} R\indices{^{k l}} \, , \\
\Delta_{13}^{(4)}=& + R\indices{^a_{\mu \rho \nu}} R\indices{_a^{\mu}_{\sigma}^{\nu}} R\indices{^{\rho \sigma}} - R\indices{_{a \mu b \nu}} R\indices{^{a \nu}} R\indices{^{b \mu}} + R\indices{^a_{\mu a \nu}} R\indices{^{\mu b}} R\indices{^{\nu}_b} \\ \notag
 & - \frac{1}{8} K\indices{^{a i j}} K\indices{_{a i j}} R\indices{^b_b} R\indices{^c_c} - \frac{1}{2} K\indices{^a_i^k} K\indices{_{a j k}} R\indices{^b_b} R\indices{^{i j}} - \frac{1}{2} K\indices{^a_{i j}} K\indices{_{a k l}} R\indices{^{i k}} R\indices{^{j l}} \, , \\
\Delta_{12}^{(4)}=& + R\indices{_{\mu \nu}^{\rho \sigma}} R\indices{_{\rho \sigma}^{\delta \gamma}} R\indices{_{\delta \gamma}^{\mu \nu}} + 3 R R\indices{_{a b \mu \nu}} R\indices{^{a b \mu \nu}} \\ \notag
 & - 6 K\indices{_{a i}^k} K\indices{_{b j k}} \left( R\indices{^{a i b j}} - R\indices{^{b i a j}} \right) R - 6 K\indices{^a_i^k} K\indices{_{a j k}} R\indices{^{b i}_b^j} R + 3 K\indices{_{a i}^j} K\indices{_{b j}^k} K\indices{^a_k^l} K\indices{^b_l^i} R \\ \notag
 & - 6 K\indices{_{a i}^j} K\indices{^a_j^k} K\indices{_{b k}^l} K\indices{^b_l^i} R \, , \\
\Delta_{11}^{(4)}=& + R\indices{_{\mu}^{\rho}_{\nu}^{\sigma}} R\indices{_{\rho}^{\delta}_{\sigma}^{\gamma}} R\indices{_{\delta}^{\mu}_{\gamma}^{\nu}} + \frac{3}{2} R R\indices{^{a \mu}_{a \nu}} R\indices{^{b \nu}_{b \mu}} - \frac{3}{2} R R\indices{^{a \mu b \nu}} R\indices{_{b \mu a \nu}} \\ \notag
 & - \frac{3}{2} K\indices{^a_{i j}} K\indices{_{a k l}} R\indices{^{i k j l}} R - 3 K\indices{_{a i}^k} K\indices{_{b j k}} R\indices{^{a b i j}} R + \frac{3}{4} K\indices{^{a i j}} K\indices{_{a i j}} R\indices{^{b c}_{bc}} R - \frac{3}{8} K\indices{^{a i j}} K\indices{_{a i j}} K\indices{^{b k l}} K\indices{_{b k l}} R \\ \notag
 & - \frac{3}{4} K\indices{_{a i j}} K\indices{_{b k l}} K\indices{^{b i j}} K\indices{^{a k l}} R + \frac{9}{4} K\indices{_{a i}^j} K\indices{_{b j}^k} K\indices{^a_k^l} K\indices{^b_l^i} R - \frac{3}{2} K\indices{_{a i}^j} K\indices{^a_j^k} K\indices{_{b k}^l} K\indices{^b_l^i} R \, , \\ \notag
\Delta_{10}^{(4)}=& + \frac{1}{2} R\indices{^a_{\mu a \nu}} R\indices{^{\mu}_{\rho \sigma \delta}} R\indices{^{\nu \rho \sigma \delta}} + 2 R\indices{^{\mu \nu}} R\indices{_{\mu}^a_{\nu}^{\rho}} R\indices{^b_{a b \rho}} + \frac{1}{2} R\indices{^a_a} R\indices{_{b \mu \nu \rho}} R\indices{^{b \mu \nu \rho}} - \frac{1}{2} R\indices{^a_b} R\indices{_{a \mu \nu \rho}} R\indices{^{b \mu \nu \rho}} \\
 & + K\indices{_{a i j}} K\indices{_{b k l}} R\indices{^{i[a}} R\indices{^{b]k j l}} - K\indices{_{a i j}} K\indices{^{a}_{k l}} \left( R\indices{^{i j}} R\indices{_b^{k b l}} - \frac{1}{2} R\indices{^{i b}} R\indices{_b^{k j l}} \right) \\ \notag
 & - K\indices{_{a i}^k} K\indices{_{b j k}} R\indices{^{i [a}} R\indices{_c^{b] c j}} - \frac{1}{2} K\indices{^{a i j}} K\indices{_{a i j}} \left( R\indices{^{k l}} R\indices{^{b}_{k b l}} + R\indices{^{b k}} R\indices{^{c}_{b c k}} + \frac{1}{2} R\indices{^b_b} R\indices{^{c d}_{c d}} \right) \\ \notag
 & + K\indices{_{a i}^k} K\indices{^a_{j k}} \left( R\indices{_{b l}} R\indices{^{b i j l}} - \frac{1}{2} R\indices{_b^i} R\indices{_c^{b c j}} + \frac{2}{3} R\indices{_{b c}} R\indices{^{b i c j}} - \frac{1}{6} R\indices{_b^b} R\indices{_c^{i c j}} - R\indices{_{l m}} R\indices{^{i l j m}} \right) \\ \notag
 & + \frac{1}{2} K\indices{_{a l}^i} K\indices{_{b i}^j} K\indices{^b_j^k} K\indices{^a_{k m}} R\indices{^{l m}} - K\indices{_{a i}^j} K\indices{^a_j^k} K\indices{_{b k}^i} K\indices{^b_{l m}} R\indices{^{lm}} - \frac{1}{4} K\indices{^{a i j}} K\indices{_{a i j}} K\indices{_{b l}^k} K\indices{^b_{m k}} R\indices{^{l m}} \\ \notag
 & + \frac{1}{8} K\indices{^{a i j}} K\indices{_{a i j}} K\indices{^{b k l}} K\indices{_{b k l}} R\indices{^c_c} - \frac{1}{4} K\indices{_{a i}^j} K\indices{^a_j^k} K\indices{_{b k}^l} K\indices{^b_l^i} R\indices{^c_c} \, , \\
\Delta_{9}^{(4)}=& + \frac{1}{2} R\indices{^{\mu \nu \rho \sigma}} R\indices{_{\mu \nu}^{\xi a}} R\indices{_{\rho \sigma \xi a}} + R\indices{^{\mu \nu}} R\indices{^{a b \rho}_{\mu}} R\indices{_{a b \rho \nu}} - 2 R\indices{^{\mu a}} R\indices{^{\nu \rho b}_{\mu}} R\indices{_{\nu \rho a b}} \\ \notag
 & + 2 K\indices{_{a i j}} K\indices{_{b k l}} R\indices{^{i k}} R\indices{^{j [a b] l}} - K\indices{_{a i j}} K\indices{^a_{k l}} R\indices{^{i k}} R\indices{_b^{j b l}} + K\indices{_{a i}^k} K\indices{_{b j k}} \left( - 4 R\indices{^{i l}} R\indices{^{j [a b]}_l} + R\indices{^{i c}} R\indices{^{a b j}_c} \right. \\ \notag
 & \left. - 2 R\indices{^{a l}} R\indices{^{b [i j]}_l} + 6 R\indices{^{c [a}} R\indices{^{b] j i}_c} \right) + K\indices{_{a i}^k} K\indices{^a_{j k}} \left( - 2 R\indices{^{i l}} R\indices{^{b j}_{b l}} - R\indices{_b^{i}} R\indices{_c^{b c j}} + R\indices{_{b l}} R\indices{^{b i j l}} \right. \\ \notag
 & \left. - \frac{2}{3} R\indices{_{b c}} R\indices{^{b i c j}} - \frac{7}{6} R\indices{_b^b} R\indices{_c^{i c j}} \right) - \frac{3}{2} K\indices{_{a l}^i} K\indices{_{b i}^j} K\indices{^b_j^k} K\indices{^a_{k m}} R\indices{^{l m}} + \frac{3}{2} K\indices{_{a l}^i} K\indices{_{b i}^j} K\indices{^a_j^k} K\indices{^b_{k m}} R\indices{^{l m}} \\ \notag
 & - \frac{3}{2} K\indices{_{a l}^i} K\indices{^a_i^j} K\indices{_{b j}^k} K\indices{^b_{k m}} R\indices{^{l m}} + \frac{3}{4} K\indices{_{a i}^j} K\indices{_{b j}^k} K\indices{^a_k^l} K\indices{^b_l^i} R\indices{^c_c} - \frac{3}{2} K\indices{_{a i}^j} K\indices{^a_j^k} K\indices{_{b k}^l} K\indices{^b_l^i} R\indices{^c_c} \, , \\
\Delta_{8}^{(4)}=& + \frac{1}{2} R\indices{^{\mu \nu \rho \sigma}} R\indices{_{\mu}^a_{\rho}^{\xi}} R\indices{_{\nu a \sigma \xi}} + R\indices{^{\mu \nu}} R\indices{^{a \rho}_{[a| \mu}} R\indices{^b_{\rho |b] \nu}} + 2 R\indices{^{a \mu}} R\indices{_{\mu}^{\nu}_{[a|}^{\rho}} R\indices{^b_{\nu |b] \rho}} \\ \notag
 & - \frac{1}{2} K\indices{_{a i j}} K\indices{_{b k l}} \left( R\indices{^{i[a}} R\indices{^{b] k j l}} + R\indices{^{i k}} R\indices{^{a b j l}} \right) + K\indices{_{a i j}} K\indices{^a_{k l}} \left( R\indices{^{i m}} R\indices{^{j k l}_m} - \frac{1}{4} R\indices{_b^i} R\indices{^{b k j l}} - \frac{1}{4} R\indices{_b^b} R\indices{^{i k j l}} \right) \\ \notag
 & - K\indices{_{a i}^k} K\indices{_{b j k}} \left( R\indices{^{j l}} R\indices{^{a b i}_l} - \frac{1}{2} R\indices{^{a l}} R\indices{^{i j b}_l} + \frac{1}{2} R\indices{^{i [a}} R\indices{_c^{b] c j}} + \frac{3}{4} R\indices{_c^c} R\indices{^{a b i j}} \right) \\ \notag
 & + \frac{1}{4} K\indices{_{a i}^k} K\indices{^a_{j k}} \left( R\indices{^{i j}} R\indices{^{b c}_{b c}} - R\indices{_b^i} R\indices{_c^{b c j}}  \right) + \frac{1}{4} K\indices{^{a i j}} K\indices{_{a i j}} \left( R\indices{^{b k}} R\indices{^c_{b c k}} + R\indices{^b_b} R\indices{^{c d}_{c d}} \right) \\ \notag
 & - \frac{1}{2} K\indices{_{a l}^i} K\indices{_{b i}^j} K\indices{^b_j^k} K\indices{^a_{k m}} R\indices{^{l m}} + \frac{5}{4} K\indices{_{a l}^i} K\indices{_{b i}^j} K\indices{^a_j^k} K\indices{^b_{k m}} R\indices{^{l m}} - \frac{1}{4} K\indices{_{a l}^i} K\indices{^a_{i}^j} K\indices{_{b j}^k} K\indices{^b_{k m}} R\indices{^{l m}} \\ \notag
 & - \frac{1}{2} K\indices{_a^{i j}} K\indices{_{b i j}} K\indices{^a_l^k} K\indices{^b_{m k}} R\indices{^{l m}} - \frac{1}{8} K\indices{^{a i j}} K\indices{_{a i j}} K\indices{_{b l}^k} K\indices{^b_{m k}} R\indices{^{l m}} - \frac{1}{8} K\indices{_{a i j}} K\indices{_{b k l}} K\indices{^{b i j}} K\indices{^{a k l}} R\indices{^c_c} \\ \notag
 & - \frac{1}{8} K\indices{^{a i j}} K\indices{_{a i j}} K\indices{^{b k l}} K\indices{_{b k l}} R\indices{^c_c} + \frac{1}{2} K\indices{_{a i}^j} K\indices{_{b j}^k} K\indices{^a_k^l} K\indices{^b_l^i} R\indices{^c_c} - \frac{3}{8} K\indices{_{a i}^j} K\indices{^a_j^k} K\indices{_{b k}^l} K\indices{^b_l^i} R\indices{^c_c} \, , \\
\Delta_{7}^{(4)}=& + 4 R\indices{_{\mu \nu \rho \sigma}} R\indices{^{\mu \nu \rho \sigma}} R\indices{^{a b}_{a b}} \\ \notag
 & + \frac{64}{3} K\indices{_{a i j}} K\indices{_{b k l}} R\indices{^{i a j [b|}} R\indices{^{k |c] l}_c} - \frac{8}{3} K\indices{_{a i j}} K\indices{^a_{k l}} R\indices{^{i b j}_b} R\indices{^{k c l}_c} - 4 K\indices{^{a i j}} K\indices{_{a i j}} \left( R\indices{^{b c}_{b c}} R\indices{^{d e}_{de}} + 2 R\indices{^{b c d k}} R\indices{_{b c d k}} \right. \\ \notag 
 & \left. + 2 R\indices{^{b c k l}} R\indices{_{b c k l}} + \frac{8}{3} R\indices{^{b k c l}} R\indices{_{b k c l}} - \frac{4}{3} R\indices{^{b k c l}} R\indices{_{c k b l}} + \frac{4}{3} R\indices{_b^{k b l}} R\indices{^c_{k c l}} + 2 R\indices{^{b k l m}} R\indices{_{b k l m}} + R\indices{^{k l m n}} R\indices{_{k l m n}} \right) \\ \notag
 & - 8 K\indices{^{a i j}} K\indices{_{a i j}} K\indices{_{b k l}} K\indices{^b_{m n}} R\indices{^{k m l n}} - 16 K\indices{^{a i j}} K\indices{_{a i j}} K\indices{_{b k}^m} K\indices{_{c l m}} \left( R\indices{^{[b| k |c] l}} + R\indices{^{b c k l}} \right) \\ \notag
 & - 8 K\indices{^{a i j}} K\indices{_{a i j}} K\indices{_{b k}^m} K\indices{^b_{l m}} R\indices{_c^{k c l}} + 4 K\indices{^{a i j}} K\indices{_{a i j}} K\indices{^{b k l}} K\indices{_{b k l}} R\indices{^{c d}_{c d}} \\ \notag
 & + \frac{32}{3} K\indices{^{a i j}} K\indices{_{a i j}} K\indices{_{b k}^l} K\indices{_{c l}^m} K\indices{^b_m^n} K\indices{^c_n^k} - \frac{32}{3} K\indices{^{a i j}} K\indices{_{a i j}} K\indices{_{b k}^l} K\indices{^b_l^m} K\indices{_{c m}^n} K\indices{^c_n^k} \\ \notag
 & - \frac{8}{3} K\indices{^{a i j}} K\indices{_{a i j}} K\indices{_{b k l}} K\indices{_{c m n}} K\indices{^{c k l}} K\indices{^{b m n}} - \frac{4}{3} K\indices{^{a i j}} K\indices{_{a i j}} K\indices{^{b k l}} K\indices{_{b k l}} K\indices{^{c m n}} K\indices{_{c m n}} \, , \\
\Delta_{6}^{(4)}=& + 4 R\indices{^{\mu \nu \rho \sigma}} R\indices{^a_{\nu \rho \sigma}} R\indices{^b_{a b \mu}} \\ \notag
 & + 2 K\indices{_{a i j}} K\indices{_{b k l}} \left( R\indices{^{[b| i j m}} R\indices{^{|a] k l}_m} + \frac{4}{3} R\indices{^{i a j [b|}} R\indices{^{k |c] l}_c} + 2 R\indices{^{i k j [a|}} R\indices{^{l c |b]}_c} \right) \\ \notag
 & + K\indices{_{a i j}} K\indices{^a_{k l}} \left( \frac{1}{3} R\indices{^{i b k}_{b}} R\indices{^{j c l}_{c}} - \frac{2}{3} R\indices{^{i b k c}} R\indices{^{(j}_b^{l)}_c} - \frac{7}{3} R\indices{^{i b j}_b} R\indices{^{k c l}_c} - R\indices{^{i b j m}} R\indices{^k_b^l_m} + 2 R\indices{^{i k j b}} R\indices{^{l c}_{b c}} \right) \\ \notag
 & + K\indices{_{a i}^k} K\indices{_{b j k}} \left( R\indices{^{i [b| c d}} R\indices{^{j |a]}_{c d}} + \frac{4}{3} R\indices{^{i [b| c l}} R\indices{^{j |a]}_{c l}} + \frac{4}{3} R\indices{^{i [a b] l}} R\indices{^{j c}_{c l}} \right) \\ \notag
 & + K\indices{_{a i}^k} K\indices{^a_{j k}} \left( - \frac{3}{2} R\indices{^{i b c d}} R\indices{^j_{b c d}} + 2 R\indices{^{i b}_{l [c}} R\indices{^{j c l}_{b]}} - 3 R\indices{^{i b l c}} R\indices{^j_{b l c}} - 2 R\indices{^{i l b c}} R\indices{^j_{l b c}} - R\indices{^{i b l m}} R\indices{^j_{b l m}} \right. \\ \notag
 & \left. - 2 R\indices{^{i l b m}} R\indices{^j_{l b m}} - 2 R\indices{^{i l m n}} R\indices{^j_{l m n}} \right) + K\indices{^{a i j}} K\indices{_{a i j}} \left( - R\indices{^{b c}_{b c}} R\indices{^{d e}_{d e}} - \frac{3}{2} R\indices{^{b c d k}} R\indices{_{b c d k}} \right. \\ \notag
 & \left. - R\indices{^{b c k l}} R\indices{_{b c k l}} - \frac{4}{3} R\indices{^{b k c l}} R\indices{_{b k c l}} + \frac{2}{3} R\indices{^{b k c l}} R\indices{_{c k b l}} - \frac{2}{3} R\indices{_b^{k b l}} R\indices{^c_{k c l}} - \frac{1}{2} R\indices{^{b k l m}} R\indices{_{b k l m}} \right) \\ \notag
 & - 4 K\indices{_{a i}^k} K\indices{_{b k}^l} K\indices{^b_{l j}} K\indices{^a_{m n}} R\indices{^{i m j n}} - 4 K\indices{_{a i}^k} K\indices{_{b k}^l} K\indices{_{c l}^m} K\indices{^c_{m j}} \left( R\indices{^{a b i j}} + R\indices{^{[a| i |b] j}} \right) \\ \notag
 & - 2 K\indices{_{a i}^k} K\indices{^a_k^l} K\indices{_{b l}^m} K\indices{^b_{m j}} R\indices{_c^{i c j}} - 2 K\indices{^{a i j}} K\indices{_{a i j}} K\indices{_{b k}^m} K\indices{_{c l m}} \left( R\indices{^{b c k l}} + R\indices{^{[b| k |c] l}} \right) \\ \notag
 & - K\indices{^{a i j}} K\indices{_{a i j}} K\indices{_{b k}^m} K\indices{^b_{l m}} R\indices{_c^{k c l}} - 2 K\indices{_{a i}^j} K\indices{^a_j^k} K\indices{_{b k}^i} K\indices{^b_{l m}} R\indices{_c^{l c m}} + K\indices{^{a i j}} K\indices{_{a i j}} K\indices{^{b k l}} K\indices{_{b k l}} R\indices{^{c d}_{c d}} \\ \notag
 & + \frac{10}{3} K\indices{_{a i}^j} K\indices{^a_j^k} K\indices{_{b k}^l} K\indices{_{c l}^m} K\indices{^b_m^n} K\indices{^c_n^i} - 2 K\indices{_{a i}^j} K\indices{^a_j^k} K\indices{_{b k}^l} K\indices{_{c l}^m} K\indices{^c_m^n} K\indices{^b_n^i} \\ \notag 
 & - \frac{2}{3} K\indices{_{a i}^j} K\indices{^a_j^k} K\indices{_{b k}^l} K\indices{^b_l^m} K\indices{_{c m}^n} K\indices{^c_n^i} - \frac{2}{3} K\indices{_{a i}^j} K\indices{^a_j^k} K\indices{_{b k}^i} K\indices{^b_l^m} K\indices{_{c m}^n} K\indices{^c_n^l} \\ \notag
 & - \frac{4}{3} K\indices{_a^{i j}} K\indices{_{b i j}} K\indices{^a_k^l} K\indices{^b_l^m} K\indices{_{c m}^n} K\indices{^c_n^k} + K\indices{^{a i j}} K\indices{_{a i j}} K\indices{_{b k}^l} K\indices{_{c l}^m} K\indices{^b_m^n} K\indices{^c_n^k} \\ \notag
 & - \frac{4}{3} K\indices{^{a i j}} K\indices{_{a i j}} K\indices{_{b k}^l} K\indices{^b_l^m} K\indices{_{c m}^n} K\indices{^c_n^k} - \frac{1}{3} K\indices{^{a i j}} K\indices{_{a i j}} K\indices{^{b k l}} K\indices{_{b k l}} K\indices{^{c m n}} K\indices{_{c m n}} \, , \\
\Delta_{5}^{(4)}=& + 4 R\indices{^{\mu \nu \rho \sigma}} R\indices{_{\mu \nu}^{a b}} R\indices{_{\rho \sigma a b}} \\ \notag
 & + 16 K\indices{_{a i j}} K\indices{_{b k l}} R\indices{^{i [a b] k}} R\indices{^{j c l}_c} + 4 K\indices{_{a i j}} K\indices{^a_{k l}} \left( 2 R\indices{^{i b k c}} R\indices{^j_{[b}^l_{c]}} - R\indices{^{i b k}_b} R\indices{^{j c l}_c} \right) \\ \notag
 & + 8 K\indices{_{a i}^k} K\indices{_{b j k}} \left( R\indices{^{i [b| c d}} R\indices{^{j |a]}_{c d}} + \frac{4}{3} R\indices{^{i [b| c l}} R\indices{^{j |a]}_{c l}} + \frac{8}{3} R\indices{^{i [a b] l}} R\indices{^{j c}_{l c}} + R\indices{^{[b| i l m}} R\indices{^{|a] j}_{l m}} \right) \\ \notag
 & - 4 K\indices{_{a i}^k} K\indices{^a_{j k}} \left( R\indices{^{i b c d}} R\indices{^j_{b c d}} + \frac{8}{3} R\indices{^{i b l c}} R\indices{^j_{b l c}} - \frac{8}{3} R\indices{^{i b l}_{[c|}} R\indices{^{j c}_{l |b]}} + R\indices{^{i b l m}} R\indices{^j_{b l m}} \right) \\ \notag
 & + 48 K\indices{_{a i}^k} K\indices{_{b k}^l} K\indices{_{c l}^m} K\indices{^c_{m j}} R\indices{^{i [a b] j}} + 24 K\indices{_{a i}^k} K\indices{_{c k}^l} K\indices{_{b l}^m} K\indices{^c_{m j}} R\indices{^{a i b j}} \\ \notag 
 & - 12 K\indices{_{a i}^k} K\indices{_{c k}^l} K\indices{^c_{l}^m} K\indices{_{b m j}} R\indices{^{a i b j}} - 12 K\indices{_{c i}^k} K\indices{_{a k}^l} K\indices{_{b l}^m} K\indices{^c_{m j}} R\indices{^{a i b j}} \\ \notag
 & - 12 K\indices{_{a i}^k} K\indices{^a_k^l} K\indices{_{b l}^m} K\indices{^b_{m j}} R\indices{_c^{i c j}} + 12 K\indices{_{a i}^j} K\indices{^a_j^k} K\indices{_{b k}^l} K\indices{_{c l}^m} K\indices{^b_m^n} K\indices{^c_n^i} \\ \notag
 & - 12 K\indices{_{a i}^j} K\indices{^a_j^k} K\indices{_{b k}^l} K\indices{_{c l}^m} K\indices{^c_m^n} K\indices{^b_n^i} - 4 K\indices{_{a i}^j} K\indices{^a_j^k} K\indices{_{b k}^l} K\indices{^b_l^m} K\indices{_{c m}^n} K\indices{^c_n^i} \, , \\
\Delta_{4}^{(4)}=& + 4 R\indices{^{a b \mu \nu}} R\indices{_{a \mu}^{\rho \sigma}} R\indices{_{b \nu \rho \sigma}} \\ \notag
 & + 4 K\indices{_{a i j}} K\indices{_{b k l}} \left( 2 R\indices{^{i [a b] k}} R\indices{^{j c l}_c} - R\indices{^{[a| i k m}} R\indices{^{|b] l j}_m} \right) + 2 K\indices{_{a i j}} K\indices{^a_{k l}} \left( \frac{4}{3} R\indices{^{i b k}_{[c}} R\indices{^{l c j}_{b]}} \right. \\ \notag  
 & \left. - \frac{4}{3} R\indices{^{i b k c}}R\indices{^l_b^j_c} - R\indices{^{i b k m}}R\indices{^l_b^j_m} \right) + 2 K\indices{_{a i}^k} K\indices{_{b j k}} \left( - 2 R\indices{^{a b i j}} R\indices{^{c d}_{c d}} + \frac{4}{3} R\indices{^{i [a| c l}} R\indices{^{j |b]}_{c l}} \right. \\ \notag
 & \left. + \frac{4}{3} R\indices{^{i [a b] l}} R\indices{^{j c}_{l c}} - 4 R\indices{^{i j a l}} R\indices{^{b c}_{l c}} + R\indices{^{[a| i l m}} R\indices{^{|b] j}_{l m}} - 2 R\indices{^{a b l m}} R\indices{^{i j}_{l m}} \right) \\ \notag
 & + K\indices{_{a i}^k} K\indices{^a_{j k}} \left( - 2 R\indices{^{i b c d}} R\indices{^j_{b c d}} - \frac{14}{3} R\indices{^{i b l c}} R\indices{^j_{b l c}} + \frac{20}{3} R\indices{^{i b l}_{[c|}} R\indices{^{j c}_{l |b]}} - R\indices{^{i b l m}} R\indices{^j_{b l m}} \right) \\ \notag
 & - 4 K\indices{_{a i}^m} K\indices{_{b j m}} K\indices{^a_k^n} K\indices{^b_{l n}} R\indices{^{i j k l}} + 4 K\indices{_{a i}^k} K\indices{_{b k}^l} K\indices{_{c l}^m} K\indices{^c_{m j}} \left( R\indices{^{a b i j}} + 4 R\indices{^{i [a b] j}} \right) \\ \notag
 & + 4 K\indices{_{a i}^k} K\indices{_{c k}^l} K\indices{_{b l}^m} K\indices{^c_{m j}} \left( R\indices{^{a i b j}} - R\indices{^{a b i j}} \right) - 2 K\indices{_{a i}^k} K\indices{_{c k}^l} K\indices{^c_l^m} K\indices{_{b m j}} R\indices{^{a i b j}} \\ \notag
 & - 2 K\indices{_{c i}^k} K\indices{_{a k}^l} K\indices{_{b l}^m} K\indices{^c_{m j}} R\indices{^{a i b j}} - 6 K\indices{_{a i}^k} K\indices{^a_k^l} K\indices{_{b l}^m} K\indices{^b_{m j}} R\indices{_c^{i c j}} \\ \notag
 & + 2 K\indices{^{c l m}} K\indices{_{c l m}} K\indices{_{a i}^k} K\indices{_{b j k}} R\indices{^{a b i j}} + 2 K\indices{_{a i}^j} K\indices{_{b j}^k} K\indices{^a_k^l} K\indices{^b_l^i} R\indices{^{c d}_{c d}} \\ \notag
 & - 2 K\indices{_{a i}^j} K\indices{^a_j^k} K\indices{_{b k}^l} K\indices{^b_l^i} R\indices{^{c d}_{c d}} + \frac{2}{3} K\indices{_{a i}^j} K\indices{^a_j^k} K\indices{_{b k}^l} K\indices{_{c l}^m} K\indices{^b_m^n} K\indices{^c_n^i} \\ \notag
 & + 2 K\indices{_{a i}^j} K\indices{^a_j^k} K\indices{_{b k}^l} K\indices{_{c l}^m} K\indices{^c_m^n} K\indices{^b_n^i} - \frac{14}{3} K\indices{_{a i}^j} K\indices{^a_j^k} K\indices{_{b k}^l} K\indices{^b_l^m} K\indices{_{c m}^n} K\indices{^c_n^i} \\ \notag
 & - \frac{4}{3} K\indices{^{a i j}} K\indices{_{a i j}} K\indices{_{b k}^l} K\indices{_{c l}^m} K\indices{^b_m^n} K\indices{^c_n^k} + \frac{4}{3} K\indices{^{a i j}} K\indices{_{a i j}} K\indices{_{b k}^l} K\indices{^b_l^m} K\indices{_{c m}^n} K\indices{^c_n^k} \, , \\
\Delta_{3}^{(4)}=& + 2 R\indices{^{a b \mu \nu}} R\indices{_a^{\rho}_{\mu}^{\sigma}} R\indices{_{b \rho \nu \sigma}} + R\indices{^{\mu a \rho \sigma}} R\indices{^{\nu}_{a \rho \sigma}} R\indices{^b_{\mu b \nu}} - R\indices{^{\mu a \rho \sigma}} R\indices{^{\nu b}_{\rho \sigma}} R\indices{_{\mu b \nu a}} \\ \notag
 & - 2 K\indices{_{a i j}} K\indices{_{b k l}} \left( R\indices{^{i k j [a|}} R\indices{^{l c |b]}_c} + R\indices{^{i k a b}} R\indices{^{j c l}_c} + 2 R\indices{^{i [a b] m}} R\indices{^{j k l}_m} + R\indices{^{[a| i k m}} R\indices{^{j l |b]}_m} \right) \\ \notag
 & + K\indices{_{a i j}} K\indices{^a_{k l}} \left( 2 R\indices{^{i k b c}} R\indices{^j_b^l_c} - \frac{1}{2} R\indices{^{i k b c}} R\indices{^{j l}_{b c}} - R\indices{^{i k j b}} R\indices{^{l c}_{b c}} + R\indices{^{i k b m}} R\indices{^j_b^l_m} - \frac{1}{2} R\indices{^{i k b m}} R\indices{^{j l}_{b m}} \right. \\ \notag
 & \left. + 2 R\indices{^{i b m}_b} R\indices{^{j k l}_m} - \frac{1}{2} R\indices{^{i k m n}} R\indices{^{j l}_{m n}} \right) - K\indices{_{a i}^k} K\indices{_{b j k}} \left( R\indices{^{i [a b] j}} R\indices{^{c d}_{c d}} + R\indices{^{a b i j}} R\indices{^{c d}_{c d}} \right. \\ \notag
 & \left. + \frac{1}{2} R\indices{^{i [a| c d}} R\indices{^{j |b]}_{c d}} + 2 R\indices{^{i j a l}} R\indices{^{b c}_{l c}} + R\indices{^{[a| l i m}} R\indices{^{|b]}_m^j_l} + R\indices{^{a b l m}} R\indices{^{i j}_{l m}} + 2 R\indices{^{a l b m}} R\indices{^i_{[l}^j_{m]}} \right) \\ \notag
 & + K\indices{_{a i}^k} K\indices{^a_{j k}} \left( \frac{1}{2} R\indices{_b^{i b j}} R\indices{^{c d}_{c d}} - \frac{1}{4} R\indices{^{i b c d}} R\indices{^j_{b c d}} - R\indices{^b_{l b m}} R\indices{^{i l j m}} + \frac{1}{2} R\indices{^{i l b m}} R\indices{^j_{m b l}} \right) \\ \notag
 & + \frac{1}{4} K\indices{^{a i j}} K\indices{_{a i j}} \left( R\indices{^{b c}_{b c}} R\indices{^{d e}_{d e}} + R\indices{^{b c d k}} R\indices{_{b c d k}} + R\indices{^{b c k l}} R\indices{_{b c k l}} \right) \\ \notag
 & + K\indices{_{a i}^m} K\indices{_{b j m}} K\indices{^a_k^n} K\indices{^b_{l n}} \left( \frac{1}{2} R\indices{^{i l j k}} - R\indices{^{i j k l}} - \frac{3}{2} R\indices{^{i k j l}} \right) - \frac{1}{2} K\indices{_{a i}^m} K\indices{^a_{j m}} K\indices{_{b k}^n} K\indices{^b_{l n}} R\indices{^{i k j l}} \\ \notag
 & + K\indices{_{a i j}} K\indices{_{b k}^m} K\indices{^a_m^n} K\indices{^b_{n l}} R\indices{^{i k j l}} - 2 K\indices{_{a i j}} K\indices{^a_k^m} K\indices{_{b m}^n} K\indices{^b_{n l}} R\indices{^{i k j l}} \\ \notag
 & - K\indices{_{a i}^k} K\indices{_{b k}^l} K\indices{_{c l}^m} K\indices{^c_{m j}} \left( 2 R\indices{^{b i a j}} + 3 R\indices{^{a i b j}} \right) + K\indices{_{a i}^k} K\indices{_{c k}^l} K\indices{_{b l}^m} K\indices{^c_{m j}} \left( 3 R\indices{^{a i b j}} + R\indices{^{b i a j}} - 2 R\indices{^{a b i j}} \right) \\ \notag
 & + \frac{1}{2} K\indices{_{a i}^k} K\indices{_{c k}^l} K\indices{^c_l^m} K\indices{_{b m j}} R\indices{^{a i b j}} + \frac{1}{2} K\indices{_{c i}^k} K\indices{_{a k}^l} K\indices{_{b l}^m} K\indices{^c_{m j}} R\indices{^{a i b j}} + K\indices{_a^{l m}} K\indices{_{b l m}} K\indices{_{c i}^k} K\indices{^c_{j k}} R\indices{^{a i b j}} \\ \notag
 & - 2 K\indices{^{c l m}} K\indices{_{b l m}} K\indices{_{a i}^k} K\indices{_{c j k}} R\indices{^{a i b j}} + K\indices{^{c l m}} K\indices{_{c l m}} K\indices{_{a i}^k} K\indices{_{b j k}} \left( R\indices{^{a b i j}} + \frac{3}{4} R\indices{^{b i a j}} + \frac{1}{4} R\indices{^{a i b j}} \right) \\ \notag
 & + \frac{3}{2} K\indices{_{a i}^k} K\indices{^a_k^l} K\indices{_{b l}^m} K\indices{^b_{m j}} R\indices{_c^{i c j}} - \frac{1}{2} K\indices{_{a k}^l} K\indices{_{b l}^m} K\indices{^b_m^k} K\indices{^a_{i j}} R\indices{_c^{i c j}} - \frac{5}{4} K\indices{^{b l m}} K\indices{_{b l m}} K\indices{_{a i}^k} K\indices{^a_{j k}} R\indices{_c^{i c j}} \\ \notag
 & + \frac{1}{4} K\indices{_{a i}^j} K\indices{_{b j}^k} K\indices{^a_k^l} K\indices{^b_l^i} R\indices{^{c d}_{c d}} - \frac{1}{4} K\indices{^{a i j}} K\indices{_{a i j}} K\indices{^{b k l}} K\indices{_{b k l}} R\indices{^{c d}_{c d}} \\ \notag
 & + \frac{2}{3} K\indices{_{a i}^j} K\indices{^a_j^k} K\indices{_{b k}^l} K\indices{_{c l}^m} K\indices{^b_m^n} K\indices{^c_n^i} + 2 K\indices{_{a i}^j} K\indices{^a_j^k} K\indices{_{b k}^l} K\indices{_{c l}^m} K\indices{^c_m^n} K\indices{^b_n^i} \\ \notag
 & - \frac{4}{3} K\indices{_{a i}^j} K\indices{^a_j^k} K\indices{_{b k}^l} K\indices{^b_l^m} K\indices{_{c m}^n} K\indices{^c_n^i} - \frac{1}{3} K\indices{_{a i}^j} K\indices{_{b j}^k} K\indices{_{c k}^i} K\indices{^a_l^m} K\indices{^b_m^n} K\indices{^c_n^l}  \\ \notag
 & - \frac{1}{3} K\indices{_{a i}^j} K\indices{^a_j^k} K\indices{_{b k}^i} K\indices{_{c l}^m} K\indices{^c_m^n} K\indices{^b_n^l} - \frac{2}{3} K\indices{_a^{i j}} K\indices{_{b i j}} K\indices{^a_k^l} K\indices{^b_l^m} K\indices{_{c m}^n} K\indices{^c_n^k} \\ \notag
 & - \frac{1}{6} K\indices{^{a i j}} K\indices{_{a i j}} K\indices{_{b k}^l} K\indices{^b_l^m} K\indices{_{c m}^n} K\indices{^c_n^k} + \frac{1}{12} K\indices{^{a i j}} K\indices{_{a i j}} K\indices{^{b k l}} K\indices{_{b k l}} K\indices{^{c m n}} K\indices{_{c m n}} \, , \\
\Delta_{2}^{(4)}=& + 4 R\indices{^{\mu \nu \rho \sigma}} R\indices{^{[a}_{\mu a \rho}} R\indices{^{b]}_{\nu b \sigma}} \\ \notag
 & + 2 K\indices{_{a i j}} K\indices{_{b k l}} \left( \frac{2}{3} R\indices{^{i a j [c|}} R\indices{^{k |b] l}_c} + 2 R\indices{^{i k j [a|}} R\indices{^{l c |b]}_c} - R\indices{^{i [a b] k}} R\indices{^{j c l}_c} \right) \\ \notag
 & + K\indices{_{a i j}} K\indices{^a_{k l}} \left( R\indices{^{i k j l}} R\indices{^{b c}_{b c}} - 2 R\indices{^{i k j b}} R\indices{^{l c}_{b c}} - \frac{1}{6} R\indices{^{i b j}_b} R\indices{^{k c l}_c} - \frac{1}{2} R\indices{^{i b k}_b} R\indices{^{j c l}_c} - \frac{4}{3} R\indices{^{i b j c}} R\indices{^k_b^l_c} \right. \\ \notag
 & \left. + R\indices{^{i b k c}} R\indices{^{[j}_b^{l]}_c} + R\indices{^{i k b c}} R\indices{^{j l}_{b c}} - 2 R\indices{^{i b j m}} R\indices{^k_b^l_m} - 2 R\indices{^{i m j n}} R\indices{^k_m^l_n} \right) \\ \notag
 & - 2 K\indices{_{a i}^k} K\indices{_{b j k}} \left( \frac{8}{3} R\indices{^{i [a b] l}} R\indices{^{j c}_{l c}} + \frac{2}{3} R\indices{^{i [a| c l}} R\indices{^{j |b]}_{c l}} + R\indices{^{[a| l i m}} R\indices{^{|b]}_l^j_m} \right) \\ \notag
 & - K\indices{_{a i}^k} K\indices{^a_{j k}} \left( R\indices{^{i b c d}} R\indices{^j_{b c d}} + \frac{4}{3} R\indices{^{i b l c}} R\indices{^j_{b l c}} + \frac{4}{3} R\indices{^{i b l}_{[b|}} R\indices{^{j c}_{l |c]}} + 2 R\indices{^{i l b c}} R\indices{^j_{l b c}} + R\indices{^{i l b m}} R\indices{^j_{l b m}} \right) \\ \notag
 & - \frac{1}{2} K\indices{^{a i j}} K\indices{_{a i j}} \left( R\indices{^{b c}_{b c}} R\indices{^{d e}_{d e}} + R\indices{^{b c d k}} R\indices{_{b c d k}} + \frac{2}{3} R\indices{^{b k c l}} R\indices{_{(b| k |c) l}} - \frac{1}{3} R\indices{_b^{k b l}} R\indices{^c_{k c l}} \right) \\ \notag
 & + 2 K\indices{_{a i j}} K\indices{_{b k}^m} K\indices{^a_m^n} K\indices{^b_{n l}} R\indices{^{i k j l}} - 2 K\indices{_a^{m n}} K\indices{_{b m n}} K\indices{^a_{i j}} K\indices{^b_{k l}} R\indices{^{i k j l}} - \frac{1}{2} K\indices{^{a m n}} K\indices{_{a m n}} K\indices{_{b i j}} K\indices{^b_{k l}} R\indices{^{i k j l}} \\ \notag
 & - 2 K\indices{_{a i}^k} K\indices{_{b k}^l} K\indices{_{c l}^m} K\indices{^c_{m j}} \left( 3 R\indices{^{a b i j}} + R\indices{^{i (a b) j}} \right) + K\indices{_{a i}^k} K\indices{_{c k}^l} K\indices{_{b l}^m} K\indices{^c_{m j}} \left( 2 R\indices{^{a b i j}} - R\indices{^{a i b j}} - 2 R\indices{^{b i a j}} \right) \\ \notag
 & + \frac{1}{2} K\indices{_{a i}^k} K\indices{_{c k}^l} K\indices{^c_l^m} K\indices{_{b m j}} R\indices{^{a i b j}} + \frac{1}{2} K\indices{_{c i}^k} K\indices{_{a k}^l} K\indices{_{b l}^m} K\indices{^c_{m j}} R\indices{^{a i b j}} - \frac{3}{2} K\indices{_{a i}^k} K\indices{^a_k^l} K\indices{_{b l}^m} K\indices{^b_{m j}} R\indices{_c^{i c j}} \\ \notag
 & - K\indices{_{a k}^l} K\indices{_{b l}^m} K\indices{^b_m^k} K\indices{^a_{i j}} R\indices{_c^{i c j}} - \frac{1}{2} K\indices{_{a i}^j} K\indices{_{b j}^k} K\indices{^a_k^l} K\indices{^b_l^i} R\indices{^{c d}_{c d}} + \frac{1}{2} K\indices{_a^{i j}} K\indices{_{b i j}} K\indices{^{a k l}} K\indices{^b_{k l}} R\indices{^{c d}_{c d}} \\ \notag
 & + \frac{1}{2} K\indices{^{a i j}} K\indices{_{a i j}} K\indices{^{b k l}} K\indices{_{b k l}} R\indices{^{c d}_{c d}} + \frac{5}{6} K\indices{_{a i}^j} K\indices{^a_j^k} K\indices{_{b k}^l} K\indices{_{c l}^m} K\indices{^b_m^n} K\indices{^c_n^i} \\ \notag
 & - \frac{7}{2} K\indices{_{a i}^j} K\indices{^a_j^k} K\indices{_{b k}^l} K\indices{_{c l}^m} K\indices{^c_m^n} K\indices{^b_n^i} + \frac{3}{2} K\indices{_{a i}^j} K\indices{^a_j^k} K\indices{_{b k}^l} K\indices{^b_l^m} K\indices{_{c m}^n} K\indices{^c_n^i} \\ \notag
 & - \frac{1}{3} K\indices{_{a i}^j} K\indices{^a_j^k} K\indices{_{b k}^i} K\indices{_{c l}^m} K\indices{^c_m^n} K\indices{^b_n^l} + \frac{4}{3}  K\indices{_a^{i j}} K\indices{_{b i j}} K\indices{^a_k^l} K\indices{^b_l^m} K\indices{_{c m}^n} K\indices{^c_n^k} \\ \notag
 & + K\indices{^{a i j}} K\indices{_{a i j}} K\indices{_{b k}^l} K\indices{_{c l}^m} K\indices{^b_m^n} K\indices{^c_n^k} - \frac{2}{3} K\indices{^{a i j}} K\indices{_{a i j}} K\indices{_{b k}^l} K\indices{^b_l^m} K\indices{_{c m}^n} K\indices{^c_n^k} \\ \notag
 & - \frac{4}{3} K\indices{^{a i j}} K\indices{_{a i j}} K\indices{_b^{k l}} K\indices{_{c k l}} K\indices{^{b m n}} K\indices{^c_{m n}} + \frac{1}{6} K\indices{^{a i j}} K\indices{_{a i j}} K\indices{^{b k l}} K\indices{_{b k l}} K\indices{^{c m n}} K\indices{_{c m n}} \, , \\
\Delta_{1}^{(4)}=& + 4 R\indices{^{a \mu \rho \nu}} R\indices{^{\sigma}_{\nu [a| \mu}} R\indices{^b_{\rho |b] \sigma}} \\ \notag
 & + K\indices{_{a i j}} K\indices{_{b k l}} \left( \frac{4}{3} R\indices{^{i a j [c|}} R\indices{^{k |b] l}_c} + 2 R\indices{^{i k j [a|}} R\indices{^{l c |b]}_c} + 4 R\indices{^{i [a b] m}} R\indices{^{j k l}_m} + R\indices{^{[a| i j m}} R\indices{^{|b] k l}_m} \right) \\ \notag
 & + K\indices{_{a i j}} K\indices{^a_{k l}} \left( - R\indices{^{i k j b}} R\indices{^{l c}_{b c}} - \frac{1}{3} R\indices{^{i b k}_b} R\indices{^{j c l}_c} - \frac{2}{3} R\indices{^{i b j c}} R\indices{^k_b^l_c} - \frac{2}{3} R\indices{^{i b k c}} R\indices{^j_b^l_c} + \frac{1}{3} R\indices{^{i b k c}} R\indices{^l_b^j_c} \right. \\ \notag
 & \left. + R\indices{^{i k b c}} R\indices{^{j l}_{b c}} + 2 R\indices{_b^{i b m}} R\indices{^{j k l}_m} - \frac{1}{2} R\indices{^{i b j m}} R\indices{^k_b^l_m} - R\indices{^{i b k m}} R\indices{^j_b^l_m} + \frac{1}{2} R\indices{^{i k b m}} R\indices{^{j l}_{b m}} - R\indices{^{i m k n}} R\indices{^j_m^l_n} \right) \\ \notag
 & + K\indices{_{a i}^k} K\indices{_{b j k}} \left( 3 R\indices{^{i [a b] j}} R\indices{^{c d}_{c d}} + \frac{1}{2} R\indices{^{i [a| c d}} R\indices{^{j |b]}_{c d}} + \frac{2}{3} R\indices{^{i [a b] l}} R\indices{^{j c}_{l c}} - 4 R\indices{^{i [a| j l}} R\indices{^{|b] c}_{l c}} + \frac{2}{3} R\indices{^{i [a| c l}} R\indices{^{j |b]}_{c l}} \right. \\ \notag
 & \left. - 2 R\indices{^{a l b m}} R\indices{^i_{[l}^j_{m]}} \right) + K\indices{_{a i}^k} K\indices{^a_{j k}} \left( \frac{1}{2} R\indices{_b^{i b j}} R\indices{^{c d}_{c d}} - \frac{3}{4} R\indices{^{i b c d}} R\indices{^j_{b c d}} + \frac{1}{6} R\indices{^{i b l}_b} R\indices{^{j c}_{l c}} + \frac{1}{3} R\indices{^{i b l c}} R\indices{^j_{(b c) l}}  \right. \\ \notag
 & \left. - R\indices{^{i l b c}} R\indices{^j_{l b c}} - R\indices{^{i l j m}} R\indices{^b_{l b m}} \right) + \frac{1}{4} K\indices{^{a i j}} K\indices{_{a i j}} \left( 2 R\indices{^{b k c l}} R\indices{_{[b| k |c] l}} - R\indices{_b^{k b l}} R\indices{^c_{k c l}} \right) \\ \notag
 & + K\indices{_{a i}^m} K\indices{_{b j m}} K\indices{^a_k^n} K\indices{^b_{l n}} R\indices{^{i (k l) j}} - \frac{1}{2} K\indices{_{a i}^m} K\indices{^a_{j m}} K\indices{_{b k}^n} K\indices{^b_{l n}} R\indices{^{i k j l}} - 2 K\indices{_{a i j}} K\indices{_{b k}^m} K\indices{^a_m^n} K\indices{^b_{n l}} R\indices{^{i k j l}} \\ \notag
 & - 2 K\indices{_{a i}^k} K\indices{_{b k}^l} K\indices{_{c l}^m} K\indices{^c_{m j}} \left( 2 R\indices{^{a b i j}} + R\indices{^{b i a j}} \right) + K\indices{_{a i}^k} K\indices{_{c k}^l} K\indices{_{b l}^m} K\indices{^c_{m j}} \left( 2 R\indices{^{a b i j}} + 3 R\indices{^{b i a j}} - R\indices{^{a i b j}} \right) \\ \notag
 & + K\indices{_a^{l m}} K\indices{_{b l m}} K\indices{_{c i}^k} K\indices{^c_{j k}} R\indices{^{a i b j}} - 2 K\indices{_a^{l m}} K\indices{_{c l m}} K\indices{_{b i}^k} K\indices{^c_{j k}} R\indices{^{b i a j}} \\ \notag
 & + \frac{1}{2} K\indices{^{c l m}} K\indices{_{c l m}} K\indices{_{a i}^k} K\indices{_{b j k}} \left( 5 R\indices{^{a i b j}} - 3 R\indices{^{b i a j}} \right) + K\indices{_{a i}^k} K\indices{^a_k^l} K\indices{_{b l}^m} K\indices{^b_{m j}} R\indices{_c^{i c j}} \\ \notag
 & - K\indices{_{a k}^l} K\indices{_{b l}^m} K\indices{^b_m^k} K\indices{^a_{i j}} R\indices{_c^{i c j}} - \frac{3}{2} K\indices{^{b l m}} K\indices{_{b l m}} K\indices{_{a i}^k} K\indices{^a_{j k}} R\indices{_c^{i c j}} + \frac{3}{4} K\indices{_{a i}^j} K\indices{_{b j}^k} K\indices{^a_k^l} K\indices{^b_l^i} R\indices{^{c d}_{c d}} \\ \notag
 & - \frac{1}{2} K\indices{_{a i}^j} K\indices{^a_j^k} K\indices{_{b k}^l} K\indices{^b_l^i} R\indices{^{c d}_{c d}} + \frac{13}{6} K\indices{_{a i}^j} K\indices{^a_j^k} K\indices{_{b k}^l} K\indices{_{c l}^m} K\indices{^b_m^n} K\indices{^c_n^i} \\ \notag
 & + \frac{1}{2} K\indices{_{a i}^j} K\indices{^a_j^k} K\indices{_{b k}^l} K\indices{_{c l}^m} K\indices{^c_m^n} K\indices{^b_n^i} - \frac{7}{6} K\indices{_{a i}^j} K\indices{^a_j^k} K\indices{_{b k}^l} K\indices{^b_l^m} K\indices{_{c m}^n} K\indices{^c_n^i} \\ \notag
 & - \frac{1}{3} K\indices{_{a i}^j} K\indices{_{b j}^k} K\indices{_{c k}^i} K\indices{^a_l^m} K\indices{^b_m^n} K\indices{^c_n^l} - \frac{1}{2} K\indices{_{a i}^j} K\indices{^a_j^k} K\indices{_{b k}^i} K\indices{_{c l}^m} K\indices{^c_m^n} K\indices{^b_n^l} \\ \notag
 & - K\indices{_a^{i j}} K\indices{_{b i j}} K\indices{^a_k^l} K\indices{^b_l^m} K\indices{_{c m}^n} K\indices{^c_n^k} - \frac{13}{12} K\indices{^{a i j}} K\indices{_{a i j}} K\indices{_{b k}^l} K\indices{_{c l}^m} K\indices{^b_m^n} K\indices{^c_n^k} \\ \notag
 & + \frac{5}{6} K\indices{^{a i j}} K\indices{_{a i j}} K\indices{_{b k}^l} K\indices{^b_l^m} K\indices{_{c m}^n} K\indices{^c_n^k} \, .
\end{align}%
Again, we observe that the greater the number of Riemann tensors involved in the corresponding density, the more complicated the expressions. In particular, for theories with zero or one Riemann tensors, the contribution comes completely from the Wald piece. For densities with two Riemanns we get contributions which are quadratic in extrinsic curvatures, for those with three Riemanns, we get terms which are quartic, and for densities with four Riemann tensors there are terms involving up to six extrinsic curvatures. %The functionals presented above are ready to be used in particular cases (although in the explicit calculations performed in the following sections we will
%restrict ourselves to terms involving up to cubic densities).

%For the sake of clarity, we will present the HEE functionals one by one, turning off all the others in each case. The full $\see^{\rm Riem^4}$ will be just the addition of all the different terms. We have
%\begin{equation}
%\see^{\rm Riem^4}=\frac{A}{4G}+\frac{L^4}{4G} \int \rd^{d-1}y \sqrt{h}   \sum_{i=1}^{27}\gamma_i \Delta_i^{(4)} + \mathcal{O}(\gamma_i^2)\, ,
%\end{equation} 
% where now we find

\subsection{$\mathcal{L}(g_{\mu\nu},R_{\rho\sigma})$ gravities}\label{subsec:functional_Ricci}
Let us now consider  densities constructed from general contractions of the Ricci tensor, \ie of the form
\begin{equation}\label{lricci}
I_E^{\rm \mathcal{L}(Ricci)}=-\frac{1}{16\pi G} \int \rd^{d+1}x \sqrt{|g|} \left[\frac{d(d-1)}{L^2}+R+\lambda \mathcal{L}(g_{\mu\nu},R_{\rho\sigma})\right]\, ,
\end{equation}
where $\lambda$ is some constant.
By looking at the quadratic, cubic and quartic densities of this kind, we observe that no contribution from the anomaly part arises in the HEE functional when those terms are considered perturbatively. As we show now, this is in fact a general property which holds for all theories of the form (\ref{lricci}). 

%For these, we can show that the anomaly part does not contribute to the HEE functional when considered perturbatively.
 The proof goes as follows. For the anomaly term, we need to compute the second derivative of the Lagrangian with respect to $R_{z i z j}$ and $R_{\bz k \bz l}$. Let us consider first  the one with $R_{z i z j}$. Since the Lagrangian is a contraction of $n$ Ricci tensors for an $n$-th order theory, we can expand the derivative as
\begin{equation} \label{functional_Ricci:Lagrangian_expansion}
\frac{\partial \mathcal{L}}{\partial R_{z i z j}} = \sum_{k=1}^{n} \frac{\partial R_{\mu \nu}}{\partial R_{z i z j}} T_{(k)}^{\mu \nu} \, ,
\end{equation}
where $T_{(k)}^{\mu \nu}$ represents the remaining part of the Lagrangian contracted with each of the Ricci tensors ---this  can include metric tensors, so the previous expansion is also valid when there are Ricci scalars in the Lagrangian. Now, it can be shown from \req{SymmetryFactors:RiemannDerivative} that
\begin{equation} \label{functional_Ricci:derivative_Ricci}
\frac{\partial R_{\mu \nu}}{\partial R_{\alpha \beta \gamma \delta}} = \delta_{(\mu}^{[\beta} g^{\alpha] [\gamma} \delta_{\nu)}^{\delta]}\quad \Rightarrow \quad \frac{\partial R_{\mu \nu}}{\partial R_{z i z j}} = \frac{1}{4} h^{ij} \delta_{\mu}^{z} \delta_{\nu}^z \, ,
\end{equation}
since $g^{zz} = g^{z i} = 0$. Therefore, \eqref{functional_Ricci:Lagrangian_expansion} is proportional to $h^{ij}$. An analogous argument with the other derivative shows that it is proportional to $h^{kl}$. The conclusion is that the anomaly term is then some expression containing curvature tensors in which we have to perform the $\alpha$-expansion, times the following contraction of extrinsic curvatures:
\begin{equation} \label{functional_Ricci:K_contraction}
h^{ij}h^{kl} K_{z i j} K_{\bz k l} = K_z K_{\bz} = \frac{1}{4} K^a K_a \, ,
\end{equation}
which vanishes when evaluated for the RT surface. Hence, the anomaly part of the functional does not contribute perturbatively for theories constructed from general contractions of the Ricci tensor. Note that this is actually true irrespective of the splitting being used.

Hence, for theories of this kind one finds
\begin{equation}
S_{\rm \ssc HEE}^{\rm \mathcal{L}(Ricci)}=\frac{\mathcal{A}(\Gamma_A)}{4G}+\frac{\lambda}{8G} \int_{\Gamma_A} \rd^{d-1}y\sqrt{h}\, \frac{\partial \mathcal{L}}{\partial R_{\mu \nu}} \bot_{\mu\nu}  + \mathcal{O}(\lambda^2)\, .
\end{equation} 
%where we have emphasized that the expression is valid at leading order in the corresponding coupling(s). 
We emphasize that this formula holds for general-order densities of the form $\mathcal{L}(g_{\mu\nu},R_{\rho\sigma})$. Hence, we observe that, at least perturbatively in the higher-curvature couplings, the purely-Wald nature of the $f(R)$ functional actually extends to the much greater family of densities  constructed from arbitrary contractions of the Ricci tensor and the metric. 
%\varepsilon^{\gamma}_{\mu } \varepsilon_{\gamma \nu}

%%%%%%%%%%%%%%%%%%
%%%%%%%%%%%%%%%%%%
%%%%%%%%%%%%%%%%%%
\subsection{General structure depending on the number of Riemann tensors}
\label{subsec:functional_Riemann}
%\comment{tbe}
The observations made in the previous subsections suggest a more general pattern which we explore here. 
The starting point is the observation made in subsection \ref{subsec:functional_Ricci} that whenever one of the two derivatives appearing in \eqref{NewFunctional:ShorthandSecondDerivative} hits a Ricci tensor, the contraction of the resulting intrinsic metric with the extrinsic curvature produces a trace, $K\indices{_z}$ or $K\indices{_{\bz}}$, which is zero for the RT surface (and therefore also for the perturbative functional). Consider then an $n$-th order curvature density containing $n_R$ Riemann tensors and $n - n_R$ Ricci tensors or scalars. After the two derivatives are taken, the only non-vanishing pieces will be of the form %after the two derivatives are taken the only non-vanishing pieces will be of the form:
\begin{equation} \label{GeneralStructureDerivatives:SecondDerivative}
\D2R \sim \sum K^2{\rm Ricci}_1 \dots {\rm Ricci}_{n-n_R} {\rm Riem}_1 \dots {\rm Riem}_{n_R-2} \, .
\end{equation}
In this expression, we use the symbol $\sim$ to represent the structure of the object in terms of the curvature tensors appearing, ignoring the particular components. The sum means that several terms with this structure will show up in general. Each  ${\rm Ricci}_k$ represents a particular component of the Ricci tensor or scalar and, analogously, ${\rm Riem}_k$ represents a component of the Riemann tensor.

Observe now the following property. Writing explicitly the Ricci tensor and scalar in terms of Riemann tensor components, we get
\begin{align} \label{GeneralStructureDerivatives:RicciComponents}
\nonumber R\indices{_{z z}} & = h\indices{^{i j}} R\indices{_{z i z j}}  ~, & R\indices{_{z \bz}} & = - 2 R\indices{_{z \bz z \bz}} + h\indices{^{ij}} R\indices{_{z i \bz j}} \, , \\
\nonumber R\indices{_{z i}} & = - 2 R\indices{_{z \bz z i}} + h\indices{^{j k}} R\indices{_{z j i k}} \, , & R\indices{_{i j}} & = 2 R\indices{_{z i \bz j}} + 2 R\indices{_{z j \bz i}} + h\indices{^{k l}} R\indices{_{i k j l}} \, , \\
R & = 4 R\indices{_{z \bz}} + h\indices{^{i j}} R\indices{_{i j}} \, ,
\end{align}
plus the ones obtained by complex conjugation. Then, the differential operators defined in \req{NewFunctional:OperatorA} and \req{NewFunctional:OperatorB} act on these components as follows
\begin{align} \label{GeneralStructureDerivatives:OperatorsAction}
\mathcal{K}\indices{_{AI}} \hd^{AI} R\indices{_{z z}} & = R\indices{_{z z}} \, , & \mathcal{K}\indices{_{BI}} \hd^{BI} R\indices{_{z z}} & = 0 \, , \\
\mathcal{K}\indices{_{AI}} \hd^{AI} R\indices{_{z \bz}} & = 0 \, , & \mathcal{K}\indices{_{BI}} \hd^{BI} R\indices{_{z \bz}} & = 0 \, , \\
\mathcal{K}\indices{_{AI}} \hd^{AI} R\indices{_{z i}} & = 0 \, , & \mathcal{K}\indices{_{BI}} \hd^{BI} R\indices{_{z i}} & = R\indices{_{z i}} \, , \\
\mathcal{K}\indices{_{AI}} \hd^{AI} R\indices{_{i j}} & = - K\indices{_{a i j}} K\indices{^a} \, , & \mathcal{K}\indices{_{BI}} \hd^{BI} R\indices{_{i j}} & = 0 \, , \\
\mathcal{K}\indices{_{AI}} \hd^{AI} R & = -K\indices{_a} K\indices{^a} \, , & \mathcal{K}\indices{_{BI}} \hd^{BI} R & = 0 \,  .
\end{align}
Notice also that if the Ricci components are acted upon with several powers of the differential operators in normal order, like in the functional \eqref{NewFunctional:FinalForm}, the remaining powers would not act on the curvature tensors appearing in the right-hand side of the previous expressions. In any case, the relevant observation is that after applying the differential operator, any Ricci factor in \eqref{GeneralStructureDerivatives:SecondDerivative} generates either something proportional to the very same component or something proportional to $K\indices{^a}$. When evaluated at the RT surface, this second possibility gives zero, so in a perturbative functional no Ricci tensor component can ever generate powers of the extrinsic curvature. This is not the case with Riemann tensor components, for which the differential operator generates non-vanishing contractions of extrinsic curvatures in general.\footnote{This is not true for \emph{all} Riemann tensor components. As shown in subsection \ref{subsec:MixedTypes}, some components do not generate extrinsic curvatures, and a second derivative monomial of the form \eqref{MixedTypes:ExampleTerm} produces only something with the structure ${\rm Riem}^3 + {\rm Riem}^2 K^2$, as in \eqref{MixedTypes:ResultDerivatives}.} The conclusion is that the expression which results from applying the full differential operator of the anomaly term to a second derivative of the form \eqref{GeneralStructureDerivatives:SecondDerivative} has the structure
\begin{align}
&\sum_{\alpha} \frac{1}{1 + q_{\alpha}} \left. \D2R \right._{\alpha} \\ \notag & \quad \sim \sum{\rm Ricci}^{n-n_R} \left( {\rm Riem}^{n_R-2} K^2 + {\rm Riem}^{n_R-3} K^4 + \dots + {\rm Riem}\, K^{2n_R - 4} +K^{2n_R - 2}\right) \, .
\end{align}
One can verify that this is indeed the case for all quadratic, cubic, and quartic Lagrangian densities presented in the previous sections.

%\comment{HEEEEREEEEE}

In summary, we have shown that densities containing $n_R$ Riemann curvatures can contain terms involving extrinsic curvatures up to the power $2n_R-2$. In particular, this implies that densities with zero or one Riemann tensors have no anomaly piece. We already studied the former case in the previous subsection. As for the latter,  for a theory of the form
\begin{equation}\label{lricci}
 -\frac{1}{16\pi G} \int \rd^{d+1}x \sqrt{|g|} \left[\frac{d(d-1)}{L^2}+R+\lambda R^{\mu \nu \rho \sigma} T_{\mu \nu \rho \sigma} ({\rm Ricci})\right]\, ,
\end{equation}
where $T_{\mu \nu \rho \sigma} ({\rm Ricci})$ is some tensorial structure involving Ricci tensors and metrics,
the corresponding functional reads
\begin{equation}
 \frac{\mathcal{A}(\Gamma_A)}{4G}+\frac{\lambda}{8G} \int_{\Gamma_A} \rd^{d-1}y\sqrt{h}\, \left[2 T^{\mu \nu \rho \sigma}  \bot_{\mu[\rho} \bot_{\nu |\sigma]} +R^{\mu\nu\rho\sigma} \frac{\partial T_{\mu\nu\rho\sigma}}{\partial R_{\alpha \beta}} \bot_{\alpha \beta}\right]  + \mathcal{O}(\lambda^2)\, .
\end{equation} 
On the other hand, densities with two Riemann tensors have terms with up to two extrinsic curvatures, those with three have terms with up to four extrinsic curvatures, and so on.

\section{Universal terms }\label{unite}
In this section we study how the universal coefficients appearing in the EE of various symmetric entangling regions get modified in the presence of quadratic and cubic corrections. Some of these coefficients can be computed from alternative methods, and in that case we verify that the results agree with them. In other cases, like for strip regions, the corresponding universal coefficients do not have a known alternative interpretation beyond entanglement entropy. Universal terms for various types of regions have been previously computed for particular higher-curvature theories in certain dimensions in several papers such as \cite{Buchel:2009sk,Myers:2010jv,deBoer:2011wk,Hung:2011xb,Bueno2,Safdi:2012sn,Miao:2015iba,Bhattacharyya:2014yga,Cano:2018ckq}. Our results reproduce the ones found in those papers in the appropriate cases.   

We will restrict ourselves to the vacuum state. This means that all expressions involving intrinsic bulk curvatures will be evaluated on pure AdS$_{d+1}$, for which $R_{\mu\nu}^{\rho\sigma}=-1/\Ls^{2}\cdot \left[\delta_{\mu}^{\rho}\delta_{\nu}^{\sigma}-\delta_{\mu}^{\sigma}\delta_{\nu}^{\rho} \right]$. On such a background ---more generally, on any maximally symmetric background--- one can show that the variation of any higher-curvature Lagrangian with respect to the Riemann tensor is given by 
\begin{equation} \label{spheres:lagrangian_derivative_AdS}
 \left. \frac{\partial \mathcal{L}}{\partial R_{\mu \nu \rho \sigma}} \right|_{\rm AdS} =  k_0 \left[g^{\mu \rho} g^{\sigma \nu}-g^{\mu \sigma} g^{\rho \nu} \right]\, , 
\end{equation}
where the constant $k_0$ is fixed by imposing AdS$_{d+1}$ to be a solution of the equations of motion of the theory as \cite{Aspects}
\begin{equation}\label{k00}
k_0= - \frac{\Ls^2}{4d} \left.\mathcal{L} \right|_{\rm AdS}\, ,
\end{equation}
where $\left.\mathcal{L} \right|_{\rm AdS}$ is the on-shell Lagrangian of the theory evaluated on AdS$_{d+1}$.
Now, it has been argued using different arguments \cite{Imbimbo:1999bj,Schwimmer:2008yh,Myers:2010tj,Myers:2010xs,HoloECG} that $\mathcal{L}
|_{{\rm AdS}}$ is actually related to the universal coefficient $a^{\star(d)}$ appearing in the EE across spherical regions in general dimensions. For a general CFT in $d$-dimensions, this is given by
\begin{equation}\label{asta}
\left. \see^{(d)} \right|_{\rm sphere}\supset \begin{cases}
(-)^{\frac{d-2}{2}} 4 a^{\star(d)} \log\left(\tfrac{ R}{\delta} \right) \quad &\text{for even } d \, , \\
 (-)^{\frac{d-1}{2}}2\pi a^{\star(d)} \quad &\text{for odd } d\, .
\end{cases}
\end{equation}
The exact relation for holographic higher-curvature gravities reads
\begin{equation}\label{astar}
a^{\star(d)}=-\frac{\pi^{d/2} \Ls^{d+1}}{d \Gamma(d/2)}\mathcal{L}
|_{{\rm AdS}}\, , \quad \text{so} \quad k_0=\frac{\Gamma\left[ \tfrac{d}{2}\right] a^{\star(d)}}{4\pi^{d/2} \Ls^{d-1}}\, .
\end{equation}
%holds for general higher-curvature gravities. 
As a consequence, Wald's piece in the HEE formula becomes proportional to the Ryu-Takayanagi functional in that case, with an overall coefficient controlled by $a^{\star(d)}$. One has %\comment{check relative factor}
\begin{equation} \label{spheres:functional_Wald_AdS}
S_{\rm \ssc HEE} = \frac{2\Gamma\left[\tfrac{d}{2} \right]  a^{\star(d)}}{\pi^{\frac{d-2}{2}}\Ls^{d-1}} \int \rd^{d-1}y\, \sqrt{h} + S_{\rm \ssc Anomaly} \, .
 %\int \rd^{d-1}y\, \sqrt{h}  \sum_{\alpha} \left( \frac{\partial^2 \mathcal{L}}{\partial R_{z i z j} \partial R_{\bz k \bz l}} \right)_\alpha \frac{16\pi K_{zij} K_{\bz k l}}{q_\alpha + 1} \, .
\end{equation}
Hence, for theories for which the anomaly piece is absent, all possible universal terms are proportional to the coefficient $a^{\star(d)}$. As we saw above this includes, at the perturbative level, all $\mathcal{L}(g_{\mu\nu},R_{\rho\sigma})$ densities as well as those including a single Riemann tensor.  For them, all the different universal coefficients we will consider in this section will modify the Einstein gravity result by the same overall factor given by $a^{\star(d)}/a^{\star(d)}_{\rm E}$, where 
\cite{Ryu:2006bv,Ryu:2006ef}
\begin{equation}
a^{\star(d)}_{\rm E}=\frac{\pi^{\frac{d-2}{2}}}{8\Gamma \left[\tfrac{d}{2} \right]} \frac{L_{\star}^{d-1}}{G}\, .
\end{equation}
The coefficient $a^{\star(d)}$ can be easily computed for quadratic and cubic theories, yielding
\begin{align}
a^{\star(d)}_{\rm Riem^2}=& \left[1-2d(d+1)\alpha_1-2d\alpha_2-4\alpha_3\right]  a^{\star(d)}_{\rm E}\, ,\\  \notag 
a^{\star(d)}_{\rm Riem^3}=&\left[1+3(d-1)\beta_1+12\beta_2+6d\beta_3+6d(d+1)\beta_4+3d^2\beta_5+3d^2\beta_6 \right. \\  \notag  &\left.+3d^2(d+1)\beta_7+3d^2(d+1)^2\beta_8 \right]a^{\star(d)}_{\rm E} \, .
\end{align}
For the reasons explained above, the corrections corresponding to $\alpha_1$, $\alpha_2$, $\beta_5$, $\beta_6$, $\beta_7$, $\beta_8$ will appear as overall corrections to the Einstein gravity result with precisely the above coefficients for all possible entangling regions. Particularizing to the Gauss-Bonnet and cubic Lovelock cases, one finds  
 \begin{align}
a^{\star(d)}_{ \mathcal{X}_4}&=[1-2 (d-2)(d-1) \lambda_2]  a^{\star(d)}_{ \rm E}\, , \\
a^{\star(d)}_{ \mathcal{X}_6}&=[1+3 (d-4)(d-3)(d-2)(d-1)\lambda_3]  a^{\star(d)}_{ \rm E}  \, .
\end{align}
In both cases, the corrections are zero below the critical dimensions, as they should, since in those cases the corresponding contributions to the JM functional (\ref{jm}) identically vanish. For a general Lovelock theory of the form (\ref{lovel}), one would have 
\begin{equation}
a^{\star(d)}_{ \rm Lovelock}=[1-\sum_n n \, (-1)^n\, \prod_{k=1}^{2(d-1)} (d-k)\lambda_n]  a^{\star(d)}_{ \rm E}  \, .
\end{equation}
The result for the charges $a^{\star(d)}$ for Quasi-topological gravity and Einsteinian cubic gravity reads in each case
\begin{align}
a^{\star(4)}_{\rm QTG}&=[1+9\mu_{\rm QTG}]a^{\star (4)}_{ \rm E}  \, , \\ 
a^{\star(3)}_{\rm ECG}&=[1+3\mu_{\rm ECG}]a^{\star (3)}_{ \rm E} \, .
\end{align}
%The above results reduce to those appeared in \cite{Hung:2011xb,,holoECG} in the corresponding subcases.

\subsection{Spherical regions}
Let us see how the above results for $a^{\star(d)}$ can be obtained from an explicit calculation for spherical entangling surfaces,  $\partial A=\mathbb{S}^{d-2}$, using the corresponding HEE functionals. Across spheres, the universal contribution to the entanglement entropy is given, for a general CFT in $d$-dimensions by \req{asta}.
%\begin{equation}\label{asta}
%\see^{(d)}\supset \begin{cases}
%(-)^{\frac{d-2}{2}} 4a^\star \log\left(\tfrac{ R}{\delta} \right) \quad &\text{for even } d \, , \\
% (-)^{\frac{d-1}{2}}2\pi a^\star \quad &\text{for odd } d\, .
%\end{cases}
%\end{equation}

In the even-dimensional case, the corresponding logarithmic term for a general smooth region is a linear combination of local integrals over the entangling surface weighted by the different trace-anomaly charges \cite{Solodukhin:2008dh,Fursaev:2012mp,Safdi:2012sn,Miao:2015iba} ---see \req{see4} and \req{see6} below. One of the integrals involves the Euler density of the entangling surface and the corresponding trace-anomaly coefficient which appears in front is customarily denoted by  ``$a$'' (or ``$A$'' in $d\geq 6$). The rest of integrals involve various combinations of the extrinsic curvature of $\partial A$, and therefore all of them vanish for a spherical entangling surface. Hence, the sphere isolates the $a$-type coefficient, and we have simply $a^\star =a$ for even $d$. 

The nature of  $a^\star$ is very different in odd dimensions. In that case, it appears as a constant contribution to the EE, and it has an intrinsically non-local nature.  In fact, as shown in \cite{CHM}, $a^\star$ is proportional to the free energy, $F=-\log Z$, of the corresponding theory evaluated on $\mathbb{S}^{d}$, namely 
$
F_{\mathbb{S}^{d}}= (-)^{\frac{d+1}{2}} 2\pi a^\star $ %. In addition, the EE across a spherical region is also equivalent 
or, alternatively,
to the thermal entropy of the corresponding CFT at a temperature  $T= 1/(2\pi R)$ on the hyperbolic cylinder $\mathbb{R}\times \mathbb{H}^{d-1}$ \cite{CHM}.
From an holographic perspective, this means that $a^\star$ can be obtained, besides from a direct entanglement entropy calculation, like the one we perform here, either from the Euclidean
on-shell action of pure AdS$_{(d+1)}$ with $\mathbb{S}^d$ boundary or from the Wald entropy of AdS$_{(d+1)}$ with $\mathbb{R} \times \mathbb{H}^{d-1}$ boundary ---see also \cite{Fonda:2015nma,Anastasiou:2020smm}.

%The above comments apply to general CFTs in arbitrary dimensions. In the case of holographic higher-curvature theories, \comment{blah blah}

We write the AdS$_{d+1}$ metric as
\begin{equation} \label{spheres:bulk_metric_D}
\rd s^2 = \frac{\Ls^2}{z^2} \left[ \rd \tau^2+\rd z^2 + \rd r^2 + r^2 \rd \Omega^2_{d-2}   \right]\, ,
\end{equation}
where $\rd \Omega^2_{d-2}$ is the metric of the usual round sphere. Our entangling surface is a sphere $\mathbb{S}^{d-2}$ of radius $r=\ell$ centered at $r=0$.  Let us parametrize the RT surface  as: $\tau=0$, $z=Z(r)$. Then,
%
%where $\Ls = L$ if the Lagrangian of the theory is written as:
%
%\begin{equation} \label{spheres:lagrangian_GR}
%\mathcal{L}_{GR} = \frac{1}{16 \pi G_N} \left( R + \frac{d (d-1)}{L^2} \right) ~ .
%\end{equation}
%
%Consider now a surface at fixed $\tau$ ending on a disk of radius $r = \ell$ at the boundary $z = 0$. Taking advantage of the symmetry in the angular directions, we parametrize it  in terms of a function $z = Z(r)$ and the $d-2$ angular variables of the $\mathbb{S}^{d-2}$. The coordinate $r$ in the surface takes values $r \in [0, \ell]$. %Tangent vectors to the surface are:
%
%\begin{equation} \label{spheres:tangent_vectors}
%\partial_i \equiv \partial_{\phi^i} ~ , \qquad \partial_r + Z' \partial_z ~ ,
%\end{equation}
%
%where $\phi^i$ are the angular variables (here, $i = 1, \dots, d-2$). We can also obtain the two unit normals:
%
unit normals to the surface are given by
\begin{equation} \label{spheres:unit_normals}
n_1 = \frac{z}{\Ls} \partial_{\tau} ~ , \qquad n_2 = \frac{z}{\Ls \sqrt{1 + Z'^2}} \left( Z' \partial_r - \partial_z \right) \, .
\end{equation}
We have already extended these vector fields to a neighborhood of the surface while keeping them normalized. 
On the surface, one fixes $z = Z(r)$, and $Z'(r)$ is well-defined for any $(r,z)$ with $r \in (0, \ell)$. The induced metric on the surface is given by
\begin{equation} \label{spheres:induced_metric}
h_{\mu \nu} \rd x^{\mu} \rd x^{\nu} = \frac{\Ls^2}{Z^2} \left[\frac{1}{1+Z'^2} \left( \rd r + Z' \rd z \right)^2 + r^2 \rd \Omega^2_{d-2}  \right] \, .
\end{equation}
With these results one can compute in full generality the components of the extrinsic curvatures, %$K^a{}_{\mu \nu} = h^{\rho}_{\mu} h^{\sigma}_{\nu} \nabla_{\rho} n^a{}_{\sigma}$, obtaining: \comment{conventions K} %\textbf{NOTE: THIS IS MINUS THE DEFINITION IN THE INTRODUCTION}
\begin{align}
K^1{}_{\mu \nu} & = 0 ~, \\
K^2{}_{rr} & = \frac{\Ls}{Z^2 (1 + Z'^2)^{5/2}} \left(1 + Z'^2 + Z Z'' \right) ~ , \quad K^2{}_{rz} = Z' K^2{}_{rr} ~ ,  \\
K^2{}_{zz} &= Z'^2 K^2{}_{rr} ~ ,\quad K^2{}_{mn}  = \frac{\Ls}{Z^2 \sqrt{1 + Z'^2}} \left( Z Z' + r \right) r \hat{g}_{mn} \, ,
\end{align}
where $\hat{g}_{mn}$ is the metric of the unit $\mathbb{S}^{d-2}$. Obtaining the traces is now easy. $K^1 = 0$ trivially, whereas
\begin{equation} \label{spheres:trace_extrinsic}
K^2 = \frac{1}{\Ls r (1 + Z'^2)^{3/2}} \left[ r Z Z'' + (d-2) Z Z' (1 + Z'^2) + (d-1) r (1 + Z'^2) \right] \, .
\end{equation}
The vanishing of this trace is exactly the differential equation for the surface one would obtain by minimizing the RT functional, which in this case reads%
\begin{equation} \label{spheres:RT_functional}
S_{\rm \ssc HEE}^{\rm E} = \frac{\Ls^{d-1} \pi^{(d-1)/2}}{2 G \Gamma \left[ \frac{d-1}{2} \right]} \int_0^{\ell} \rd r \, \frac{r^{d-2}}{Z^{d-1}} \sqrt{1 + Z'^2} \, .
\end{equation}
The solution for this differential equation satisfying the boundary condition $Z(\ell) = 0$ is $r^2 + Z^2 = \ell^2$. %(alternatively, writing explicitly the function $Z$ we have $Z(r) = \sqrt{\ell^2 - r^2}$). 
The simplicity of this RT surface has another important consequence: since $ZZ' = -r$ and $ZZ'' = - (1 + Z'^2)$, the extrinsic curvature $K^2{}_{\mu \nu}$ vanishes. Thus, both $K^1{}_{\mu \nu}$ and $K^2{}_{\mu \nu}$ are zero for the RT surface. Now, since the anomaly term in the general higher-curvature functional is quadratic in extrinsic curvatures of the surface, when minimizing the functional, the RT surface will also be extremal for the full functional if we were to consider it fully non-perturbatively.\footnote{This is true irrespective of the splitting used.}

In order to compute the universal contribution to the HEE the last step is to regulate \req{spheres:RT_functional}, \eg by writing
%The final step of the argument is then computing the universal part of the RT functional. First of all, we rewrite \eqref{spheres:RT_functional} for the RT surface as:
%
\begin{equation} \label{spheres:RT_reparametrized}
S_{\rm \ssc HEE}^{\rm E} = \frac{\Ls^{d-1} \pi^{(d-1)/2}}{2 G \Gamma \left[ \frac{d-1}{2} \right]} \int_{\delta/\ell}^{1} \rd y \, \frac{(1-y^2)^{(d-3)/2}}{y^{d-1}} \, ,
\end{equation}
where we introduced a cutoff at $z = \delta$. Integrating by parts, it is easy to show that for odd $d$ we get a constant term while for even $d$ we get a logarithmic one. The final result takes the form \req{asta}, plus a series of non-universal divergent pieces of the form $(\ell/\delta)^{(d-2k)}$ with $k=1,2,\dots,(d-1)/2$ for odd $d$ and $k=1,2,\dots,(d-2)/2$ for even $d$ ---see \eg \cite{Ryu:2006bv} for the numerical coefficients. When higher-curvature terms are included, the vanishing of $K^1{}_{\mu \nu}$ and $K^2{}_{\mu \nu}$ makes the result reduce to the corresponding Wald piece, which in turn reduces to an overall constant proportional to $\mathcal{L}|_{\rm AdS}$ via \req{k00} times the Einstein gravity result. Hence, we are left again with \req{asta} where $a^\star$ is given by \req{astar} in each case.

\subsection{Slab regions}
Let us consider now an entangling region consisting of a slab of width $\ell$ along a particular dimension, $x \in \left[\ell/2, \ell/2 \right]$, and infinite along the remaining $(d-2)$. For general theories, the EE in that case takes the form
\begin{equation}\label{slabs}
\see=\xi \frac{L_y^{d-2}}{\delta^{d-2}} - \kappa^{(d)} \frac{L_y^{d-2}}{\ell^{d-2}}\, ,
\end{equation}
where $\xi$ is a non-universal constant.  As opposed to other universal EE contributions considered here, $ \kappa^{(d)}$ does not have any (known) alternative interpretation beyond EE. For instance, it is not expected to be related to charges characterizing simple local correlators. Previous papers where $ \kappa^{(d)}$ was computed for certain holographic higher-curvature gravities include \cite{Bueno2}, where it was evaluated for quadratic theories in $d=3$, and \cite{deBoer:2011wk}, where it was computed for Gauss-Bonnet gravity in $d=4$ fully nonperturbatively using the JM functional.

We write the AdS$_{d+1}$ metric  as
\begin{equation} \label{spheres:bulk_metric_D}
\rd s^2 = \frac{\Ls^2}{z^2} \left[ \rd \tau^2+\rd z^2 + \rd x^2+\rd \vec{y}^2_{d-2}  \right]\, .
\end{equation}
The RT surface will be invariant under translations along the  $(d-2)$ transverse directions, so we can parametrize it by $z=Z(x)$. Unit normals to the surface will be given by
%
%where $\Ls = L$ if the Lagrangian of the theory is written as:
%
%\begin{equation} \label{spheres:lagrangian_GR}
%\mathcal{L}_{GR} = \frac{1}{16 \pi G_N} \left( R + \frac{d (d-1)}{L^2} \right) ~ .
%\end{equation}
%
%Consider now a surface at fixed $\tau$ ending on a disk of radius $r = \ell$ at the boundary $z = 0$. Taking advantage of the symmetry in the angular directions, we parametrize it  in terms of a function $z = Z(r)$ and the $d-2$ angular variables of the $\mathbb{S}^{d-2}$. The coordinate $r$ in the surface takes values $r \in [0, \ell]$. %Tangent vectors to the surface are:
%
%\begin{equation} \label{spheres:tangent_vectors}
%\partial_i \equiv \partial_{\phi^i} ~ , \qquad \partial_r + Z' \partial_z ~ ,
%\end{equation}
%
%where $\phi^i$ are the angular variables (here, $i = 1, \dots, d-2$). We can also obtain the two unit normals:
%
%Unit normals
\begin{equation} \label{spheres:unit_normals}
n_1 = \frac{z}{\Ls} \partial_{\tau} ~ , \qquad n_2 = \frac{z}{\Ls \sqrt{1 + Z'^2}} \left( Z' \partial_x - \partial_z \right) \, .
\end{equation}
%
%We have already extended these vector fields to a neighborhood of the surface while keeping them normalized. 
%On the surface, one fixes $z = Z(x)$, and $Z'(x)$ is well-defined for any $(x,z)$ with $x \in (-\ell/2, \ell/2)$. 
The induced metric on the surface is now given by
\begin{equation} \label{spheres:induced_metric}
h_{\mu \nu} \rd x^{\mu} \rd x^{\nu} = \frac{\Ls^2}{Z^2} \left[\frac{1}{1+Z'^2} \left( \rd x + Z' \rd z \right)^2 + \rd \vec{y}^2_{d-2}  \right] \, .
\end{equation}
The non-vanishing components of the extrinsic curvatures $K_{\mu\nu}^a$ read %$K^a{}_{\mu \nu} = h^{\rho}_{\mu} h^{\sigma}_{\nu} \nabla_{\rho} n^a{}_{\sigma}$, obtaining: \comment{conventions K} %\textbf{NOTE: THIS IS MINUS THE DEFINITION IN THE INTRODUCTION}
\begin{align}
K^2{}_{xx} = \frac{\Ls  \left(1 + Z'^2 + Z Z'' \right)}{Z^2 (1 + Z'^2)^{5/2}} =\frac{ K^2{}_{xz}}{  Z'}=\frac{
K^2{}_{zz}}{ Z'^2} \, , \quad K^2{}_{mn} = \frac{\Ls \delta_{mn}}{Z^2 \sqrt{1 + Z'^2}}\,  ,
\end{align}
whereas all components of $K_{\mu\nu}^1$ vanish. 

Projectors on the surface are given by
\begin{equation}
t_x=Z' \partial_z + \partial_x\, , \qquad t_{m}=\partial_m \, ,\quad  \forall m=1,\dots,d-2\, .
\end{equation}
Using these, we find
\begin{equation}
h_{ij}  \rd y^i \diff y^j= \frac{\Ls^2}{Z^2}  \left[(1+Z'^2)\diff x^2+\rd \vec{y}^2_{d-2}  \right] \, .
\end{equation}
Also, the non-vanishing components of $K_{ij}^2$ read (note the slight abuse of notation)
\begin{equation} \label{spheres:trace_extrinsic}
K^2_{xx} = \frac{\Ls (1+Z'^2+Z Z'')}{Z^2\sqrt{1+Z'^2}}\, , \quad K_{mn}^2=\frac{\Ls}{Z^2\sqrt{1+Z'^2}}\delta_{mn}\, .
\end{equation}
With these building blocks we can compute all the different pieces appearing in the corresponding EE functionals. For instance, the relevant expressions for the quadratic ones read 
%and from this
\begin{equation}
K_{ij}^aK^{aij}=\frac{(d-1)(1+Z'^2)^2+Z^2Z''^2+2(1+Z'^2)ZZ''}{\Ls^2 (1+Z'^2)^3}\, , %\quad K_a K^a=\frac{\left[(d-1) (1+Z'^2)+Z Z''  \right]^2}{\Ls^2 (1 + Z'^2)^{3}}\, ,
\end{equation}
\begin{equation}
R^{ab}\,_{ab}=-\frac{2}{\Ls^2}\, , \quad R=-\frac{d(d+1)}{\Ls^2}\, ,\quad R^a\,_a=-\frac{2d}{\Ls^2}\, .
\end{equation}
%\begin{equation}
% \mathcal{R}=\frac{(d-2) [2ZZ''-(d-1)Z'^2(1+Z'^2)]}{\Ls^2(1+Z'^2)^2}
%\end{equation}

%
Now, the Ryu-Takayanagi surface is determined by the condition $K^2=0$, where in this case we have
\begin{equation} \label{spheres:trace_extrinsic}
K^2 = \frac{(d-1) (1+Z'^2)+Z Z''}{\Ls (1 + Z'^2)^{3/2}} \, .
\end{equation}
A first integral can be shown to exist so that  
\begin{equation}
Z'=-\frac{\sqrt{z_{\star}^{2(d-1)} - Z^{2(d-1)}}}{Z^{(d-1)}}\, , \quad \text{where} \quad   z_\star=\frac{\Gamma\left[\tfrac{1}{2(d-1)} \right]}{2\sqrt{\pi} \Gamma \left[ \tfrac{d}{2(d-1)}\right]}\ell  %\quad  \rightarrow \quad Z''=-(d-1)\frac{z_{\star}^{2(d-1)}}{Z^{2d-1}}
\end{equation}
%holds on-shell. The constant $z_{\star}$ 
is the value of $z$ corresponding to the turning point of the surface.
%\begin{equation}
%\frac{\ell}{2} =\frac{ \sqrt{\pi} \Gamma \left[ \tfrac{d}{2(d-1)}\right]}{\Gamma\left[\tfrac{1}{2(d-1)} \right]} z_\star \, .
%\end{equation}
Now, after some massaging, the EE for Einstein gravity can be seen to be given by \cite{Ryu:2006ef,Ryu:2006bv}
\begin{equation}
S_{\rm \ssc HEE}^{\rm E}=\frac{\Ls^{d-1} L_y^{d-2}}{2G z_\star^{d-2}} \int^1_{\delta} \frac{\rd y }{y^{d-1}\sqrt{1-y^{2(d-1)}}}= \xi_{\rm E}\frac{L_y^{d-2}}{\delta^{d-2}} - \kappa^{(d)}_{\rm E} \frac{L_y^{d-2}}{\ell^{d-2}}  \, ,
\end{equation}
where $L_y$ are IR regulators for the $(d-2)$ transverse directions. The universal and non-universal constants $\kappa^{(d)}_{\rm E}$ and $\xi_{\rm E}$ read respectively
\begin{equation}
\kappa^{(d)}_{\rm E}=\frac{2^{d-3 } \pi^{\frac{d-1}{2}} \Gamma\left[ \tfrac{d}{2(d-1)}\right]^{d-1}  }{(d-2) \Gamma\left[ \tfrac{1}{2(d-1)}\right]^{d-1}} \frac{\Ls^{d-1}}{G}\, , \quad \xi_{\rm E}=\frac{\Ls^{d-1}}{2(d-2)G} \, .
\end{equation}
%is the universal piece characteristic of the slab region.
Let us see how these generalize when quadratic and cubic terms are introduced. For a general quadratic theory of the form (\ref{quaact}) one finds
%In the case of the quadratic  the EE for a general quadratic gravity reads in turn
\begin{equation}
S_{\rm \ssc HEE}^{\rm Riem^2}=\frac{\Ls^{d-1} L_y^{d-2}}{2G z_\star^{d-2}} \int_{\delta}^1 \frac{\left[1-2d (d+1) \alpha_1-2d \alpha_2 -2\alpha_3\left[2+(d-1)(d-2) y^{2(d-1)}\right] \right] }{y^{d-1}\sqrt{1-y^{2(d-1)}}}  \rd y\, ,
\end{equation}
and from this,
\begin{equation}
S_{\rm \ssc HEE}^{\rm Riem^2}=\xi_{\rm Riem^2}\frac{L_y^{d-2}}{\delta^{d-2}} - \kappa^{(d)}_{\rm Riem^2} \frac{L_y^{d-2}}{\ell^{d-2}}  \, ,
\end{equation}
where now $\xi_{\rm Riem^2}$ gets a factor identical to the one of $a^{\star (d)}_{\rm Riem^2}$ whereas the universal coefficient reads 
\begin{align}
\kappa^{(d)}_{\rm Riem^2} &=\left[1-2d(d+1)\alpha_1-2d \alpha_2+2(d-3) \left[2+d(d-2) \right]\alpha_3\right] \kappa^{(d)}_{\rm E} \, .
\end{align}
Note that there are two kinds of terms in the integrand. On the one hand, pieces arising from purely intrinsic curvatures are proportional to the Einstein gravity one, which is of the form $\sim 1/(y^{d-1}\sqrt{1-y^{2(d-1)}})$. On the other hand, the contribution which involves two extrinsic curvatures has an extra $\sim y^{2(d-1)}$ factor. It is easy to see that 
$\xi_{\rm Riem^2}$ is unaffected by the second type of terms, which explains why the same prefactor as for $a^*_{\rm Riem^2}$ appears in that case. Nevertheless, recall that $\xi$ is not a universal quantity (we can modify it by changing the regulator), so its interest is very limited. On the other hand, the universal constant $\kappa^{(d)}_{\rm Riem^2} $ does get affected by the extrinsic curvature term. The result for $\kappa^{(3)}_{\rm Riem^2}$ agrees with the one obtained in \cite{Bueno2}, as it should.

We find a similar kind of behavior for the cubic theories. Wald-like terms produce contributions proportional to the Einstein gravity result, and the non-universal constant $\xi_{\rm Riem^3}$ is proportional to $a^{\star(d)}_{\rm Riem^3}$, namely, $\xi_{\rm Riem^3}/\xi_{\rm E}  = a^{\star(d)}_{\rm Riem^3}/a^{\star(d)}_{\rm E}$.
%\begin{align}
%\xi_{\rm Riem^3}=&\left[1+3(d-1)\beta_1+12\beta_2+6d\beta_3+6d(d+1)\beta_4  \right. \\ \notag & \left.+3d^2\beta_5+3d^2\beta_6+3d^2(d+1)\beta_7+3d^2(d+1)^2\beta_8\right]  \xi_{\rm E} \, .
%\end{align}
On the other hand, terms with two extrinsic curvatures have an extra factor  $\sim y^{2(d-1)}$ in the integrand, and those with four, one of the form $\sim y^{4(d-1)}$. Both types of terms affect the universal coefficient.\footnote{Effectively every extra factor  $y^{2(d-1)}$ can be replaced by a $(2-d)$ factor as far as $\kappa^{(d)}$ is concerned, and every extra $y^{4(d-1)}$ can be replaced by a $d(2-d)/(2d-1)$. If we repeated the calculation including general order-$n$ higher-curvature pieces, we would obtain extra factors generally involving all even powers of $y$ up to $y^{2(n-1)(d-1)}$.} The final result reads
\begin{align}\label{kaappa}
\kappa^{(d)}_{\rm Riem^3} =&\Bigg[1+ 3\left[d-1+(d-1)(d-2)^2+\frac{d(d-2)^2(8+d(d^2-9))}{8(2d-1)} \right]\beta_1  \\ \notag &  \left. + 12\left[1-(d-1)(d-2)^2+\frac{d(d-1)(d-2)^2(7+d(d-5))}{4(2d-1)} \right]\beta_2 \right. \\ \notag &  \left.+2d\Big[3-(d-1)(d-2)^2\Big]\beta_3+2d(d+1)\Big[3-(d-1)(d-2)^2\Big]\beta_4 \right. \\ \notag & +3d^2\beta_5+3d^2\beta_6+3d^2(d+1)\beta_7+3d^2(d+1)^2\beta_8\Bigg] \kappa^{(d)}_{\rm E} \, .
\end{align}
A check of these results for $\kappa^{(d)}_{\rm Riem^2}$ and $\kappa^{(d)}_{\rm Riem^3}$ can be performed by particularizing them to Lovelock theories, for which the JM formula in \req{jm} can be alternatively used. We find
\begin{align}
\kappa^{(d)}_{ \mathcal{X}_4}&=[1+2 (d-3)(d-2)(d-1) \lambda_2]  \kappa^{(d)}_{ \rm E}\, , \\
\kappa^{(d)}_{ \mathcal{X}_6}&=[1-\frac{3 (d-5)(d-4)(d-3)(d-2)(d-1)^2}{(2d-1)} \lambda_3]  \kappa^{(d)}_{ \rm E}  \, ,
\end{align}
which precisely agree with the ones obtained using \req{jm}. Observe that the corrections to the Einstein gravity result vanish in dimensions lower or equal to the critical one, \ie for $d+1\leq 2n$. One can also verify that $\kappa^{(4)}_{ \mathcal{X}_4}$ agrees with the nonperturbative result found in \cite{deBoer:2011wk} at leading order in $\lambda_2$. For Quasi-topological and Einsteinian cubic gravity we find, respectively
\begin{align}
\kappa^{(4)}_{\rm QTG}&= \left[1+9\mu_{\rm QTG} \right] \kappa^{(4)}_{\rm E}\, , \\
\kappa^{(3)}_{\rm ECG}&=\left[1+3\mu_{\rm ECG} \right] \kappa^{(3)}_{\rm E}\, .
\end{align}
As mentioned above, the coefficient $\kappa^{(d)}$ does not have an alternative interpretation beyond entanglement entropy, which is manifest in this case from the fact that in all cases in which various coefficients characterizing the dual theory have been computed for some of the above theories, all the corresponding values differ from the ones obtained here for $\kappa^{(d)}$.\footnote{The exception is the sharp-limit corner coefficient $\kappa$, which can be shown to coincide with $\kappa^{(3)}$ on general grounds, as explained below.} This includes, in particular, all the rest of coefficients computed in this paper ($c$, $a$ in $d=4$; $A$, $B_1$, $B_2$, $B_3$ in $d=6$; $a^{\star (d)}$ in general $d$; the corner charge $\sigma$ in $d=3$) as well as others like the stress-tensor two-point function charge $\ctt$, the coefficient $C_S$ relating the thermal entropy of a plasma to its temperature, as well as others arising in the context of holographic complexity 
\cite{Buchel:2009sk,Myers:2010jv,deBoer:2011wk,Hung:2011xb,Bueno2,Safdi:2012sn,Miao:2015iba,Bhattacharyya:2014yga,Cano:2018ckq}.
%\comment{This includes, in particular, all the rest of coefficients computed in this paper ($c$, $a$, $B_1$, $B_2$, $B_3$, $a^{\star(d)}$), as well as others like $\ctt$ $C_S$, complexity related \cite{Cano:2018ckq}}

%\comment{coefficient not matching others}

\subsection{Cylinder regions}
Let us now consider (hyper)cylindrical entangling surfaces. We will be mostly interested in the universal logarithmic piece arising for such regions in $d=4$ and $d=6$ theories.
We write the Euclidean AdS$_{(d+1)}$ metric as
\begin{equation}
\diff s^2=\frac{\Ls^2}{z^2} \left[\diff \tau^2 +\rd z^2+\rd \vec{y}^2_{(d-3-j)}+\rd r^2+ r^2 \rd \Omega^2_{(j+1)} \right] \, ,
\end{equation}
where $\rd \Omega^2_{(j+1)}$ is the metric of a round $(j+1)$-dimensional sphere.
Our entangling regions will be parametrized by $\tau=0$, $r=R_0$, with $j$ taking values $j=0,\dots,d-3$, which correspond to entangling surfaces $\partial A=\mathbb{S}^1\times \mathbb{R}^{d-3},\mathbb{S}^2\times \mathbb{R}^{d-4},\dots,\mathbb{S}^{d-3}\times \mathbb{R}^1,\mathbb{S}^{d-2}$, respectively. %Note that the cases $j=-1$ and $j=d-3$ correspond to the slab and sphere cases, treated separately in other sections.

We parametrize the RT surface as $r=R(z)$. Unit normals and projectors on the surface read
\begin{equation}
n_1=\frac{z}{\Ls} \partial_{\tau}\, , \quad n_2=\frac{z}{\Ls \sqrt{1+R'^2}} \left(R' \partial_z - \partial_r \right)\, , \quad t_z=\partial_z+ R' \partial_r\, , t_m=\partial_m\, , t_{\phi}=\partial_{\phi} \, ,
\end{equation}
where $m=1,\dots,d-3-j$ and $\phi=1,\dots,j+1$.
The induced metric reads
\begin{equation}
h_{ij}  \rd y^i \diff y^j= \frac{\Ls^2}{z^2}  \left[(1+R'^2)\diff z^2+\rd \vec{y}^2_{(d-3-j)}+R^2 d\Omega^2_{(j+1)}  \right] \, .
\end{equation}
The non-vanishing components of $K_{ij}^2$ read now 
\begin{align} \label{spheres:trace_extrinsic}
K^2_{zz} &= \frac{-\Ls (R'+R'^3-z R'')}{z^2\sqrt{1+R'^2}}\, , \quad K_{mn}^2=\frac{-\Ls R' \delta_{mn} }{z^2\sqrt{1+R'^2}}\, , \\ K^2_{\phi_j \phi_k}&=\frac{-\Ls (z+R R') R}{z^2\sqrt{1+R'^2}} \prod_{l=1}^{j-1} \sin^2\phi_l \delta_{jk}\, .
\end{align}
The equation for the RT surface is, as usual, $K^{2}=0$, where
\begin{equation}\label{k2}
K^{2}=\frac{(R R''-(j+1))z - (d-1) R R' (1+R'^2)-(j+1)z R'^2}{\Ls  R (1+R'^2)^{3/2}}\, .
\end{equation}
In the case of Einstein gravity, the RT functional reduces to
\begin{equation}\label{eee}
\see^{\rm E}=\frac{\Ls^{d-1} L_y^{d-3-j} \Omega_{(j+1)}}{4G } \int^{z_{\rm max}}_{\delta} \rd z \frac{R^{j+1}}{z^{d-1}}\sqrt{1+R'^2}  \, ,
\end{equation}
where $\Omega_{(j+1)} \equiv 2\pi^{(j+2)/2}/ \Gamma[(j+2)/2]$. As anticipated, we are interested in the logarithmic contribution to the entanglement entropy in even dimensional theories. Such a contribution is local in the entangling surface $\partial A$ so, from the holographic perspective, it suffices to consider a perturbative solution to $K^{2}=0$ near the boundary. The result reads\footnote{When performing this expansion, it does not seem to be possible to solve the equation beyond quadratic order for $d=4$ and beyond quartic order in $d=6$. While this does not affect our calculations, it would be interesting to better understand the origin of this issue.  }
\begin{equation}\label{exp}
R(z)=R_0 - \frac{(j+1)}{2R_0 (d-2)}z^2+\mathcal{O}(z^4) \, ,
\end{equation}
which we need to plug back into our functionals.

\subsubsection{Four dimensions}
For general CFTs in four dimensions, the universal contribution to the entanglement entropy for a smooth entangling surface characterized by some scale $\ell$ is given by Solodukhin's formula \cite{Solodukhin:2008dh,Fursaev:2012mp}
\begin{equation} \label{see4}
\see^{(4)} \supset -\frac{1}{2\pi } \int_{\partial A} \diff^2y \sqrt{\gamma}\left[ a {\cal R} +c \left( \tr k^2-\frac12 k^2\right)
\right] \log \left( \frac{\ell }{\delta} \right)\, ,
\end{equation}
where $\mathcal{R}$ is the Ricci scalar of the induced  metric induced on $\partial A$, $\gamma_{ij}$,  and here and in the next subsection we use the notation $k \equiv \gamma^{ij} k_{ij}$  and $\tr k^n\equiv k_{i_1}^{i_2} k_{i_2}^{i_3}\dots k_{i_n}^{i_1}$, where $k_{ij}$ is the extrinsic curvature. $a$ and $c$ are the coefficients appearing in the usual trace-anomaly expression \cite{Duff:1977ay}
\begin{equation}
\braket{T_a^a} = -\frac{a}{16\pi^2} \mathcal{X}_4+\frac{c}{16\pi^2} C_{abcd}C^{abcd}\, ,
\end{equation}
where $\mathcal{X}_4$ and $ C_{abcd}$ are the Euler density and Weyl tensor of the curved manifold in which the CFT is considered.

Let us then start considering our holographic functionals for $d=4$ and $j=0$. For \req{eee} one finds
\begin{equation}\label{egs}
S_{\rm \ssc HEE}^{\rm E}=\frac{\pi \Ls^{3} L_y  }{2G } \int^{z_{\rm max}}_{\delta} \rd z \left[\frac{R_0}{z^3}- \frac{1}{8R_0 z}+\dots \right]=\dots -  \frac{c_{\rm E}}{2}\frac{L_y}{R_0} \log\left(R_0/\delta \right)  +\dots \, ,
\end{equation}
where
\begin{equation}
c_{\rm E}=\frac{\pi \Ls^3}{8G}\, .
\end{equation}
This takes the form expected for a cylinder region in general CFTs, where the value of $c_{\rm E}$ matches the corresponding trace anomaly charge. In our conventions, this is in turn related to the stress-tensor two-point function charge\footnote{This is defined as the only theory-dependent content of the stress-tensor correlator, which otherwise is completely fixed by conformal symmetry \cite{Osborn:1993cr}. For a general CFT in $d$-dimensions one finds $\braket{T_{ab}(x) T_{cd}(0)}=\ctt I_{ab,cd}(x)/x^{2d}$, where  $I_{ab,cd}(x)$ is a fixed tensorial structure.} $\ctt$ through $c=\pi^4 \ctt/40 $ for general theories ---compare with $\ctt^{\rm E}$ in \req{ctee}.

Performing the analogous calculations for quadratic and cubic theories, we observe that introducing the expansion \req{exp} in the corresponding functionals there are three kinds of terms which appear multiplying the Einstein gravity integrand in \req{egs}: terms coming from the Wald pieces, which are constant; terms involving products of two extrinsic curvatures, which are $\sim z^2$;
 and terms involving products of four extrinsic curvatures, which go with $\sim z^4$. Terms of the latter kind do not contribute to $c$, which is a manifestation of the splitting-independent nature of this coefficient. The final result for $c_{\rm Riem^2}$ and $c_{\rm Riem^3}$ reads
\begin{align}
c_{\rm Riem^2}&= \left[1-40\alpha_1-8\alpha_2+4\alpha_3 \right]c_{\rm E}\, , \\
c_{\rm Riem^3}&=\left[1+21\beta_1-36\beta_2-8\beta_3-40\beta_4+48\beta_5+48\beta_6 +240\beta_7+1200\beta_8 \right] c_{\rm E} \, .
\end{align}
These are again in agreement with the general relation with $\ctt$. Indeed, for general quadratic and cubic theories in $d$-dimensions one finds
\begin{align}\label{ctee23}
\ctt^{\rm Riem^2}=& \left[1-2d(d+1)\alpha_1-2d\alpha_2+4(d-3)\alpha_3\right]  \ctt^{\rm E} \\  \notag 
\ctt^{\rm Riem^3}=&\left[1+3(3d-5)\beta_1-12(2d-5)\beta_2-2d(2d-7)\beta_3-2d(2d-7)(d+1)\beta_4+3d^2\beta_5\right. \\  \notag  &\left.+3d^2\beta_6 +3d^2(d+1)\beta_7+3d^2(d+1)^2\beta_8 \right]  \ctt^{\rm E}  \, ,
\end{align}
where the Einstein gravity result reads
\begin{equation}\label{ctee}
\ctt^{\rm E}=\frac{\Gamma[d+2]}{8(d-1)\Gamma[\tfrac{d}{2}]\pi^{\tfrac{d+2}{2}}} \frac{\Ls^{d-1}}{G}\, .
\end{equation}
These results for $\ctt$ can be obtained in different ways. A simple one consists in computing the linearized equations of the theory around an AdS background. For a general higher-curvature gravity, these are fourth-order equations which describe the dynamics of a massive scalar mode and a ghost-like massive graviton in addition to the usual general relativity massless graviton. The resulting equations can be characterized in terms of the masses of the new two modes as well as an effective Newton constant \cite{Tekin1,Aspects}. This generically takes the form $G_{\rm eff}=G/ \gamma$,  where $\gamma$ depends on the higher-curvature couplings. Via holography, a rescaling of $G$ is equivalent to a rescaling of the stress-tensor charge $\ctt$, which becomes $\gamma C_{\ssc T}^{\rm E}$. %, where
%\begin{equation}\label{ctee}
%\ctt^{\rm E}=\frac{\Gamma[d+2]}{8(d-1)\Gamma[\tfrac{d}{2}]\pi^{\tfrac{d+2}{2}}} \frac{\Ls^{d-1}}{G}\, .
%\end{equation}
% is the Einstein gravity result. 
$G_{\rm eff}$ was computed in \cite{Aspects} explicitly for general quadratic, cubic and quartic gravities in general dimensions, so we can easily obtain the values of $\ctt$ shown above. %They read
% \begin{align}\label{ctee23}
%\ctt^{\rm Riem^2}=& \left[1-2d(d+1)\alpha_1-2d\alpha_2+4(d-3)\alpha_3\right]  \ctt^{\rm E} \\  \notag 
%\ctt^{\rm Riem^3}=&\left[1+3(3d-5)\beta_1-12(2d-5)\beta_2-2d(2d-7)\beta_3-2d(2d-7)(d+1)\beta_4+3d^2\beta_5\right. \\  \notag  &\left.+3d^2\beta_6 +3d^2(d+1)\beta_7+3d^2(d+1)^2\beta_8 \right]  \ctt^{\rm E}  \, ,
%\end{align}
In the particular cases of Lovelock, Quasi-topological and Einsteinian cubic gravity densities, they reduce to
\begin{align}
\ctt^{\mathcal{X}_4}=& \left[1-2 (d-2)(d-3)\lambda_2\right]   \ctt^{\rm E} \\ 
\ctt^{\mathcal{X}_6}=&\left[1+3(d-2)(d-3)(d-4)(d-5)\right] \ctt^{\rm E} \, , \\ 
\ctt^{\rm QTG}=&\left[ 1-3 \mu_{\rm QTG}\right] \ctt^{\rm E} \, , \\ 
\ctt^{\rm ECG}=&\left[1-3 \mu_{\rm ECG} \right] \ctt^{\rm E} \, . 
\end{align}
Note that all these differ from the slab coefficients $\kappa^{(d)}$ computed in the previous subsection.

\subsubsection{Six dimensions}
Let us now turn to six dimensions. In this case, a similar expression for the logarithmic term involving the trace anomaly coefficients holds for general CFTs, and is given by \cite{Safdi:2012sn,Miao:2015iba}
\begin{align}\label{see6}
\see^{(6)} \supset   \int_{\partial A} \diff^4 y \sqrt{\gamma} \Big[2 A \mathcal{X}_4+\frac{3\pi}{2} B_1 (3T_1-2T_2)-12\pi B_2 T_2 \\ \notag  + 6\pi B_3 (T_3+9T_1-12T_2) \Big] \log\left(\frac{\ell}{\delta}\right)\, ,
\end{align}
%Here $\diff \sigma\equiv \diff^4y \sqrt{\gamma}$, where 
where $\mathcal{X}_4$ is the Euler density associated to the induced metric $\gamma_{ij}$ and now
\begin{align}
T_1 & \equiv (\tr k^2)^2- \frac{1}{2}k^2\tr k^2+\frac{1}{16}k^4 \, , \\
T_2 & \equiv \tr k^4-k \tr k^3+\frac{3}{8}k^2\tr k^2-\frac{3}{64}k^4 \, ,  \\
T_3 &\equiv (\nabla_i k)^2-\frac{25}{16}k^4+11 k^2\tr k^2-6(\tr k^2)^2-16 k \tr k^3+12 \tr k^4\, .
\end{align}
Similarly, the coefficients $A$, $B_1$, $B_2$ and $B_3$ are the ones appearing in the trace anomaly, which in this case takes the form \cite{Bonora:1985cq,Deser:1993yx,Henningson:1998gx,Bastianelli:2000hi}
\begin{equation}
\braket{T_a^a}=\sum_{i=1}^3 B_i I_i + 2 A \mathcal{X}_6\, ,
\end{equation}
where $\mathcal{X}_6$ is the Euler density and the $I_i$ are cubic conformal invariants given by 
\begin{align}
I_1\equiv C_{d abc}C^{a ef b}C\indices{_e ^{dc} _f}\, , \quad I_2\equiv C\indices{_{ab} ^{cd}}C\indices{_{cd} ^{ef}} C\indices{_{ef} ^{ab}}\, , \\ I_3\equiv C\indices{_{aceg}} (\nabla^2 \delta^a_{b}+ 4R^a_b-\frac{6}{5} R \delta^a_b) C^{bceg}\, .
\end{align}

For the entangling regions we are considering here, the induced metric on $d=6$ Minkowski space  reads
\begin{equation}
ds^2_{\gamma}= \rd \vec{y}^2_{(3-j)}+R_0^2 \rd \Omega^2_{(j+1)} \, .
\end{equation}
The relevant expressions for the extrinsic curvature invariants read
\begin{equation}
k=\frac{(j+1)}{R_0}\, , \quad \tr k^n=\frac{(j+1)}{R_0^n}\, ,
\end{equation}
and from this, one finds
\begin{align}
\mathcal{X}_4&=\frac{(j-2)(j-1)j(j+1)}{R_0^4} \, , \\
T_1&=\frac{(j-3)^2(j+1)^2}{16 R_0^4} \, , \\ \quad T_2&=-\frac{(j-3)(j+1)(7+3j(j-2))}{64 R_0^4}\, ,\\  \quad T_3&=\frac{(j-3)(j+1)(3+j(26-25j)}{16 R_0^4}\, ,
\end{align} 
where, for completeness, we also included the value of $\mathcal{X}_4$ which vanishes for all the cylinder-like regions ($j=0,1,2$). Then, the entanglement entropy universal term reduces, for general CFTs, to
\begin{align}\label{redu}
\see \supset  \frac{(j+1) \Omega_{(j+1)}}{64} & \left[128 A j (j-1)(j-2)+3\pi (j-3) \Big[B_1 (9j (j-2)-11)\right.  \\ \notag & \left. +4B_2(3j(j-2)+7)-8B_3(j+1)(3+7j) \Big] \right]\frac{L_y^{3-j}}{R_0^{3-j}} \log\left(\frac{\ell}{\delta}\right) \, .
\end{align}
%where we also used $\mathcal{X}_4=0$, which holds for all these regions.

On the other hand, the holographic result for Einstein gravity reads%\footnote{If we wanted to incorporate the case $j=3$, corresponding to $\partial A=\mathbb{S}^4$, we would need to include an extra piece coming from the $2A\mathcal{X}_4$ term. This would add a contribution $-32(j-2)(j-1)j(j+1)$ to the $(1+j)^3(7j-9)$ appearing in the numerator of \req{eede}. }
\begin{align}\notag
S_{\rm \ssc HEE}^{\rm E}&=\frac{\Ls^{5} L_y^{3-j} \Omega_{(j+1)}}{4G } \int^{z_{\rm max}}_{\delta} \rd z \left[\frac{R_0^{j+1}}{z^5}-\frac{3(j+1)^2R_0^{j-1}}{32z^3}+\frac{(j+1)^3(7j-9)}{2048 R_0^{3-j} z}+\dots \right]  \, , \\ \label{eede}
 &=\dots + \frac{(1+j)^3(7j-9) \Omega_{(j+1)}\Ls^5 }{8192 G }\frac{L_y^{3-j}}{R_0^{3-j}} \log\left(\frac{\ell}{\delta}\right)+\cdots
\end{align}
Comparing with \req{redu} for $j=0,1,2,3$ we can obtain the Einstein gravity values of $A$, $B_1$, $B_2$, $B_3$. The results read
\begin{equation}
A_{\rm E}=\frac{\Ls^5}{512 G} \, , \quad B_1^{\rm E}=-\frac{\Ls^5}{256 \pi G} \, , \quad B_2^{\rm E}=-\frac{\Ls^5}{1024 \pi G} \, , \quad B_3^{\rm E}=\frac{\Ls^5}{3072 \pi G} \, , 
\end{equation}
in agreement with previous calculations \cite{deBoer:2009pn,Safdi:2012sn}. In particular, the value of the $A$ charge satisfies $A_{\rm E}=a^{\star (6)}_{\rm E}/ (32\pi^2)$, a relation which holds for general theories in the present conventions. In particular, the values of $A$ for all the rest of holographic higher-curvature theories are proportional to the corresponding coefficients $a^{\star (6)}$.

Moving to quadratic theories, the contributions without anomaly piece modify the charges in the same way as $a^{\star (6)}$, whereas the term involving two Riemanns contains an extra piece coming from a  contraction of extrinsic curvatures, which in this case reads
\begin{equation}
K_{aij} K^{aij}=-\frac{(j-3)(j+1)}{4L_{\star}^2 R_0^2} z^2 + \frac{(j-3)^2(j+1)^2}{64L_{\star}^2R_0^4}z^4 +\dots
\end{equation}
Putting the pieces together in the quadratic functional \req{seerie2} and again comparing with \req{redu}  we find
\begin{align}
B_1^{\rm Riem^2}&=\left[1-84\alpha_1-12\alpha_2+\frac{4}{3}\alpha_3\right]B_1^{\rm E}\, , \\
B_2^{\rm Riem^2}&=\left[1-84\alpha_1-12\alpha_2-\frac{28}{3}\alpha_3\right]B_2^{\rm E}\, , \\
B_3^{\rm Riem^2}&=\left[1-84\alpha_1-12\alpha_2+12\alpha_3\right]B_3^{\rm E}\, .
\end{align}
We have verified that these results reduce to the ones  found in \cite{Miao:2013nfa} for seven-dimensional Critical Gravity \cite{Lu,Deser:2011xc}. In that case, $\alpha_1=-1/240$, $\alpha_2=1/20$, $\alpha_3=-1/16$ and the charges read $B_1^{\rm CG}=2/3$, $B_1^{\rm CG}=4/3$, $B_3^{\rm CG}=0$. It is also easy to verify that the resulting charges satisfy the relation $3B_3=(B_2-B_1/2)$, which holds for theories that are unaffected by the splittings choice, as argued in \cite{Miao:2015iba}.

Proceeding analogously with the cubic densities, we obtain
\begin{align}
B_1^{\rm Riem^3}&=\left[1+39\beta_1-20\beta_2+4\beta_3+28\beta_4+108\beta_5+108\beta_6+756\beta_7+5292\beta_8\right]B_1^{\rm E}\, , \\
B_2^{\rm Riem^3}&=\left[1+7\beta_1-20\beta_2+68\beta_3+476\beta_4+108\beta_5+108\beta_6+756\beta_7+5292\beta_8\right]B_2^{\rm E}\, , \\
B_3^{\rm Riem^3}&=\left[1+39\beta_1-84\beta_2-60\beta_3-420\beta_4+108\beta_5+108\beta_6+756\beta_7+5292\beta_8\right]B_3^{\rm E}\, .
\end{align}
We can check, at this order, which theories satisfy the  $3B_3-(B_2-B_1/2)=0$ condition. Evaluating the quantity in the left-hand side, one obtains
\begin{equation}
3B_3-(B_2-B_1/2)=-\frac{(\beta_1+2\beta_2) \Ls^2}{32\pi G}\, .
\end{equation}
Hence, such a combination vanishes for all theories for which $\beta_1=-2\beta_2 $. This includes, in particular, the cubic Lovelock density, in agreement with the result of \cite{Safdi:2012sn}. The explicit expressions for the quadratic and cubic theories read
\begin{align}
B_1^{\mathcal{X}_4}&=\left[1-\frac{104}{3}\lambda_2\right]B_1^{\rm E}\, , \quad B_2^{\mathcal{X}_4}=\left[1-\frac{136}{3}\lambda_2\right]B_2^{\rm E}\, , \quad B_3^{\mathcal{X}_4}=\left[1-24\lambda_2\right]B_3^{\rm E}\, ,\\
B_1^{\mathcal{X}_6}&=\left[1+136\lambda_3\right]B_1^{\rm E}\, , \quad B_2^{\mathcal{X}_6}=\left[1+200\lambda_3\right]B_2^{\rm E}\, , \quad B_3^{\mathcal{X}_6}=\left[1+72\lambda_3\right]B_3^{\rm E}\, .
\end{align}

\subsection{Corner regions}
In this subsection we construct the universal function characteristic of corner regions for general holographic cubic gravities using the perturbative HEE functionals. We show that the introduction of such terms in the bulk Lagrangian modifies the angular dependence of the Einstein gravity function, as opposed to previously considered quadratic and $f(R)$ theories. We compute the new functions explicitly and perform some comparisons with the analogous ones corresponding to free scalars and fermions.
\subsubsection*{General aspects of corner entanglement}
The structure of divergences and universal terms in the entanglement entropy gets modified when the entangling surface $\partial A$ contains geometric singularities ---see \eg \cite{Myers:2012vs,Bueno:2019mex} for some general accounts of this phenomenon in various dimensions. Here, we will focus on the prototypical example of (straight) corners in $d=3$ CFTs. Given a fixed time slice, the entanglement entropy corresponding to a corner region of opening angle $\theta$ in the ground state of a CFT regulated by a UV cutoff $\delta$ takes the form 
\begin{equation}\label{geneco}
\see = b_1 \frac{H }{\delta} - a(\theta) \log(H/\delta)+ b_0\, .
\end{equation}
Here, $H$ is an IR regulator and $b_1$ is a non-universal coefficient. On the other hand, $b_0$ is a coefficient which generically contains a universal non-local contribution and a non-universal part of intrinsically local nature induced by possible redefinitions of the regulator $\delta$.

With respect to the case of smooth regions, the novelty here is the appearance of a new logarithmic divergence controlled by the corner function $a(\theta)$, of universal nature. By now, many aspects of this function have been studied in a plethora of contexts %---see \eg \cite{Casini:2006hu,Casini:2008as,fradkin,Hirata:2006jx,Casini:2009sr,2011PhRvB..84p5134K,Kallin:2014oka,Bueno1,Bueno2,Bueno3,Fonda:2014cca,Bueno:2015ofa} and references therein.
---\eg for free fields \cite{Casini:2006hu,Casini:2008as,Casini:2009sr,Bueno3,Dowker:2015pwa,Dowker:2015tma,Elvang:2015jpa}, for large-$N$ vector models \cite{Whitsitt2017}, for holographic theories \cite{Hirata:2006jx,Bueno2,Fonda:2015nma,Alishahiha:2015goa,Miao:2015dua,Pang:2015lka,Bianchi:2016xvf,Mozaffar:2015xue,Pastras:2017fsy,Ghasemi:2017pke,Bakhshaei:2017qud,Caceres:2018luq,Ghasemi:2019mif,Dorn:2018als}, in interacting lattice models \cite{2011PhRvB..84p5134K,PhysRevLett.110.135702,sahoo15,Kallin:2014oka,Laflorencie:2015lwa,Helmes:2015mwa,DeNobili:2016nmj,Helmes:2016fcp}, and for general CFTs \cite{Bueno1,Faulkner:2015csl,Bueno:2015ofa,Witczak-Krempa:2016jhc,Chu:2016tps}.  As a result of this thorough study, the function $a(\theta)$ has been shown to satisfy a number of properties, universal relations and bounds which we summarize now. 

On the one hand, the purity of the ground state, which implies the well-known relation $\see(A)=\see(\bar A)$, requires $a(\theta)=a(2\pi-\theta)$. Besides, using strong subadditivity and Lorentz invariance one can show that  \cite{Casini:2008as}
\begin{equation}\label{atht}
a(\theta)\geq 0 \, ,\quad \partial_{\theta} a(\theta) \leq 0\, , \quad \partial^2_{\theta} a(\theta) \geq -\frac{\partial_{\theta }a(\theta)}{\sin \theta}\, , \quad \text{for} \quad \theta \in [0,\pi]\, .
\end{equation}
In particular, this implies that $a(\theta)$ is a positive, monotonously-decreasing and convex function of the opening angle as we vary it from $\theta\sim 0$, corresponding to a very sharp corner, to $\theta\sim \pi$, corresponding to a very open, almost-smooth, corner. In those two limits, the function behaves, respectively, as \cite{Casini:2006hu,Casini:2009sr,Casini:2008as}
\begin{equation}\label{athth}
a(\theta \simeq 0) = \frac{\kappa}{ \theta} + \mathcal{O}(\theta)\, , \quad  \quad a(\theta \simeq \pi) = \sigma \cdot (\theta - \pi)^2+\sum_{p=2} \sigma^{(p-1)} \cdot (\theta- \pi)^{2p}\, .
\end{equation}
In the first expression, $\kappa$ is a constant which can be shown to coincide with the slab coefficient $\kappa^{(3)}$ ---see \req{slabs} above--- for general theories \cite{Myers:2012vs,Bueno2}. In the second formula, we have made manifest the fact that only even powers appear in the expansion. The leading coefficient, $\sigma$, turns out to be related to the stress-energy tensor two-point function coefficient $\ctt$ through
%\footnote{For any CFT in $d$ dimensions, the stress-tensor correlator behaves as $\braket{T_{ab}(x)T_{cd}(0)}=\ctt I_{ab,cd}(x)/|x|^{2d}$  where $I_{ab,cd}(x)$ is a fixed tensorial structure, and the only theory-dependent quantity is the charge $\ctt$ \cite{Osborn:1993cr}. }
\begin{equation}\label{sigmact}
\sigma=\frac{\pi^2}{24} \ctt \, ,
\end{equation}
for general CFTs. This relation was conjectured in \cite{Bueno1} based on holographic and free-field calculations and proved in full generality in \cite{Faulkner:2015csl} ---see also \cite{Miao:2015dua,Bueno4,Elvang:2015jpa} for intermediate progress and partial proofs. In fact, the full corner functions of all CFTs considered so far in the literature turn out to become very close to each other when normalized by $\ctt$  \cite{Bueno1}.

Using \req{sigmact} and the third relation in \req{atht}, a lower bound on $a(\theta)$ valid for general CFTs was obtained in \cite{Bueno:2015ofa}. This takes the form
\begin{equation} \label{amin}
a(\theta) \geq \mathfrak{a}_{\rm min}(\theta) \, , \quad \text{where} \quad  \mathfrak{a}_{\rm min}(\theta)\equiv \frac{\pi^2 \ctt}{3} \log \left[1/\sin(\theta/2) \right]\, ,
\end{equation}
where $\ctt$ is to be understood as the one corresponding to the theory we are comparing with. The bound turns out to be pretty tight for all theories considered so far, even for considerably small values of the opening angle \cite{Bueno:2015ofa} ---see also \cite{Sirois:2020zvc}. In particular, the actual values found from numerical and lattice simulations corresponding to various models for $\theta=\pi/2$, all fall within the approximate range $a(\pi/2)/\ctt \in(1.2, 1.3)$ \cite{PhysRevLett.110.135702,sahoo15,Kallin:2014oka,Helmes:2016fcp}, whereas the bound value reads $\mathfrak{a}_{\rm min}(\pi/2)/\ctt\simeq 1.1402$.

Additional lower bounds valid also for the general R\'enyi entropy versions of $a(\theta)$ can be constructed using the inequalities 
\begin{equation}
{\rm det} \left\{ \partial_{\theta}^{j+k+2} a_{n}(\theta) \right\}_{j,k=0}^{M-1} \geq 0\, ,
\end{equation}
which follow from the reflection positivity property of Euclidean QFTs \cite{Casini4}. Such bounds were explored in \cite{Bueno:2015ofa,Helmes:2016fcp} and suggest, in particular, that all coefficients in the almost-smooth expansion in \req{athth} are positive, \ie $ \sigma^{(p-1)} >0$ $\forall \, p$.\footnote{This has been shown to be true in general for $p=1,2,3,4,5$ in \cite{Bueno:2015ofa}.} In fact, for sufficiently large $p$, it was observed in \cite{Bueno:2015ofa} that those coefficients behave as
\begin{equation}
\sigma^{(p)} \simeq \frac{2\kappa}{\pi^{2p+3}}\, , \quad p\gg 1\, ,
\end{equation}
where $\kappa$ is the sharp-limit coefficient.

The results mentioned so far are valid for general CFTs. Theories for which $a(\theta)$ has been actually computed for general values of the opening angle are nonetheless scarce. For free scalars and fermions,  $a(\theta)$ was obtained numerically from a complicated set of coupled  differential and algebraic equations in \cite{Casini:2006hu,Casini:2008as,Casini:2009sr}. In addition, the Ryu-Takayanagi prescription allowed for the computation of the corresponding corner function for holographic theories dual to Einstein gravity \cite{Hirata:2006jx}. The resulting expression is shown below in \req{eisss} and is given implicitly in terms of two integrals.
%where $L$ is the AdS$_{4}$ radius and $G$ is Newton's constant and the second integral 
%
The only two cases for which a completely explicit expression for $a(\theta)$ is known correspond, respectively, to certain Lifshitz quantum critical points \cite{fradkin} and the so-called ``Extensive Mutual Information model'' \cite{Casini:2005rm,Casini:2008wt,Swingle:2010jz}. The corresponding corner functions read 
\begin{equation}
a_{\rm \ssc Lif.}(\theta)=\frac{(\theta-\pi)^2}{\theta(2\pi-\theta)}\, , \quad\quad a_{\rm \ssc EMI}(\theta)=1+ (\pi-\theta) \cot \theta\, .
\end{equation}
Using these two functions, it is possible to construct a simple approximation to the corner function of any CFT provided one knows the values of the corresponding sharp and smooth coefficients, $\kappa$ and $\sigma$. This is given by \cite{Bueno3}
\begin{equation}\label{trii}
\tilde a(\theta)=\frac{2\pi (\kappa-3\pi \sigma)}{\pi^2-6} \frac{(\theta-\pi)^2}{\theta(2\pi-\theta)}-\frac{3(2\kappa-\pi^3\sigma)}{\pi (\pi^2-6)} \left[1+ (\pi-\theta) \cot \theta \right]\, .
\end{equation}
This respects the asymptotic behavior both as $\theta\rightarrow 0$ and as $\theta \rightarrow \pi$ and produces very precise approximations to the actual free-field and Einstein gravity results. In all cases, the relative agreement is always better than 99$\%$ for all values of $\theta$.
If access to some of the subleading coefficients $\sigma^{(p)}$ is also available, improved ansatze can be constructed, as shown in \cite{Helmes:2016fcp}.

%\subsection{Holographic Einstein gravity}

%\subsection{Corner entanglement for holographic higher-curvature gravities}
\subsubsection*{Einstein gravity}
Let us quickly review how the corner function is obtained for Einstein gravity \cite{Hirata:2006jx,Drukker:1999zq}. First, it is useful to write the AdS$_3$ metric as 
\begin{equation}
\diff s^2=\frac{\Ls^2}{z^2}[\diff \tau^2+\diff z^2+\diff r^2+r^2 \diff \phi^2]\, .
\end{equation}
The corner region is defined by $\tau=0$, $r \geq 0$, $|\phi| \leq \theta/2$.  We can parametrize the bulk surface as $z=r  h(\phi)$, where $h(\phi)$ is a function satisfying $h(\phi \rightarrow \pm \theta/2)\rightarrow 0$. Unit normals to the surface are given by
\begin{equation}
n_1=\frac{z}{\Ls} \partial_{\tau} \, , \quad n_2=\frac{z}{\Ls \sqrt{1+h^2+\dot h^2}}\left[\partial_z - h \partial_r-\frac{\dot h}{r} \partial_{\phi} \right]\, .
\end{equation}
Using these we have
\begin{align}
h_{\mu\nu}\diff x^{\mu}\diff x^{\nu}=\frac{\Ls^2}{z^2} &\left[ \diff \tau^2+ \frac{1}{(1+h^2+\dot h^2)} \left[(h^2+\dot h^2)\diff z^2+ (1+\dot h^2)\diff r^2  \right. \right. \\ &\left. \left. + r^2(1+h^2) \diff \phi^2  + 2h \diff z \diff r +2 r \dot h \diff z \diff \phi -2 r h \dot h \diff r \diff \phi \right] \right]\, . 
\end{align}
Projectors on the surface are given by
\begin{equation}
t_r= h \partial_z + \partial_r \, , \quad t_{\phi}= r \dot h \partial_z+\partial_{\phi}\, ,
\end{equation}
and the projected induced metric reads
\begin{align}
h_{ij} \diff y^i \diff y^j =\frac{\Ls^2}{r^2 h^2} \left[ (1+h^2)\diff r^2+r^2 (1+\dot h^2)\diff \phi^2 + 2r h \dot h \diff r \diff \phi \right]\, .
\end{align}
The non-vanishing components of the extrinsic curvatures, $K^2_{ij}$, are in turn given by
\begin{align}
& K^2_{rr}=\frac{-\Ls (1+h^2)}{r^2 h^2\sqrt{1+h^2+\dot h^2}}\, , \quad K^2_{r \phi}=\frac{-\Ls \dot h}{r h \sqrt{1+h^2+\dot h^2}}\, , \\ & K_{\phi\phi}^2=\frac{-\Ls (1+h^2+\dot h^2+\ddot h h)}{h^2\sqrt{1+h^2+\dot h^2}}\, .
\end{align}
These are all the pieces we will need to evaluate the corner function for perturbative higher-order gravities.

For our parametrization of the holographic entangling surface, the Ryu-Takayanagi functional becomes 
\begin{equation}
S_{\rm \ssc HEE}^{\rm E}=\frac{\Ls^2}{2G}\int_{\delta/h_0}^H\frac{\diff r}{ r} \int_0^{\theta/2-\epsilon} \diff \phi \frac{\sqrt{1+h^2+\dot h^2}}{h^2}\, ,
\end{equation}
where we already made manifest the UV cutoff at $z=\delta$ and where $h_0\equiv h(0)$ is the maximum value taken by the function $h(\phi)$. Also, the angular cutoff $\epsilon$ is defined through the condition $r h(\theta/2-\epsilon)=\delta$, which means that the integral over $r$ cannot be performed without doing the angular one first. The extremal surface condition, $K^2=0$, reads 
\begin{equation}
2+3h^2+h^4+2\dot h^2+h(1+h^2)\ddot h=0\, .
\end{equation}
This has a first integral,
\begin{equation}
\frac{1+h^2}{h^2\sqrt{1+h^2+\dot h^2}}=\frac{\sqrt{1+h_0^2}}{h_0^2}\, ,
\end{equation}
which can be used to write $\dot h$ in terms of $h$ in the RT functional. Trading the integral over $\phi$ by one over $h$ and making the change of variables $y=\sqrt{1/h^2-1/h_0^2}$ we are left with
\begin{align}
S_{\rm \ssc HEE}^{\rm E}&=\frac{\Ls^2}{2G}\int_{\delta/h_0}^H\frac{\diff r}{ r}  \int_0^{\sqrt{(r/\delta)^2-1/h_0^2}}\diff y \sqrt{\frac{1+h_0^2(1+y^2)}{2+h_0^2(1+y^2)}} \\ 
&= \frac{\Ls^2}{2G}\int_{\delta/h_0}^H\frac{\diff r}{ r}  \int_0^{\infty} \diff y \left[ \sqrt{\frac{1+h_0^2(1+y^2)}{2+h_0^2(1+y^2)}} -1\right]+ \frac{\Ls^2}{2G} \int_{\delta/h_0}^H \frac{\diff r}{ r} \sqrt{\frac{r^2}{\delta^2}-\frac{1}{h_0^2}}\, .
\end{align}
Expanding this expression for small $\delta$ one finally obtains
\begin{equation}
S_{\rm \ssc HEE}^{\rm E}=\frac{\Ls^2}{2G} \frac{H}{\delta}- a_{\rm E}(\theta) \log (H/\delta)+ \mathcal{O}(\delta^0) \, ,
\end{equation}
in agreement with the general expression \req{geneco}. The result for the Einstein gravity corner function can be written as %\begin{equation}\label{eisss}
%a_{\rm \ssc E}(\theta)=\frac{L^2}{2G} \int_{g_0}^{+\infty} \frac{  g\, dg}{\sqrt{g^2-g_0^2}}\left[1- \sqrt{\frac{1+g^2}{1+g_0^2+g^2}}\right]\, , \quad \theta=\int_{g_0}^{+\infty}\frac{2 \,dg}{g \sqrt{(1+g^2)  \left[\frac{g^2(1+g^2) }{ g_0^2(1+g_0^2)}-1 \right] }}\, .
%\end{equation}
 \cite{Drukker:1999zq,Hirata:2006jx}
\begin{align}\label{eisss}
a_{\rm E}(\theta)&=\frac{\Ls^2}{2G} \int_{0}^{+\infty} \diff y \left[1-\sqrt{\frac{1+h_0^2(1+y^2)}{2+h_0^2(1+y^2)}} \right]\, , \\ \theta&=\int_{0}^{h_0} \diff h \frac{2 \sqrt{1+h_0^2}h^2}{ \sqrt{1+h^2}\sqrt{(h_0^2-h^2)(h_0^2+(1+h_0^2)h^2)}  }\, ,
\end{align}
where the dependence on the opening angle follows implicitly from the relation $h_0(\theta)$ determined by  the second integral.
The above expressions can be alternatively written in terms of elliptic functions \cite{Fonda:2014cca} as
\begin{align}
a_{\rm E}(\theta)&=\frac{\Ls^2}{2G} \sqrt{1+\frac{2}{h_0^2}} \left[ \mathbb{E}\left[\frac{1}{2+h_0^2} \right] - \frac{(1+h_0^2)}{(2+h_0^2)}  \mathbb{K}\left[\frac{1}{2+h_0^2} \right] \right] \, , \\
\theta&=-\frac{2h_0}{\sqrt{2+h_0^2(3+h_0^2)}} \left[\mathbb{K}\left[\frac{1}{2+h_0^2} \right]- \Pi \left[\frac{1+h_0^2}{2+h_0^2},\frac{1}{2+h_0^2}  \right] \right]\, .
\end{align}
It can be verified that $a_{\rm E}(\theta)$ satisfies all properties explained in the previous subsection. Values of the opening angle close to $\theta=\pi$ correspond to $h_0\rightarrow \infty$, and an expansion of the $\theta(h_0)$ integral in that case can be obtained and inverted giving 
\begin{equation}
h_0=\left(\frac{\pi}{\pi - \theta}\right)-\frac{3}{4} \left( \frac{\pi - \theta}{\pi} \right)- \frac{11}{64}  \left( \frac{\pi - \theta}{\pi} \right)^3 -\frac{17}{256}  \left( \frac{\pi - \theta}{\pi} \right)^5 -\frac{383}{16384} \left( \frac{\pi - \theta}{\pi} \right)^7+ \mathcal{O}(\pi - \theta)^9\, .
\end{equation}
Inserting this in $a_{\rm E}(\theta)$ one obtains  an expansion of the form of the second expression in \req{athth}, where the leading smooth-limit coefficients are given by \cite{Bueno2,Bueno:2015ofa}
\begin{align*}
 \sigma_{\rm E}=\frac{\Ls^2}{8\pi G}\, ,   \quad \sigma'_{\rm E}=\frac{5\Ls^2}{64\pi^3 G}\, , \quad \sigma''_{\rm E}=\frac{37\Ls^2}{512 \pi^5 G}\, , \quad \sigma'''_{\rm E}=\frac{585\Ls^2}{8192 \pi^7 G}\, , \quad \sigma^{(4)}_{\rm E}=\frac{9399\Ls^2}{131072 \pi^9 G}\, .  %&&\sigma^{(5)}_{\rm E}=\frac{75699L^2}{1048576 \pi^{11} G}\, .
\end{align*}
As many higher-order coefficients as desired can be determined analytically in the same way. 
On the other hand, the sharp limit coefficient is given by \cite{Bueno2}
\begin{equation}
\kappa_{\rm E}=\frac{\Ls^2}{2\pi G} \Gamma[3/4]^4\, . %\quad \sigma_{\rm E}=\frac{L^2}{8\pi G}\, .
\end{equation}
%\comment{blah, limits, sigma, kappa}

\subsubsection*{Quadratic theories}
As observed in \cite{Bueno2}, the only modification produced on the Einstein gravity corner function $a_{\rm E}(\theta)$ which arises from including quadratic or $f(R)$ terms in the gravitational action is an overall constant shift. In particular, for an action of the form \req{quaact} one finds
\begin{equation}
a_{\rm Riem^2}(\theta)= \left[1-24 \alpha_1-6\alpha_2 \right]\,  a_{\rm E}(\theta)\, .
\end{equation}
Hence, no new functional dependence on the opening angle is found from these gravitational interactions.
As discussed in some detail in the same paper, the reason for this can be easily understood. On the one hand, all terms involving bulk curvatures will reduce to terms proportional to the Ryu-Takayanagi functional when evaluated on the pure AdS$_4$ background we are considering. On the other hand, any term proportional to $K^aK_a$ will also be extremized by RT surfaces, since the extremal surface condition reads  $K^a=0$. As a consequence, terms proportional to $K^aK_a$  in the action will simply vanish on extremal surfaces and will not contribute. Finally, a term like $K_{aij}K^{aij}$ can also be deduced not to contribute from the fact that we can replace the $R_{\mu\nu\rho\sigma}R^{\mu\nu\rho\sigma}$ piece by the Gauss-Bonnet density (plus additional $R^2$ and $R_{\mu\nu}R^{\mu\nu}$ terms) whose contribution to the EE functional is the intrinsic Ricci scalar on the RT surface \cite{Jacobson:1993xs,Hung:2011xb}, which is a topological term in $(d-1)=2$ dimensions and therefore makes no contribution to the equations of motion. In this case, it does not even modify the Einstein gravity result by an overall constant.

%So far, no example of bulk action is known for which the corner function gets modified in its functional form with respect to the Einstein gravity. In \cite{Bueno2} it was conjectured that this situation should change in the presence of higher-order interactions, but the lack of the corresponding functionals beyond quadratic order did not allow for such explorations at that time. 
Our results here allow us to compute the corner function for cubic theories and verify that non-trivial modifications of $a_{\rm E}(\theta)$  arise in the presence of such terms.

\subsubsection*{Cubic theories}
Let us then consider a general cubic action of the form \req{cubic}. If we only turn on the couplings corresponding to $\mathcal{L}_{i}^{(3)}$ with $i=3,4,5,6,7,8$ we find that, similarly to the quadratic case, the corner function is the same as for Einstein gravity up to an overall factor. In the $i=5,6,7,8$ cases, the fact that the functionals have no anomaly contribution imply that the overall coefficient correcting the Einstein gravity result is the same as for $a^{\star (3)}$. For $i=3,4$, even though there is no modification in the functional dependence of the corner function, there is a modification to the overall coefficient coming from the anomaly terms. The result for all these densities reads
%In particular, for
%\begin{equation}\label{cubic2}
%I_{\mathcal{L}_{(3,4,5,6,7,8)}^{(3)}}= \int \frac{d^{4}x \sqrt{|g|}}{16\pi G} \left[\frac{6}{L^2}+R+L^4 \sum_{i=3}^8 \beta_i \mathcal{L}_i^{(3)} \right]\, ,
%\end{equation}
%we find \comment{comment with some more detail on the intermediate calculations} \comment{clarify $L$ vs $\Ls$ thing, we express everything in terms of $\Ls$ throughout}
\begin{equation}
a_{\mathcal{L}_{(3,4,5,6,7,8)}^{(3)}}(\theta) = [1+6\beta_3+24\beta_4+27\beta_5+27\beta_6+108\beta_7+432\beta_8] a_{\rm E}(\theta) \, .
\end{equation}
On the other hand, $\mathcal{L}_{1}^{(3)}$ and $\mathcal{L}_{2}^{(3)}$ do modify the angular dependence of $a_{\rm E}$. Keeping only those two terms in the action, we find instead
\begin{equation}
a_{\mathcal{L}_{(1,2)}^{(3)}}(\theta) = [1+6\beta_1+12\beta_2] a_{\rm E}(\theta)+ \sum_{i=1}^2\beta_i g_{i}(\theta) \, ,
\end{equation}
where
%\begin{align}
% a_{1}(\theta)&\equiv \frac{L^2}{2G} \int_{g_0}^{+\infty} \frac{3 g_0^2 (1+g_0^2)  g \left[g_0^2(1+g_0^2)+2(1+g^2)^2\right]}{(1+g^2)^{7/2} \sqrt{(g^2-g_0^2)(1+g^2+g_0^2)}}dg  \, ,\\
% a_{2}(\theta)&\equiv  \frac{L^2}{2G} \int_{g_0}^{+\infty} \frac{6 g_0^2 (1+g_0^2)  g \left[g_0^2(1+g_0^2)-4(1+g^2)^2\right]}{(1+g^2)^{7/2} \sqrt{(g^2-g_0^2)(1+g^2+g_0^2)}}dg \, ,\\
%  a_{3}(\theta)&\equiv  \frac{L^2}{2G} \int_{g_0}^{+\infty} \frac{-12 g_0^2 (1+g_0^2)  g }{(1+g^2)^{3/2} \sqrt{(g^2-g_0^2)(1+g^2+g_0^2)}}dg \, ,\\
% a_{4}(\theta)&\equiv  \frac{L^2}{2G} \int_{g_0}^{+\infty} \frac{-48 g_0^2 (1+g_0^2)  g }{(1+g^2)^{3/2} \sqrt{(g^2-g_0^2)(1+g^2+g_0^2)}}dg \, ,\\
%\end{align}
\begin{align}
 g_{1}(\theta)&\equiv + \frac{\Ls^2}{2G} \int_{0}^{+\infty} \frac{3(1+h_0^2)\left[3+h_0^2(5+4y^2)+2h_0^4(1+y^2)^2 \right]}{\left[1+h_0^2(1+y^2)\right]^{7/2} \sqrt{2+h_0^2(1+y^2)}}\diff y  \, ,\\
g_{2}(\theta)&\equiv - \frac{\Ls^2}{2G} \int_{0}^{+\infty} \frac{6(1+h_0^2)\left[3+h_0^2(7+8y^2)+4h_0^4(1+y^2)^2 \right]}{\left[1+h_0^2(1+y^2)\right]^{7/2} \sqrt{2+h_0^2(1+y^2)}}\diff y  \, .
  %a_{3}(\theta)&\equiv - \frac{L^2}{2G} \int_{0}^{+\infty} \frac{12 (1+h_0^2) }{\left[1+h_0^2(1+y^2)\right]^{3/2} \sqrt{2+h_0^2(1+y^2)}}dy \, ,\\
% a_{4}(\theta)&\equiv - \frac{L^2}{2G} \int_{0}^{+\infty} \frac{48(1+h_0^2) }{\left[1+h_0^2(1+y^2)\right]^{3/2} \sqrt{2+h_0^2(1+y^2)}}dy\, ,\\
\end{align}
%\comment{limits of new functions, plots}
Hence, at cubic order we find the first examples of holographic corner functions which modify the angular dependence of $a(\theta)$ in a nontrivial way with respect to the Einstein gravity case.

As we mentioned earlier, the almost-smooth limit of the corner function is controlled by $\ctt$ for all CFTs. %We can independently compute $\ctt$ for a general cubic theory using standard holographic methods and verify whether \req{sigmact} indeed holds. There are several ways $\ctt$ can be accessed. A simple one consists in computing the linearized equations of the theory around an AdS background. For a general higher-curvature gravity, these are fourth-order equations which describe the dynamics of a massive scalar mode and a ghost-like massive graviton in addition to the usual general relativity massless graviton \comment{rewriting here perhaps depending on what is said earlier}. The resulting equations can be characterized in terms of the masses of the new two modes as well as an effective Newton constant \cite{Tekin1,Aspects}. This generically takes the form $G_{\rm eff}=G/ \gamma$,  where $\gamma$ depends on the higher-curvature couplings. Via holography, a rescaling of $G$ is equivalent to a rescaling of the stress-tensor charge $\ctt$, which becomes $\gamma C_{\ssc T}^{\rm E}$, where
%$
%C_{\ssc T}^{\rm E}=3L^2/(\pi^3 G)
%$
 %is the Einstein gravity result. $G_{\rm eff}$ was computed in \cite{Aspects} explicitly for general quadratic, cubic and quartic gravities in general dimensions. 
For cubic theories, the result for this coefficient appears in \req{ctee23} above. In $d=3$ one finds
 \begin{equation}
C_{\ssc T}^{\rm Riem^3}=\left[1+12\beta_1-12\beta_2+6\beta_3+24\beta_4+27\beta_5+27\beta_6+108\beta_7+432\beta_8\right] C_{\ssc T}^{\rm E}\, ,
 \end{equation}
 where $C_{\ssc T}^{\rm E}=3L^2/(\pi^3 G)$.
Now, including all cubic terms in the action, we find for the smooth limit of $a_{{ \rm{Riem}}^3}(\theta)$ that indeed
\begin{equation}
\sigma_{\rm Riem^3}= \frac{\pi^2}{24} C_{\ssc T}^{\rm Riem^3}\, ,
\end{equation}
holds, as expected. This was in fact previously verified in \cite{Miao:2015dua}, where several general results regarding the behavior of $a(\theta)$ for holographic theories were discussed, including the fact that $\kappa$ is not universally related to $\ctt$, as opposed to $\sigma$. The subleading coefficients in the smooth-limit expansion are modified with respect to the Einstein gravity result in an obvious way for $\mathcal{L}_{(3,4,5,6,7,8)}^{(3)}$ but in a nontrivial one for $\mathcal{L}_{1}^{(3)}$ and $\mathcal{L}_{2}^{(3)}$. The first few of them read
\begin{align}
& \sigma_{\mathcal{L}_{(1,2)}^{(3)}}=[1+12\beta_1-12\beta_2] \sigma_{\rm E}\, ,   \quad  &&\sigma'_{\mathcal{L}_{(1,2)}^{(3)}}=[1+15\beta_1-6\beta_2]\sigma'_{\rm E}\, , \\  &\sigma''_{\mathcal{L}_{(1,2)}^{(3)}}=\left[1+\frac{1173}{74}\beta_1-\frac{189}{37}\beta_2\right] \sigma''_{\rm E}\, , \quad &&\sigma'''_{\mathcal{L}_{(1,2)}^{(3)}}=\left[1+\frac{963}{65}\beta_1-\frac{414}{65}\beta_2\right] \sigma'''_{\rm E}\, , \\
& \sigma^{(4)}_{\mathcal{L}_{(1,2)}^{(3)}}=\left[1+\frac{43946}{3133}\beta_1-\frac{24896}{3133}\beta_2\right] \sigma^{(4)}_{\rm E}\, .
% \\  &\sigma^{(4)}_{\rm E}=\frac{9399L^2}{131072 \pi^9 G}\, ,  &&\sigma^{(5)}_{\rm E}=\frac{75699L^2}{1048576 \pi^{11} G}\, .
\end{align}
%and
%\begin{align}
%& \sigma_{\mathcal{L}_2^{(3)}}=[1-12\beta_2] \sigma_{\rm E}\, ,   \quad  &&\sigma'_{\mathcal{L}_2^{(3)}}=[1-6\beta_2]\sigma'_{\rm E}\, , \\  &\sigma''_{\mathcal{L}_2^{(3)}}=\left[1-\frac{189}{37}\beta_2\right] \sigma''_{\rm E}\, , \quad &&\sigma'''_{\mathcal{L}_2^{(3)}}=\left[1-\frac{414}{65}\beta_2\right] \sigma'''_{\rm E}\, , \\
%& \sigma^{(4)}_{\mathcal{L}_2^{(3)}}=\left[1-\frac{24896}{3133}\beta_2\right] \sigma^{(4)}_{\rm E}\, .
% \\  &\sigma^{(4)}_{\rm E}=\frac{9399L^2}{131072 \pi^9 G}\, ,  &&\sigma^{(5)}_{\rm E}=\frac{75699L^2}{1048576 \pi^{11} G}\, .
%\end{align}
%\begin{align*}
%& \sigma_{\mathcal{L}_3^{(3)}}=[1+..\beta_3] \sigma_{\rm E}\, ,   \quad  &&\sigma'_{\mathcal{L}_3^{(3)}}=[1+..\beta_3]\sigma'_{\rm E}\, , \\  &\sigma''_{\mathcal{L}_3^{(3)}}=\left[1+..\beta_3\right] \sigma''_{\rm E}\, , \quad &&\sigma'''_{\mathcal{L}_3^{(3)}}=\left[1+..\beta_3\right] \sigma'''_{\rm E}\, .
% \\  &\sigma^{(4)}_{\rm E}=\frac{9399L^2}{131072 \pi^9 G}\, ,  &&\sigma^{(5)}_{\rm E}=\frac{75699L^2}{1048576 \pi^{11} G}\, .
%\end{align*}
%\begin{align*}
%& \sigma_{\mathcal{L}_4^{(3)}}=[1+..\beta_4] \sigma_{\rm E}\, ,   \quad  &&\sigma'_{\mathcal{L}_4^{(3)}}=[1+..\beta_4]\sigma'_{\rm E}\, , \\  &\sigma''_{\mathcal{L}_4^{(3)}}=\left[1+..\beta_4\right] \sigma''_{\rm E}\, , \quad &&\sigma'''_{\mathcal{L}_4^{(3)}}=\left[1+..\beta_4\right]\sigma'''_{\rm E}\, .
% \\  &\sigma^{(4)}_{\rm E}=\frac{9399L^2}{131072 \pi^9 G}\, ,  &&\sigma^{(5)}_{\rm E}=\frac{75699L^2}{1048576 \pi^{11} G}\, .
%\end{align*}
Just like $\sigma$ is controlled by the stress-tensor two-point coefficient $\ctt$ for general theories, it is tempting to speculate that $\sigma'$ may be controlled by the stress-tensor three-point coefficients, which for $d=3$ CFTs can be chosen to be $\ctt$ and an additional dimensionless coefficient, customarily denoted $t_4$ \cite{Hofman:2008ar}. This possibility was pointed out in \cite{Miao:2015dua} and explored in \cite{Bueno:2015ofa}. There, using the available results for free fields and holographic Einstein gravity it was shown that $\sigma'$ was not a linear combination of $\ctt$ and $\ctt t_4$ in general. Using the results obtained in \cite{Li:2019auk} for $t_4$ for general cubic higher-curvature theories, we verify that this is not the case either for this class of theories. In the opposite limit, we find
\begin{equation}
\kappa_{\rm Riem^3}=\left[1+\frac{69}{5}\beta_1-\frac{42}{5}\beta_2+6\beta_3+24\beta_4+27\beta_5+27\beta_6+108\beta_7+432\beta_8\right] \kappa_{\rm E}\, .
\end{equation}
Obviously, the coefficients for $\mathcal{L}_{i}^{(3)}$ with $i=3,\dots,8$ are the same as those appearing in $\sigma_{\rm Riem^3}$, but that is not the case for $\mathcal{L}_{1}^{(3)}$ and $\mathcal{L}_{2}^{(3)}$. On the other hand, as expected on general grounds \cite{Myers:2012vs,Bueno2}, $\kappa_{\rm Riem^3}$ matches the coefficient of the slab EE computed above ---compare with \req{kaappa} for $d=3$.

We would like to perform some more comparisons of our new corner functions. For the sake of conciseness, from now on we restrict the discussion to Einsteinian cubic gravity, whose Lagrangian we introduced in \req{cubic2}. %This theory was originally discovered as the simplest higher-curvature extension to Einstein gravity which only propagates the usual transverse and traceless graviton on maximally symmetric backgrounds in general dimensions while being nontrivial in four (bulk) dimensions ---recall that all higher-curvature Lovelock densities \cite{Lovelock1,Lovelock2}  are either topological or trivial in four dimensions. Later on \cite{Hennigar:2016gkm,PabloPablo2}, it was realized that four-dimensional ECG also admitted a non-hairy generalization of the Schwarzschild black hole characterized by a single metric function, $g_{tt}g_{rr}=-1$, controlled by a second-order differential equation. These properties are characteristic of a broader class of theories which, besides ECG, also includes Lovelock \cite{Lovelock1,Lovelock2} and Quasi-topological gravities \cite{Quasi2,Quasi,Dehghani:2011vu,Cisterna:2017umf} as particular cases, and which goes by the name of {\it Generalized Quasi-topological gravities} \cite{Hennigar:2017ego,Bueno:2017sui,Bueno:2019ycr}. 
\begin{figure}[t] \vspace{-0.1cm} \centering
	\includegraphics[scale=0.61]{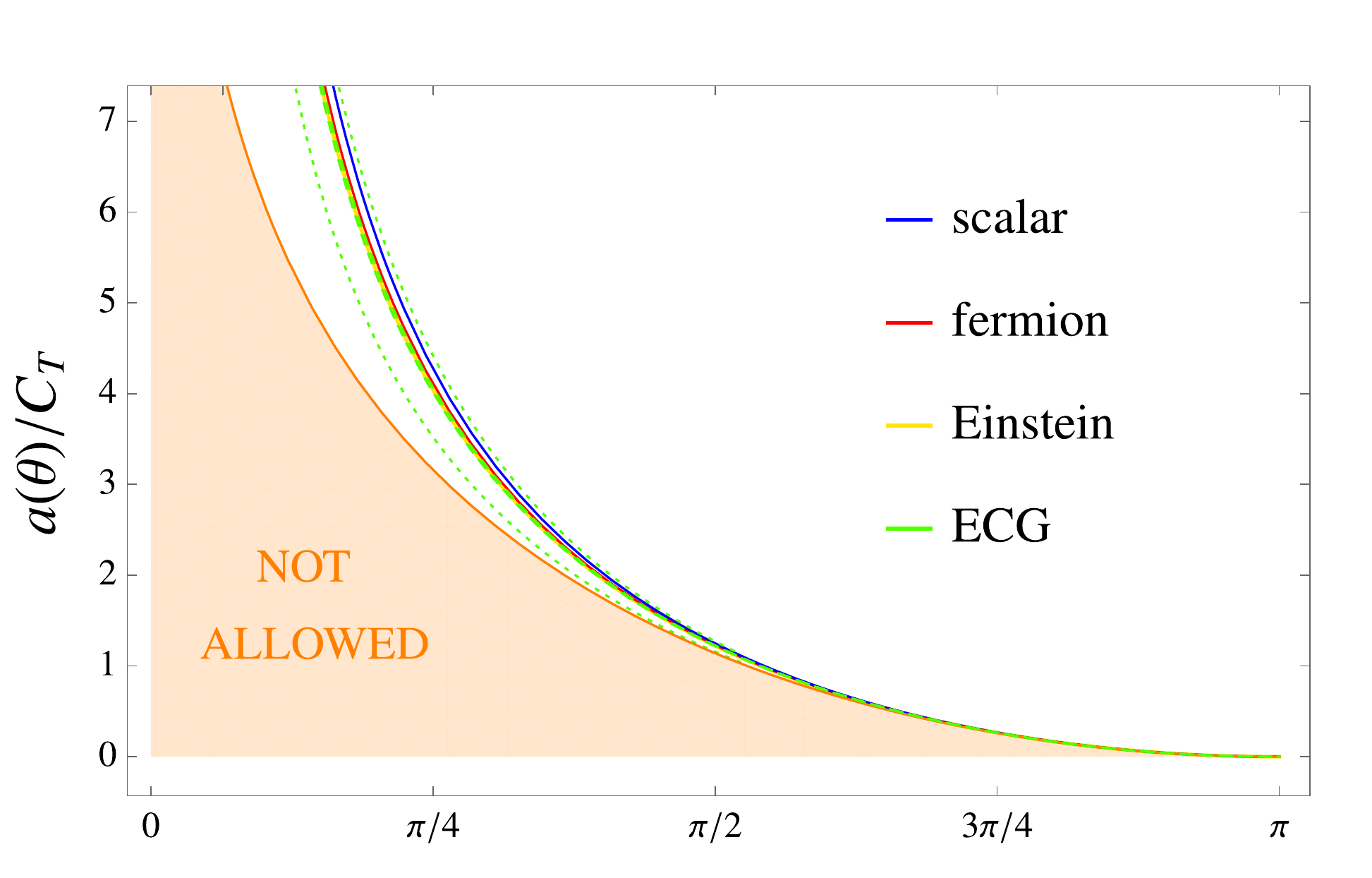} \includegraphics[scale=0.42]{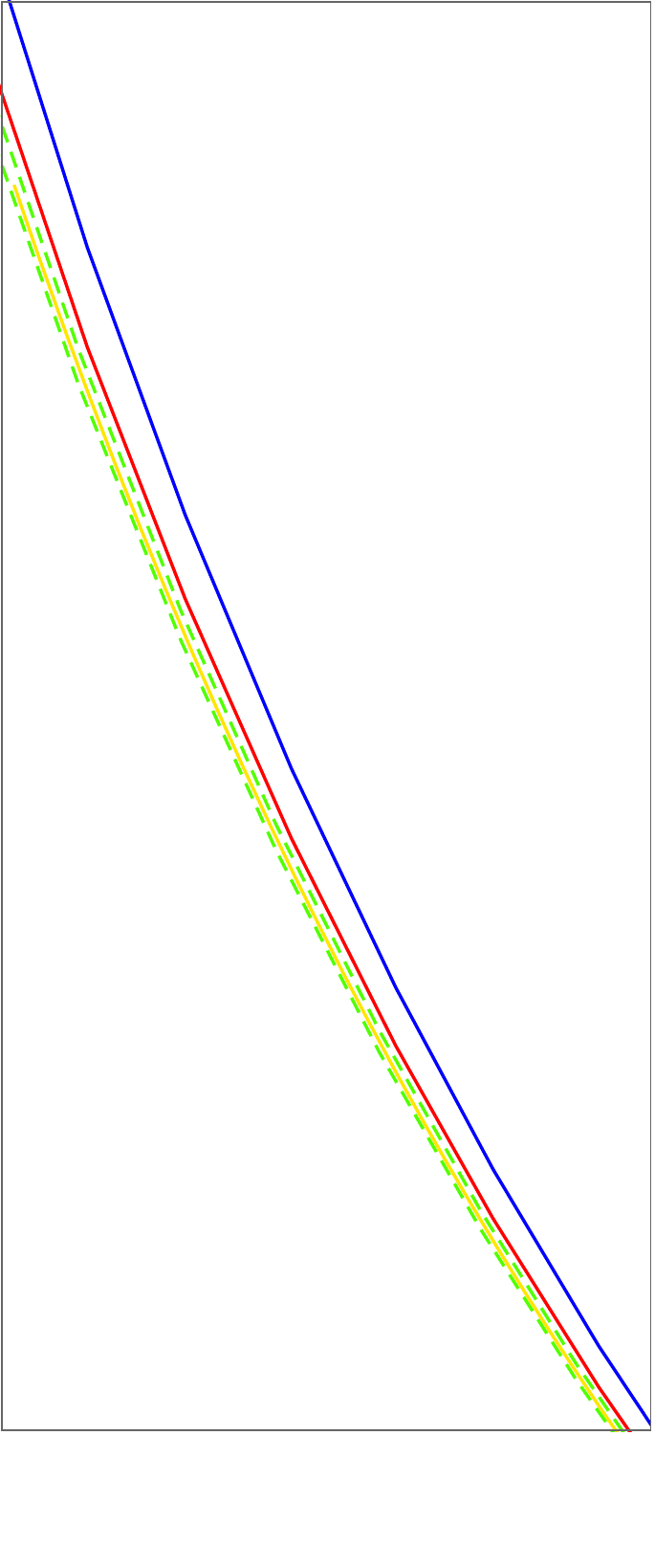}
	\caption{We plot the corner functions (normalized by their respective charges $\ctt$) for a free scalar (blue), a free fermion (red), holographic Einstein gravity (yellow) and holographic Einsteinian cubic gravity (green). For the limit value $\mu \simeq 0.00312$ corresponding to $t_4 =+4$ (see discussion below), the curve lies very close but slightly below the Einstein gravity result (green dashed line). The case $\mu \simeq -0.00322$ corresponding to the other limit value ($t_4=-4$) lies even closer but slightly above the Einstein gravity curve and just below the fermion one. The right plot is a zoom of the curves between $\theta=\pi/4$ and $\theta=3\pi/8$. The orange region in the left plot is excluded for general theories by the inequality \req{amin}.  The green dotted curves correspond to the values $\mu=-0.05$ (upper curve) and $\mu=+0.05$ (lower curve) which we have included (only) in the left plot for visual reference.    }
	\label{corn}
\end{figure}
% The ECG Lagrangian in four dimensions is given by \cite{PabloPablo}
%\begin{equation}\label{cubic2}
%I_{\rm  ECG}= \int \frac{d^{4}x \sqrt{|g|}}{16\pi G} \left[\frac{6}{L^2}+R- \frac{\mu L^4}{8} \mathcal{P} \right]\, , \quad \text{where} \quad \mathcal{P}\equiv  12 \mathcal{L}_{1}^{(3)}+\mathcal{L}_{2}^{(3)}-12\mathcal{L}_{5}^{(3)}+8\mathcal{L}_{6}^{(3)} \, .
%\end{equation}
We find the corner function for this theory to be given by
\begin{align}
a_{\rm  ECG}(\theta)&=(1+3\mu )a_{\rm E}(\theta)-\frac{\mu \Ls^2}{2G} \int_0^{\infty} \diff y \frac{3(1+h_0^2) (15+8h_0^4(1+y^2)^2+h_0^2(23+16y^2))}{4(1+h_0^2(1+y^2))^{7/2}\sqrt{2+h_0^2(1+y^2)}}\, .
\end{align}
This can be also written in terms of elliptic functions as
\begin{align}
a_{\rm  ECG}(\theta)&=(1+3\mu )a_{\rm E}(\theta)- \frac{ \mu \Ls^2}{40Gh_0\sqrt{1+h_0^2}} \left[(-51-51h_0^2+8h_0^4) \mathbb{E} \left[-\frac{1}{1+h_0^2} \right] \right. \\ \notag & \quad \quad \quad \quad \quad \quad \quad \quad \quad \quad \quad  \quad \quad \quad \quad \quad \,    \left. +(51+47h_0^2-8h_0^4) \mathbb{K}\left[-\frac{1}{1+h_0^2} \right] \right] \, .
\end{align}
The first smooth-limit coefficients and the sharp-limit one read in this case
\begin{align}
& \sigma_{\rm  ECG}=[1-3\mu] \sigma_{\rm E}\, ,   \quad  &&\sigma'_{\rm ECG}=\left[1-\frac{33}{4}\mu \right]\sigma'_{\rm E}\, , \\  &\sigma''_{\rm  ECG}=\left[1-\frac{2673}{296}\mu\right] \sigma''_{\rm E}\, , \quad &&\sigma'''_{\rm  ECG}=\left[1-\frac{2061}{260}\mu\right] \sigma'''_{\rm E}\, , \\
& \sigma^{(4)}_{\rm  ECG}=\left[1-\frac{41023}{6266}\mu\right] \sigma^{(4)}_{\rm E}\, , \quad &&\kappa_{\rm ECG}=\left[1-\frac{123}{20}\mu \right]\kappa_{\rm E}\, .
% \\  &\sigma^{(4)}_{\rm E}=\frac{9399L^2}{131072 \pi^9 G}\, ,  &&\sigma^{(5)}_{\rm E}=\frac{75699L^2}{1048576 \pi^{11} G}\, .
\end{align}
The positivity of these coefficients impose the bound $\mu\leq  0.1107$ (coming from $\sigma''_{\rm  ECG}\geq 0$). However, as shown in \cite{HoloECG}, the general bounds on the stress-tensor three-point function coefficient $-4\leq t_4 \leq 4$ \cite{Buchel:2009sk} impose more severe constraints on the allowed values of $\mu$, namely, $-0.00322 \leq \mu \leq 0.00312$. In the perturbative analysis performed in the present paper, bounds on finite values of $\mu$ are not so relevant, but we can use them to give us an idea of how much it is sensible to deviate $\mu$ from zero when performing comparisons with other theories. In Fig.\,\ref{corn} we have plotted $a_{\rm  ECG}(\theta)$ for the limiting values $\mu\simeq-0.00322$ and $\mu\simeq 0.00312$ (all intermediate values of $\mu$ lie between the two curves) along with the Einstein gravity result and the free scalar ($t_4=+4$) and free fermion ($t_4=-4$) ones \cite{Casini:2006hu,Casini:2008as,Casini:2009sr}. We can see that all curves are remarkably close to each other, in agreement with the observation/conjecture of  \cite{Bueno1} that $a(\theta)/\ctt$ is an almost-universal quantity for general CFTs. We observe this to be the case for the whole family of theories parametrized by the continuous parameter $\mu$ lying between the limiting cases extremizing the value of $t_4$.  By making the values of $|\mu|$ greater, we can obtain curves which deviate more significantly from the Einstein and free-field curves (see dotted lines in Fig.\,\ref{corn}). However, those would correspond to toy models of CFTs which do not respect the general bounds $|t_4|\leq 4$. Hence, it is reasonable to expect that for actual CFTs the curves will indeed fall extremely close to each other in general. In fact, the ECG curves with $t_4=4$ and $t_4=-4$ lie even closer to the Einstein gravity one than the scalar and fermion curves do. This suggests that the scalar field curve may be an upper bound for general CFTs. 

On the other hand, the possibility that the Einstein gravity curve is a lower bound for general curves suggested in \cite{Bueno1} seems to be ruled out by our analysis: the introduction of higher-curvature corrections allows to go below the Einstein gravity one.\footnote{The same conclusion was previously reached in \cite{Miao:2015dua}.} Note that such conjecture was also supported by the fact that while $t_4=0$ for Einstein gravity, both the scalar and the fermion curves ---which have, respectively, the largest positive and negative values of $t_4$ allowed--- lie above it. Here we observe that, contrary to the scalar case, ECG theories with $t_4\geq 0$ lie below the Einstein gravity one.

\begin{figure}[t] \vspace{-0.1cm} \centering
	\includegraphics[scale=0.7]{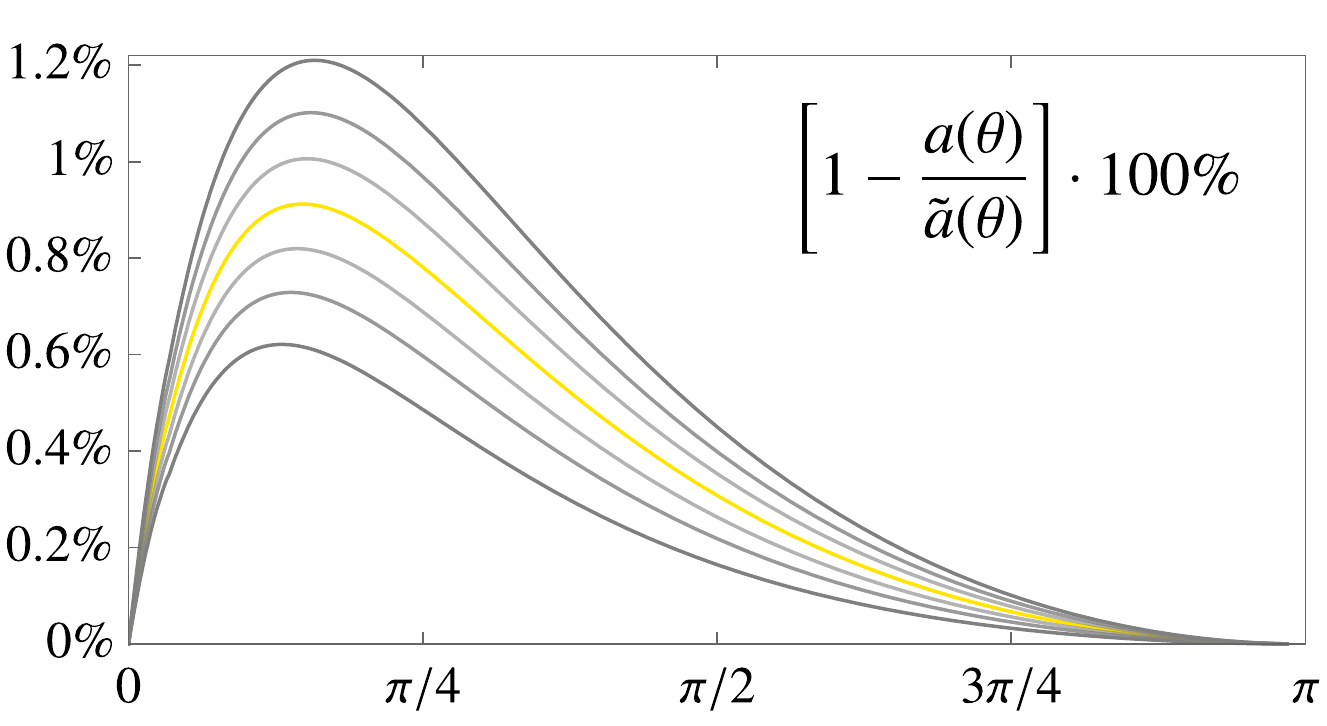} 
	\caption{We plot $1-a(\theta)/\tilde a(\theta)$ where $a(\theta)$ is the exact corner function and $\tilde a(\theta)$ the trial function defined in \req{trii} for Einstein gravity (yellow) and ECG for different values of $\mu$ (from top to bottom: $\mu=+0.00312,\,+ 0.002,\,+ 0.001,\, -0.001,\, -0.002,\, -0.00322$). The disagreement between both functions is always smaller than $\sim 1.2\%$ throughout the whole range of values of the opening angle.    }
	\label{corn2}
\end{figure}

In the previous subsection, we mentioned the possibility of approximating the function $a(\theta)$ for a given theory using the values of the almost-smooth and very-sharp limit coefficients, $\sigma$ and $\kappa$. The proposed trial function $\tilde a(\theta)$ appears in \req{trii}. We can use the new ECG corner functions to test the accuracy of such approximation beyond the free-field and Einstein cases explored in \cite{Bueno3}. In Fig.\,\ref{corn2}, we plot $1-a(\theta)/\tilde a(\theta)$ for various values of the ECG coupling falling between the limiting cases of $t_4=\pm4$. We observe that in all cases, the error in the approximation never exceeds $\sim 1.2\%$ for any value of the opening angle, the approximation being slightly better for negative values of $\mu$. This provides good evidence that $\tilde a(\theta)$ can be used as an accurate approximation to the exact corner function for general CFTs.

%\comment{ECG, trial function+ limits}

%\comment{in holography what is known}\\

%\comment{Here, first explicit HEE calculations beyond quadratic/Lovelock cases. Hopefully, first non-trivial corner entanglement function which is not Einstein, free fermion or free scalar. We can play with this a bit. We also perform several checks of the validity of the cubic functionals }

\section{Final comments}\label{finalc}
The main results of the paper appear summarized in the introduction. Let us conclude with some final comments.

In this paper we have obtained a new formula for the HEE functional valid for general higher-curvature gravities when considered as perturbative corrections to Einstein gravity ---the covariant form of the new expression appears in \req{Covariant:FullFunctional}. This formula, which gets rid of the weighted sum over $\alpha$ present in the original functional (\ref{SplittingProblem:GeneralFunctional}), is computationally much simpler to use in concrete cases beyond cubic order, and allowed us to evaluate the explicit form of the functionals for general quartic densities. If desired, it should be possible to implement it in a mathematical software and compute the analogous expressions for even higher orders. 

Besides its computational simplicity, the new form of the anomaly piece can be suggestively written in terms of the exponential of a differential operator ---this is particularly neat for Lovelock theories, see \req{OneComponent:LovelockExponential}. This form may be useful for potential applications beyond HEE, which may include new versions of the second law for higher-curvature black holes, \eg along the lines of \cite{Wall:2015raa,Bhattacharyya:2016xfs}.

As we have emphasized throughout the paper, the fact that our new expression is restricted to perturbative higher-curvature theories beyond quadratic order is related to the splitting problem, which requires the identification of the precise way in which Riemann tensor components must be decomposed into pieces of different weight $q_{\alpha}$ in the original functional for a given theory. While this could be in principle determined using the procedure developed in \cite{Dong:2017xht} on a theory by theory basis,\footnote{To the best of our knowledge, this has not been done explicitly for any non-trivial higher-curvature theory yet.} general results can be obtained at leading order in the couplings  by considering the splittings corresponding to Einstein gravity, which has been our approach in this paper. Nonetheless, we would like to stress that, in fact, our formalism should be straightforwardly adaptable to situations in which the Riemann tensor components split in a different fashion. In that case, instead of the separation into type $A$ and $B$ components one may have to introduce additional types $C$, $D$, etc., depending on the different possible weights corresponding to the different split components.  One could even think of a sort of general-splitting version of our formulas.

In Section \ref{unite} we have used our new expressions for cubic theories to evaluate several universal contributions to the EE characterizing the holographic CFTs they define. An analogous catalogue of coefficients could be obtained for quartic theories using the functionals presented in subsection \ref{quarticc}. Naturally, there are many possible additional applications within the HEE framework one could consider exploring using the new functionals presented here.

%\comment{  In the last section, universal coefficients characterizing CFT in various dimensions, an analogous catalogue of coefficients could be obtained for quartic theories using the functionals presented here... At that order, new modifications of the corner function dependence on the opening angle ... Obviously, there are many possible additional applications within the HEE framework that one can consider and for which the new functionals could be used...   }

\section*{Acknowledgements}
We thank Felix Haehl, Rong-Xin Miao, Rob Myers and  William Witczak-Krempa for useful discussions on related topics. PB and JC were supported by the Simons Foundation through the ``It From Qubit'' Simons collaboration. The work of AVL is supported by the Spanish MECD fellowship FPU16/06675, and by MINECO FPA2017-84436-P, Xunta de Galicia ED431C 2017/07, Xunta de Galicia (Centro singular de investigaci\'on de Galicia accreditation 2019-2022) and the European Union (European Regional Development Fund-ERDF), ``Mar\'ia de Maeztu'' Units of Excellence MDM-2016-0692, and the Spanish Research State Agency.

\appendix

\section{Proof of identities\,(\ref{NewFunctional:SumLambda1Expression}) and (\ref{NewFunctional:SumLambda2Expression}).}
\label{formuls}
In this appendix we present short proofs of the relations \req{NewFunctional:SumLambda1Expression} and  \req{NewFunctional:SumLambda2Expression} used in the derivation of the new HEE functional formula.

For the first, we want to show that
\begin{equation}
\sum_{\lambda=0}^{\tilde{T}} \frac{(-1)^{\lambda}}{(\lambda + n)} \frac{1}{\lambda! (\tilde{T} - \lambda)!} =\frac{(n-1)!}{(\tilde{T}+n)!}\, .
\end{equation}
Step by step, we have
\begin{align} \label{NewFunctional:SumLambda1Expression3}
\sum_{\lambda=0}^{\tilde{T}} \frac{(-1)^{\lambda}}{(\lambda + n)} \frac{1}{\lambda! (\tilde{T} - \lambda)!} & = \sum_{\lambda=0}^{\tilde{T}} \frac{(-1)^{\lambda}}{(\lambda + n)! (\tilde{T} - \lambda)!} (\lambda + n -1) \cdots (\lambda + 1) \\
\nonumber & = \frac{1}{(\tilde{T}+n)!} \partial_x^{n-1} \left[ \sum_{\lambda=0}^{\tilde{T}}  \binom{\tilde{T} + n}{\lambda + n} (-1)^{\lambda} x^{\lambda + n - 1} \right]_{x=1} \\
\nonumber & = \frac{1}{(\tilde{T}+n)!} \partial_x^{n-1} \left[ \sum_{\lambda=n}^{\tilde{T}+n}  \binom{\tilde{T} + n}{\lambda} (-1)^{\lambda-n} x^{\lambda - 1} \right]_{x=1} \\
\nonumber & = \frac{1}{(\tilde{T}+n)!} \partial_x^{n-1} \left[ \sum_{\lambda=n}^{\tilde{T}+n}  \binom{\tilde{T} + n}{\lambda} (-1)^{\lambda-n} x^{\lambda - 1} \right]_{x=1} \\
\nonumber & = \frac{(-1)^n}{(\tilde{T}+n)!} \partial_x^{n-1} \left[ \frac{1}{x} \sum_{\lambda=0}^{\tilde{T}+n} \binom{\tilde{T} + n}{\lambda} (-x)^{\lambda} - \frac{1}{x} \binom{\tilde{T}+n}{0} \right. \\ \nonumber & \quad \quad \left. - \dots - \binom{\tilde{T}+n}{n - 1} (-x)^{n-2}  \right]_{x=1} \\
 & = \frac{(-1)^n}{(\tilde{T}+n)!} \partial_x^{n-1} \left[ \frac{(1-x)^{\tilde{T}+n}}{x} - \frac{1}{x}\right]_{x=1} \\
 &  = \frac{(n-1)!}{(\tilde{T}+n)!} ~ ,
\end{align}
where the first term inside the brackets does not survive after $(n-1)$ derivatives evaluated at $x = 1$ because of the factor $(1-x)^{\tilde{T}+n}$, and we have used:

\begin{equation}
\partial_x^{n-1} \left( \frac{1}{x} \right) = \frac{(-1)^{n-1} (n-1)!}{x^n} ~ .
\end{equation}

As for the second identity, we want to show that
\begin{equation}\label{dsf}
\sum_{\lambda=0}^{T} \frac{(-1)^{\lambda}}{\lambda! (T - \lambda)!} \frac{(2 \lambda + 1 )!}{(2 \lambda + m)!}= \frac{1}{T! (m-2)!} \int_0^1 {\rm d} z \, z (1 - z^2)^T (1-z)^{m-2}\, .
\end{equation}
In hopefully self-evident steps we find
\begin{align} \label{NewFunctional:SumLambda2Expression2}
 \sum_{\lambda=0}^{T} \frac{(-1)^{\lambda}}{\lambda! (T - \lambda)!} \frac{(2 \lambda + 1 )!}{(2 \lambda + m)!} & = \sum_{\lambda=0}^{T} \frac{(-1)^{\lambda}}{\lambda! (T - \lambda)!} \frac{1}{(2 \lambda + m) \cdots (2\lambda + 2)} \\
\nonumber & = \sum_{\lambda=0}^{T} \frac{(-1)^{\lambda}}{\lambda! (T - \lambda)!} \int_0^1 {\rm d}x_1 \int_0^{x_1} {\rm d} x_2 \dots \int_0^{x_{m-2}} {\rm d} x_{m-1} x_{m-1}^{2 \lambda + 1} \\
\nonumber & = \int_0^1 {\rm d}x_1 \int_0^{x_1} {\rm d} x_2 \dots \int_0^{x_{m-2}} {\rm d} x_{m-1} \, x_{m-1} \sum_{\lambda=0}^{T} \frac{(-x_{m-1}^2)^{\lambda}}{\lambda! (T - \lambda)!} \\
\nonumber & = \frac{1}{T!} \int_0^1 {\rm d}x_1 \int_0^{x_1} {\rm d} x_2 \dots \int_0^{x_{m-3}} {\rm d} x_{m-2} \int_0^{x_{m-2}} {\rm d} z \, z (1 - z^2)^T ~ ,
\end{align}
where we have relabelled $x_{m-1} \equiv z$. We can reorder the integrals now using the following identity:
\begin{align}
&\int_0^1 {\rm d}x_1 \int_0^{x_1} {\rm d} x_2 \dots \int_0^{x_{m-3}} {\rm d} x_{m-2} \int_0^{x_{m-2}} {\rm d} z \, f(z) \\ \notag &= \int_0^1 {\rm d} z \, f(z) \int_z^{1} {\rm d}x_1 \int_z^{x_1} {\rm d} x_2 \dots \int_z^{x_{m-3}} {\rm d} x_{m-2} \, ,
\end{align}
%
%\comment{MAYBE INCLUDE PROOF?}
 and then use
\begin{equation*}
\int_z^{1} {\rm d}x_1 \int_z^{x_1} {\rm d} x_2 \dots \int_z^{x_{m-3}} {\rm d} x_{m-2} = \frac{1}{(m-2)!} \left( \int_z^1 {\rm d}x \right)^{m-2} = \frac{(1-z)^{m-2}}{(m-2)!} ~,
\end{equation*}
to finally obtain \req{dsf}.
%
%\begin{equation} \label{NewFunctional:SumLambda2FinalForm}
%\nonumber \sum_{\lambda=0}^{T} \frac{(-1)^{\lambda}}{\lambda! (\tilde{T} - \lambda)!} \frac{(2 \lambda + 1 )!}{(2 \lambda + m)!} = \frac{1}{T! (m-2)!} \int_0^1 {\rm d} z \, z (1 - z^2)^T (1-z)^{m-2} ~ .
%\end{equation}

%\renewcommand{\leftmark}{\MakeUppercase{Bibliography}}
%\phantomsection
\bibliographystyle{JHEP}
\bibliography{Gravities}
%\label{biblio}

\end{document}

%%It turns out that if one performs the radial integrals first, they can both be done exactly.
%Further, this way one gets a log(L/delta) directly. After that, 3 angle integrals remain. Mathematica doesn't
%seem to be able to perform them. I courageously Taylor expanded the integrand..
%Here's my tentative answer for the cone function of the EMI in d=5:
%        cos(W) cot(W) [ 113 + cos(2W) ]
%I'm not yet certain the Taylor expansions used to obtain it are fully justified, but it's currently my best guess.

%As a check, the expansion near W=\pi/2 yields only even powers. Further as W--> 0, I find
%\kappa/W - \kappa' W + ?
%Actually, it seems like all CFTs have a -\kappa' W subleading term, never anything else.
%It would be worthwhile to raise and investigate this question in our paper with Rongxin, aka the "sharp limit expansion".